\documentclass[11pt]{article}
\pdfoutput=1
\usepackage{putex}
\usepackage[vcentermath]{youngtab}
\usepackage{comment}

\usepackage{graphicx}
\usepackage{epstopdf}
\usepackage{enumerate}
\usepackage{cite}
\usepackage{tensor}
\usepackage{slashed}
\usepackage{feynmf}
\usepackage{multirow}
\usepackage{hyperref}
\usepackage{soul}
\usepackage[font=footnotesize,labelfont=bf]{caption}
\usepackage[table,xcdraw]{xcolor}
\numberwithin{equation}{section}
\usepackage[T1]{fontenc}
\usepackage{lscape}
\usepackage[11pt]{moresize}
\usepackage{ytableau}

\newcommand {\be} {\begin {equation}}
\newcommand {\ee} {\end {equation}}

\newcommand {\bes} {\begin {equation*}}
\newcommand {\ees} {\end {equation*}}

\newcommand{\dph}{\Delta^{\rm ph}}

\def\CH{{\cal H}}

\newcommand{\pd}[2]{\frac{\partial #1}{\partial #2}}

\newcommand{\nonn}{\nonumber}

\newcommand{\beq}{\begin{equation}}
\newcommand{\eeq}{\end{equation}}

\def\be{ \begin{equation} }
\def\ee{ \end{equation} }

\allowdisplaybreaks 
\begin{document}

\preprint{PUPT-2505}

\institution{PU}{Department of Physics, Princeton University, Princeton, NJ 08544, USA}
\institution{PCTS}{Princeton Center for Theoretical Science, Princeton University, Princeton, NJ 08544, USA}
\institution{ENS}{D\'epartement de Physique, \'Ecole Normale Sup\'erieure, 45 rue d'Ulm, 75005 Paris, France}

\title{
The ABC of Higher-Spin AdS/CFT
}

\authors{Simone Giombi,\worksat{\PU} Igor R.~Klebanov\worksat{\PU,\PCTS} and Zhong Ming Tan\worksat{\PU,\ENS}
}

\abstract{
In recent literature one-loop tests of the higher-spin AdS$_{d+1}$/CFT$_d$ correspondences were carried out. 
Here we extend these results to a more general set of theories in $d>2$.
First, we consider the Type B higher spin theories, which have been conjectured to be dual to 
CFTs consisting of the singlet sector of $N$ free fermion fields. In addition to the case of $N$ Dirac fermions, we carefully study the projections to Weyl, Majorana, symplectic, and Majorana-Weyl fermions in the dimensions where they exist.
Second, we explore theories involving elements of both Type A and Type B theories, which we call Type AB. Their spectrum includes fields of every half-integer spin, and they are expected to be related to the $U(N)/O(N)$ singlet sector of the CFT of $N$ free complex/real scalar and fermionic fields. Finally, we explore the Type C theories, which have been conjectured to be dual to the CFTs of $p$-form gauge fields, 
where $p=\frac d 2 -1$. In most cases 
we find that the free energies at $O(N^0)$ either vanish or give contributions proportional to the free-energy of a single free field in the conjectured dual CFT. Interpreting these non-vanishing values as shifts of the bulk coupling constant $G_N\sim 1/(N-k)$, we find the values $k=-1, -1/2, 0, 1/2, 1, 2$. 
Exceptions to this rule are the Type B and AB theories in odd $d$;
for them we find a mismatch between the bulk and boundary free energies that has a simple structure, but does not follow from a simple
shift of the bulk coupling constant.
}
\date{\today}
\maketitle

\tableofcontents

\section{Introduction}
Extensions of the original AdS/CFT correspondence~\cite{Maldacena:1997re,Gubser:1998bc,Witten:1998qj} to 
relations between the ``vectorial'' $d$-dimensional CFTs and the Vasiliev higher-spin theories in $(d+1)$-dimensional AdS space \cite{Fradkin:1987ks,Vasiliev:1990en,Vasiliev:1992av,Vasiliev:2003ev,Vasiliev:2004cm} have attracted considerable attention (for recent 
reviews of the higher-spin AdS$_{d+1}$/CFT$_{d}$ correspondence see \cite{Giombi:2012ms,Giombi:2016ejx}). 
The CFTs in question are quite well understood; their examples include the singlet sector of the free $U(N)/O(N)$ symmetric theories where the 
dynamical fields are in the vectorial representation (rather than in the adjoint representation), or of the vectorial interacting CFTs such as the $d=3$ Wilson-Fisher and Gross-Neveu models \cite{Klebanov:2002ja, Sezgin:2003pt,Leigh:2003gk}. 
Some years ago the singlet sectors of $U(N)/O(N)$ symmetric $d$-dimensional CFTs of scalar fields were conjectured to be dual to the Type A Vasiliev theory in AdS$_{d+1}$ \cite{Klebanov:2002ja},
while the CFTs of fermionic fields -- to the type B Vasiliev theory \cite{Sezgin:2003pt,Leigh:2003gk}. 
In $d=3$ the $U(N)/O(N)$ singlet constraint is naturally imposed by coupling the massless matter fields to the Chern-Simons gauge field \cite{Aharony:2011jz,Giombi:2011kc}. While the latter is in the
adjoint representation, it carries no local degrees of freedom so that the CFT remains vectorial.
More recently, a new similar set of dualities was proposed in even $d$ and called Type C \cite{Beccaria:2014xda,Beccaria:2014zma,Beccaria:2014qea};
it involves the CFTs consisting of some number $N$ of $\left (\frac d 2-1 \right )$-form gauge fields projected onto the $U(N)/O(N)$ singlet sector. 

The higher-spin AdS/CFT conjectures were tested through matching of three-point correlation functions of operators at order $N$, corresponding to tree level in the bulk ~\cite{Giombi:2009wh,Giombi:2012ms};
further work on the correlation functions includes \cite{Maldacena:2011jn,Maldacena:2012sf,Didenko:2012vh,Didenko:2012tv,Boulanger:2015ova,Bekaert:2015tva}. Another class of tests, which involves calculations
at order $N^0$, corresponding to the one-loop effects in the bulk, was carried out in \cite{Giombi:2013fka,Giombi:2014iua,Giombi:2014yra,Jevicki:2014mfa,Beccaria:2014xda,Beccaria:2014zma,Beccaria:2014qea}.
It concerned the calculation of one-loop vacuum energy in Euclidean AdS$_{d+1}$, corresponding to the sphere free energy $F=-\log Z_{S^d}$ in CFT$_{d}$; in even/odd $d$ this quantity enters the
$a$/$F$ theorems \cite{Zamolodchikov:1986gt,Cardy:1988cwa,Komargodski:2011vj,Myers:2010xs,Jafferis:2011zi,Klebanov:2011gs,Casini:2012ei}.
Similar tests using the thermal AdS$_{d+1}$, where the Vasiliev theory is dual to the vectorial CFT on $S^{d-1}\times S^1$, have also been conducted~\cite{Shenker:2011zf,Giombi:2014yra,Beccaria:2014xda,Beccaria:2014zma,Beccaria:2014qea}. Such calculations serve as a compact way of checking the agreement of the spectra in the two dual theories.
The quantities of interest are the formula for the thermal free energies at arbitrary temperature $\beta$, as well as the temperature-independent Casimir energy $E_c$.

In this paper we continue and extend the earlier work \cite{Giombi:2013fka,Giombi:2014iua,Giombi:2014yra,Jevicki:2014mfa,Beccaria:2014xda,Beccaria:2014zma,Beccaria:2014qea} on the one-loop tests of higher-spin AdS/CFT.
In particular, we will compare the Type B theories in various dimensions $d$ and their dual CFTs consisting of the Dirac fermionic fields (we also consider the theories with 
Majorana, symplectic, Weyl, or Majorana-Weyl fermions in the dimensions where they are admissible). Let us also comment on the Sachdev-Ye-Kitaev (SYK) 
model \cite{Sachdev:1992fk,Kitaev:2015}, which is a quantum mechanical theory of a large number
$N$ of Majorana fermions with random interactions; it has been attracting a great deal of attention recently 
\cite{Sachdev:2015efa,Polchinski:2016xgd,Jevicki:2016bwu,Maldacena:2016hyu,Maldacena:2016upp}. 
After the use of replica trick, this model has manifest $O(N)$ symmetry \cite{Sachdev:2015efa}, and
it is tempting to look for its gravity dual using some variant of type B higher spin theory. Following \cite{Almheiri:2014cka} one may 
speculate that the SYK model provides an effective IR description of 
a background of a type B Vasiliev theory asymptotic to AdS$_4$ which is dual to a theory of Majorana fermions; this background should describe RG flow from AdS$_4$ to $AdS_2$
(one could also search for RG flow from HS theory in AdS$_{d+1}$ to AdS$_2$ with $d=2, 4, \ldots$). 

Two other types of theories with no explicitly constructed Vasiliev equations are also explored. First, we consider the theories whose CFT duals are expected to consist of
both scalar and fermionic fields, with a subsequent projection onto the singlet sector. These theories, which we call of Type AB, are then expected to have half-integral spin gauge fields in addition to the integral spin gauge fields of Type A and Type B theories. Depending on the precise scalar and fermion field content, the Type AB theories may be supersymmetric in some specific dimension $d$. For example, the $U(N)$ singlet sector of one fundamental Dirac fermion and one fundamental complex scalar is supersymmetric in $d=3$, and a similar theory with one fundamental Dirac fermion and two fundamental complex scalars is supersymmetric in $d=5$.\footnote{This theory may be coupled to the $U(N)$ 5d Chern-Simons gauge theory to impose the singlet constraint.} Second, we study the Type C theories, where the CFT dual consists of some number of
$p$-form gauge fields, with $p=\frac d 2 -1$; the self-duality condition on the field strength may also be imposed. Such theories were studied in \cite{Beccaria:2014xda,Beccaria:2014zma,Beccaria:2014qea} for $d=4$ and $6$, and we extend them to more general dimensions.  



The organization of the paper is as follows.
In Section~\ref{sect:summary}, we review how the comparison of the partition functions of the higher-spin theory and the corresponding CFT allows us to draw useful conclusions about their duality.
We will also go through the various HS theories that will be examined in this paper. This will allow us to summarize our results  in Tables~\ref{Table:summary}, \ref{Table:summary2}, and \ref{Table:TypeBMin}.
In Section \ref{sect:ordinaryads}, we present our results for the free energy of Vasiliev theory in Euclidean AdS$_{d+1}$ space asymptotic to the round sphere $S^d$.
In addition, in Appendix~\ref{sect:thermalads}, we detail the calculations for the free energy of Vasiliev theory in the thermal AdS$_{d+1}$ space, which is asymptotic to $S^{d-1}\times S^1$.

{\bf Note Added:} Shortly before completion of this paper we became aware of independent forthcoming work on related topics by M. Gunaydin, E. Skvortsov and T. Tran 
\cite{Gunaydin:2016}.

\section{Review and Summary of Results}\label{sect:summary}
\subsection{Higher spin partition functions in Euclidean AdS spaces} 

According to the AdS/CFT dictionary, the CFT partition function $Z_{\rm CFT}$ on the round sphere $S^d$ has to match the partition function of the bulk theory 
on the Euclidean AdS$_{d+1}$ asymptotic to $S^d$. This is the hyperbolic space $\mathbb{H}_{d+1}$ with the metric,
$ds^2 = d\rho^2 + \sinh^2\!\rho\ d\Omega_d$,
where $d\Omega_d$ is the metric of a unit $d$-sphere. After defining the free energy $F=-\log Z$, the AdS/CFT correspondence implies $F_{\rm CFT} = F_{\rm bulk}$.

For a vectorial CFT with $U(N)$, $O(N)$ or $USp(N)$ symmetry, the large $N$ expansion is
\begin{align}
F_{\rm CFT} = Nf^{(0)}+f^{(1)}+\frac{1}{N} f^{(2)}+\ldots.
\end{align}
For a CFT consisting of $N$ free fields, one obviously has $f^{(n)}=0$ for all $n\geq 1$.

For the bulk gravitational theory with Newton constant $G_N$ the perturbative expansion of the free energy assumes the form
\begin{align}
F_{\rm bulk} = \frac{1}{G_N}F^{(0)}+F^{(1)}+G_N F^{(2)}+\ldots
\end{align}
The leading contribution is the on-shell classical action of the theory; it should match the leading term in the CFT answer which is of order $N$. 
Such a matching seems impossible at present due to the lack of a conventional action for the higher spin theories.\footnote{In the collective field approach to
the bulk theory the action does exist, and the matching of free energies works by construction \cite{Jevicki:2014mfa}. However, the precise connection of this formalism with the
Vasiliev equations remains an open problem.} 
%
However, as first noted in \cite{Giombi:2013fka}, 
the one-loop correction $F^{(1)}$ requires the knowledge of only the free quadratic actions for the higher-spin fields in AdS$_{d+1}$; it can be obtained by summing the 
logarithms of functional determinants of the relevant kinetic operators. The latter were calculated by
Camporesi and Higuchi \cite{Camporesi:1991nw,Camporesi:1993mz,Camporesi:1994ga, Camporesi:1994pf}, who derived the spectral zeta function for fields of arbitrary spin in (A)dS. 
What remains is to carry out the appropriately regularized sum over all spins present in a particular version of the higher spin theory. 


The corresponding sphere free energy in a free CFT is given by $F_{\rm CFT} = N F$, where $F$ may be extracted from the determinant 
for a single conformal field (see, for example, \cite{Klebanov:2011gs}); the examples of the latter are conformally coupled scalars, massless fermions, or $p$-form gauge fields.
For vectorial theories with double-trace interactions, such as the Wilson-Fisher and Gross-Neveu models, the CFT itself has a non-trivial $\frac{1}{N}$ expansion, and so
	$F_{\rm CFT} = N F + \mathcal{O}(N^0)$.
	To match the large $N$ scaling, the Newton constant of the bulk theory must behave as
	\begin{align}
		\frac{1}{G_N} \propto N,
	\end{align}
in the large $N$ limit.
If one assumes that 
	$\frac{1}{G_N}F^{(0)} = F_{\rm CFT}$, then all the higher-loop corrections to $F_{\rm bulk}$ must vanish for $F_{\rm CFT} = F_{\rm bulk}$ to hold.
In~\cite{Giombi:2014iua,Giombi:2013fka}, it was found that for the Vasiliev Type A theories in all dimensions $d$, the non-minimal theories containing each integer spin indeed
have a vanishing one-loop correction to $F$. However, the minimal theories with even spins only were found to have a non-vanishing one-loop contribution that matched exactly the value of the sphere free-energy of a single conformal real scalar. 
This surprising result was then interpreted as a one-loop shift
\begin{align}
\frac{1}{G_N} \sim N-1,
\end{align}
where the one-loop contribution cancels exactly the shift in the coupling constant. Such an integer shift is consistent with the quantization condition
for $\frac{1}{G_N}$ established in \cite{Maldacena:2011jn,Maldacena:2012sf}.
The rule $N\rightarrow N-1$ does not apply to all the
variants of the HS theory. In~\cite{Beccaria:2014zma,Beccaria:2014xda} it was shown that the one-loop calculations in Type C higher spin theories dual to free $U(N)/O(N)$ Maxwell fields in $d=4$ required that $\frac{1}{G_N} \sim N-1$ or $N-2$ respectively. If the Maxwell fields are taken to be self-dual then $\frac{1}{G_N} \sim N-1/2$; in view of this half-integer
shift it is not clear if such a theory is consistent. 

\subsection{Variants of Higher Spin Theories and Key Results}
\label{key-summary}

The simplest and best understood HS theory is the type A Vasiliev theory in AdS$_{d+1}$, which is known at non-linear level for any $d$ 
\cite{Vasiliev:2003ev}. The spectrum consists of a scalar with $m^2=-2(d-2)$ and a tower of totally symmetric HS gauge fields 
(in the minimal theory, only the even spins are present). This is in one to one correspondence with the 
spectrum of $O(N)$/$U(N)$ invariant ``single trace'' operators on the CFT side, which consists of the $\Delta=d-2$ scalar
\begin{equation}
J_0 = \phi_i^* \phi^i \\
\end{equation}
and the tower of conserved currents
\begin{equation}
J_{\mu_1\cdots\mu_s} = \phi^*_i\partial_{(\mu_1}\cdots\partial_{\mu_s)}\phi^i +\cdots,\qquad s\geq 1.
\end{equation}
This spectrum can be confirmed for instance by computing the tensor product of two free scalar representatios, 
which yields the result \cite{Flato:1980we,Vasiliev:2004cm,Dolan:2005wy}
\begin{equation}
\left(\frac{d}{2}-1;0\right)\otimes \left(\frac{d}{2}-1;0\right) = (d-2;0,\ldots,0)+\sum_{s=1}^{\infty} (d-2+s;s,0,\ldots,0)
\label{FF}
\end{equation}
where the notation $(\Delta;m_1,m_2,\ldots)$ indicates a representation of the conformal algebra with conformal dimension $\Delta$ and $SO(d)$ 
representation labelled by $[m_1,m_2,\ldots]$ (on the left-hand side, $(d/2-1;0)$ is a shorthand for $(d/2-1;0,\ldots,0)$). 
Equivalently, one may obtain the same result by computing the ``thermal'' partition function of the free CFT 
on $S^1\times S^{d-1}$, using a flat connection to impose the $U(N)$ singlet constraint \cite{Shenker:2011zf, Giombi:2014yra}. Similarly one 
can consider real scalars and $O(N)$ singlet constraint, where one obtains the same spectrum but with odd spins removed (this corresponds to 
symmetrizing the product in (\ref{FF})). 

Another version of the HS theory is the so-called ``type B'' theory, which is defined to be the HS gauge theory in AdS$_{d+1}$ dual to the 
free fermionic CFT$_d$ restricted to its 
singlet sector. The field content of such theories can be deduced from CFT considerations, 
by deriving the spectrum of singlet operators which are bilinears in the fermionic fields. In the case of Dirac fermions, one has the
following results for the tensor product of two free fermion representations \cite{Vasiliev:2004cm,Dolan:2005wy}: in even $d$
\begin{equation}
\begin{aligned}
&\left(\frac{d-1}{2};\frac{1}{2}\right) \otimes \left(\frac{d-1}{2};\frac{1}{2}\right)  =
2(d-1;0,\ldots,0)
+2\sum_{s=1}^{\infty}\left[(d-2+s;s,0,\ldots,0)+(d-2+s;s,1,0,\ldots,0)\right.\\
&~~~~~~~~~~~~~~\left.+(d-2+s;s,1,1,0,\ldots,0)+\ldots+(d-2+s;s,1,1,1,\ldots,1,0)\right.\\
&~~~~~~~~~~~~~~\left.+(d-2+s;s,1,1,\ldots,1,1)+(d-2+s;s,1,1,\ldots,1,-1)\right]
\label{psipsi-even}
\end{aligned}
\end{equation}
and in odd $d$
\begin{equation}
\begin{aligned}
&\left(\frac{d-1}{2};\frac{1}{2}\right) \otimes \left(\frac{d-1}{2};\frac{1}{2}\right)  =
(d-1;0,\ldots,0)+\sum_{s=1}^{\infty}\left[(d-2+s;s,0,\ldots,0)+\right.\\
&\left.+(d-2+s;s,1,0,\ldots,0)+\ldots+(d-2+s;s,1,1,\ldots,1,0)+(d-2+s;s,1,1,\ldots,1,1)\right]\,.
\label{psipsi-odd}
\end{aligned}
\end{equation}
Note that in the case $d=3$, the spectra of the type A and type B theory are the same, except for the fact that the $m^2=-2$ scalar is parity even 
in the former and parity odd in the latter (and also quantized with conjugate boundary conditions, $\Delta=1$ versus $\Delta=2$). In this special case, the fully non-linear equations for the type B HS gauge theory in AdS$_4$ are known 
and closely related to those of the type A theory \cite{Vasiliev:1992av}. For all $d>3$, however, the spectra of Type B theories differ considerably from Type A theories, since they contain towers of spins with various mixed symmetries, see (\ref{psipsi-even})-(\ref{psipsi-odd}), 
and the corresponding non-linear equations are not known. As an example, and to clarify the meaning of (\ref{psipsi-even})-(\ref{psipsi-odd}), let us 
consider $d=4$ \cite{Anselmi:1998bh,Anselmi:1999bb,Alkalaev:2012rg, Giombi:2014yra}. In this case, on the CFT side one can construct the two scalar operators
\begin{align}
J_0=\bar\psi_i \psi^i,\qquad\qquad \tilde J_0 = \bar\psi_i \gamma_5 \psi^i\ ,
\label{4d-B-sc}
\end{align}
as well as (schematically) the totally symmetric and traceless bilinear currents
 \begin{align}
 J_{\mu_1\cdots\mu_s} = \bar\psi_i \gamma_{(\mu_1}\partial_{\mu_2}\cdots\partial_{\mu_s)}\psi^i + \cdots,\qquad  \tilde{J}_{\mu_1\cdots\mu_s} = \bar\psi_i \gamma_5\gamma_{(\mu_1}\partial_{\mu_2}\cdots\partial_{\mu_s)}\psi^i + \cdots,\qquad s\geq 1\,,
\label{4d-B-sym}
\end{align}
and a tower of mixed higher symmetry bilinear current,
\begin{align}
M_{\mu_1\cdots\mu_s,\nu} = \bar\psi_i \gamma_{\nu(\mu_1}\partial_{\mu_2}\cdots\partial_{\mu_s)}\psi^i + \cdots, \qquad s\geq 1\,,
\label{4d-B-mixed}
\end{align} where $\gamma_{\nu\mu_1}=\gamma_{[\nu}\gamma_{\mu_1]}$ is the antisymmetrized product of the gamma matrices. 
These operators are dual to corresponding HS fields in AdS$_5$. In particular, in addition to two towers of Fronsdal fields and a tower of mixed symmetry gauge fields, 
there are two bulk scalar fields and a massive antisymmetric tensor dual to $\bar\psi_i\gamma_{\mu\nu}\psi^i$. Similarly, in higher dimensions 
one can construct the tower of mixed symmetry operators appearing in (\ref{psipsi-even})-(\ref{psipsi-odd}) by using the antisymmetrized product of several 
gamma matrices. In the Young tableaux notation, these operators correspond to the hook type diagrams
\begin{align}
\includegraphics[width=0.35\textwidth]{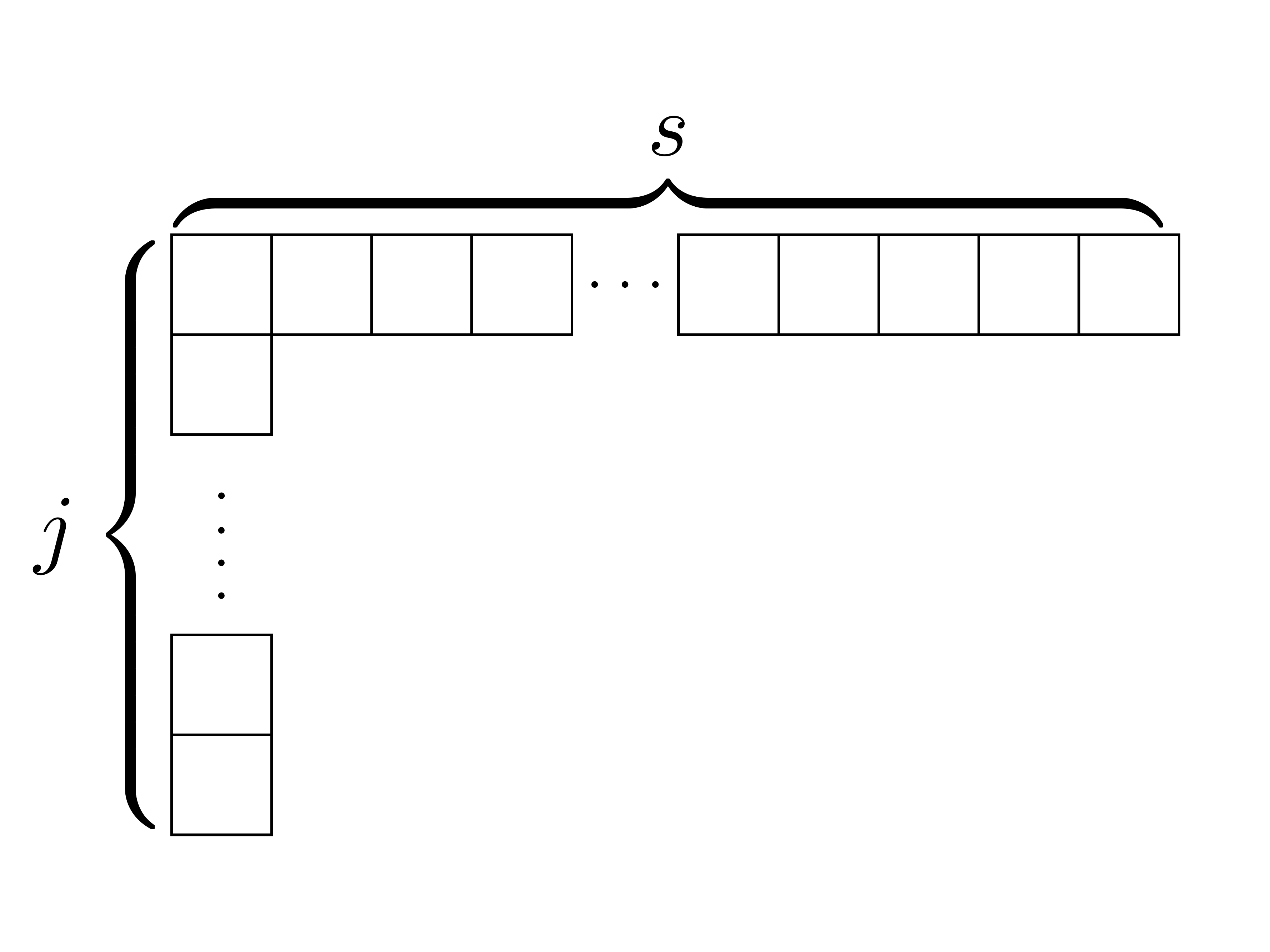}
\end{align}
where $1 < j\le p$, with $p=d/2$ for even $d$ and $p=(d-1)/2$ for odd $d$. For $s>1$, these operators are conserved currents and are dual to 
massless gauge fields in the bulk, while for $s=1$ they are dual to massive antisymmetric fields. 


For even $d$, we find evidence that the non-minimal Type B theory is exactly dual to the singlet sector of the $U(N)$ free fermionic CFT. 
The one-loop free energy of the Vasiliev theory vanishes exactly.
This generalizes the result given in~\cite{Beccaria:2014xda} for the non-minimal Type B theory in AdS$_5$; namely, there is no shift to the coupling constant in the non-minimal 
Type B theory dual to the singlet sector of Dirac fermions. 

However, for all odd $d$, the one-loop free energy does not vanish. Instead, it follows a surprising formula 
\begin{align}
F^{(1)}_{\rm type~B}=&-\frac{1}{\Gamma(d+1)}\int_0^{1/2}du\ u\sin(\pi u)\Gamma\left(\frac{d}{2}+u\right)\Gamma\left(\frac{d}{2}-u\right), 
\label{eqn:typebnonminsphere2}
\intertext{
which has an equivalent form for integer $d$}
F^{(1)}_{\rm type~B}=& \frac{1}{2\Gamma(d+1)}\int_0^1 du \cos\left(\pi u\right)\Gamma\left(\frac{d+1}{2}+u\right)\Gamma\left(\frac{d+1}{2}-u\right), \label{eqn:typebnonminsphere}
\end{align}
For example, for $d=3,5,7$ one finds 
\begin{equation}
\begin{aligned}
&F^{(1)}_{\rm type~B}=-\frac{\zeta(3)}{8\pi^2} \,,\qquad &d=3\,,\\
&F^{(1)}_{\rm type~B}=-\frac{\zeta(3)}{96\pi^2}-\frac{\zeta(5)}{32\pi^2} \,,\qquad &d=5\,,\\
&F^{(1)}_{\rm type~B}=-\frac{\zeta(3)}{720\pi^2}-\frac{\zeta(5)}{192\pi^2}- \frac{\zeta(7)}{128\pi^2}\,,\qquad &d=7\,.
\end{aligned}
\end{equation}
and similarly for higher $d$. Obviously, these complicated shifts cannot be accommodated by an integer shift of $N$. 
While the reason for this is not fully clear to us, it may be related to the fact that the imposition of the singlet constraint requires introduction of other terms in $F$.
For example, in $d=3$ the theory also contains a Chern-Simons sector, whose leading contribution to $F$ is of order $N^2$. Perhaps a detailed understanding
of these additional terms holds the key to resolving the puzzle for the fermionic theories in odd $d$.

We note that (\ref{eqn:typebnonminsphere2}) always produces only linear combinations of $\zeta(2k+1)/\pi^2$ 
with rational coefficients.  Interestingly, these formulas are related to the change in $F$ due to certain double-trace deformations \cite{Giombi:2014xxa}.
In particular, the first formula gives (up to sign) the change in free energy due to the double-trace deformation $\sim \int d^d x O_{\Delta}^2$, where $O_{\Delta}$ is a scalar operator 
of dimension $\Delta=\frac{d-1}{2}$, and the second formula is proportional to the change in free energy due to the deformation $\sim  \int d^d x \bar\Psi_{\Delta} \Psi_{\Delta}$, where $\Psi_{\Delta}$ is a fermionic operator of dimension $\Delta=\frac{d-2}{2}$. The reason for this formal relation to the double-trace flows is unclear to us.

We also consider bulk Type B theories where various truncations have been imposed on the non-minimal Type B theory and we provide evidence that they are dual to 
the singlet sectors of various free fermionic CFTs. In $d=2,3,4,8,9$ mod 8 we study the CFT of $N$ Majorana fermions with the $O(N)$ singlet constraint, 
while in $d=5,6,7$ mod 8 we study the theory of $N$ symplectic Majorana fermions with the 
$USp(N)$ singlet constraint.
We also study the CFT of Weyl fermions in even $d$, and of Majorana-Weyl fermions when $d=2$ mod 8. We will discuss these truncations in more detail in section \ref{typeB}.
For even $d$, we find that under the Weyl truncation, the Type B theories have vanishing $F$ at the one-loop level. Under the Majorana/symplectic Majorana condition, the free energy of the truncated Type B theory gives (up to sign) the free energy of one free conformal fermionic field on $S^d$. This is logarithmically divergent due to the CFT $a$-anomaly, $F_{f}^{S^d}=a_f \log(\mu R)$, where the anomaly coefficient $a_f$ is given by \cite{Giombi:2014xxa}:
	\begin{align}
		a_f & = 2^{\frac{d}{2}}\frac{(-1)^{\frac{d}{2}}}{\pi\Gamma(1+d)}\int_0^1du \cos\left(\frac{\pi u}{2}\right)\Gamma\left(\frac{1+d+u}{2}\right)\Gamma\left(\frac{1+d-u}{2}\right) \label{eqn:fermfree}\\
		 & = \left\{-\frac{1}{6},\frac{11}{180},-\frac{191}{7560},\frac{2497}{226800},-\frac{14797}{2993760},\frac{92427157}{40864824000},-\frac{36740617}{35026992000},\cdots\right\} \label{eqn:confreeenergy}
		\end{align}
		for $d=\{2,4,6,8,\ldots\}$.
Finally, under the Majorana-Weyl condition, the free energy of the corresponding truncated Type B theory reproduces half of the anomaly coefficients given in \eqref{eqn:confreeenergy}, corresponding to a single Majorana-Weyl fermion. 
		
For the odd $d$ case, the minimal type B theories dual to the Majorana (or symplectic Majorana) projections again have unexpected values 
of their one-loop free energies. They are listed in Table~\ref{Table:TypeBMin}. We did not find a simple analytic formula that reproduces these numbers, 
but we note that, as in the non-minimal type B result (\ref{eqn:typebnonminsphere2}), these values are always linear combinations of 
$\zeta(2k+1)/\pi^2$ with rational coefficients. 
It would be very interesting to understand the origin of these ``anomalous'' results in the type B theories. 
		
 
One may also consider free CFTs which involve both the conformal scalars and fermions in the fundamental of $U(N)$ (or $O(N)$), with action
\begin{align}
S = \int d^dx \sum_{i=1}^N\left[(\partial_\mu\phi_i^*)(\partial^\mu\phi^i) + \bar\psi_i(\slashed{\partial})\psi^i\right].
\label{AB-cft}
\end{align}
When we impose the $U(N)$ singlet constraint, the spectrum of single trace operators contains not only the bilinears in $\phi$ and $\psi$, which are the same 
as discussed above, but also fermionic operators of the form
\begin{align}
\Psi_{\mu_1\cdots \mu_{s}} = \bar\psi^i \partial_{(\mu_1}\cdots\partial_{\mu_{s-\frac{1}{2}})}\phi^i+ \ldots,\qquad\qquad \text{where } s = \frac{1}{2},\frac{3}{2},\cdots,\label{eqn:spintypec}
\end{align} 
The dual HS theory in AdS should then include, in addition to the bosonic fields that appear in type A and type B theories, a tower of massless 
half integer spin particles with $s=3/2,5/2,\ldots$, plus a $s=1/2$ matter field. We will call the resulting HS theory the ``type AB'' theory. Note that 
in $d=3$ this leads to a supersymmetric theory, but in general $d$ the action (\ref{AB-cft}) is not supersymmetric. One may also truncate the model 
to the $O(N)/USp(N)$ by imposing suitable reality conditions. There is no qualitative difference in the spectrum of the half-integer operators in the 
truncated models, 
with the only quantitative difference being a doubling of the degrees of freedom of each half-integer spin particle when going from $O(N)/USp(N)$ to $U(N)$ 
in the dual CFT. 


The partition function for the Type AB theory is,
\begin{align}
Z = e^{-F} = e^{-\frac{1}{G_N}F^{(0)}+F^{(1)} + G_NF^{(2)}+\cdots}
,\qquad\qquad\text{where }
F^{(1)} = F_f^{(1)}+F_b^{(1)} , \label{eqn:typeABFsplit}
\end{align}
with $F_b$ being for the contributions from bosonic higher-spin fields, which arise from purely Type A and purely Type B contributions, and 
$F_f^{(1)}$ is the contribution of the HS fermions dual to (\ref{eqn:spintypec}). Up to one-loop level, the bosonic and fermionic contributions are decoupled, as indicated in \eqref{eqn:typeABFsplit}. A similar decoupling of the Casimir energy occurs at the one-loop level, i.e. $E_c^{(1)} = E_{c,f}^{(1)} + E_{c,b}^{(1)}$.

Our calculations for the Euclidean-AdS higher spin theory shows that $F_f^{(1)}=0$ at the one-loop level for both $U(N)/O(N)$ theories for all $d$. Similarly, 
the Casimir energies are found to vanish: $E_{c,f} = 0$. In even $d$, from our results on the Type B theories and the earlier results on Vasiliev Type A theories, we see that $F_b^{(1)} = 0$ for the non-minimal Type AB theory, and this suggests that Type AB theories at one-loop have vanishing $F^{(1)}$. For odd $d$, $F^{(1)}$ is non-vanishing with the non-zero contribution coming from the Type B theory's free energy, as discussed above. 

Finally, we consider the Type C higher-spin theories, which are conjectured to be dual to the singlet sector of massless $p$-forms, where $p=(\frac{d}{2}-1)$.\footnote{The choice of the $p$-form is made to ensure that the current operators satisfy the unitary bound, as well as conformal invariance.} 
The first two examples of these theories are the $d=4$ case discussed in \cite{Beccaria:2014xda,Beccaria:2014zma}, where the dynamical fields 
are the $N$ Maxwell fields, and the $d=6$ case \cite{Beccaria:2014qea} where the dynamical fields are $N$ 2-form gauge fields with field strength $H_{\mu\nu\rho}$. In these theories, there are also an infinite number of totally symmetric conserved higher-spin currents, in addition to various fields of mixed symmetry. 
We will extend these calculations to even $d>6$.

As for type B theories in $d>3$, there are no known equations of motion for type C theories, but we can still infer their free field spectrum from 
CFT considerations, using the results of \cite{Dolan:2005wy}. The non-minimal theory is obtained by taking $N$ complex $(d/2-1)$-form gauge fields $A$, and imposing 
a $U(N)$ singlet constraint. One may further truncate these models by taking real fields and $O(N)$ singlet constraint, which results in the ``minimal type C'' theory. In addition, one can further impose a self-duality condition on the $d/2$-form field strength $F=dA$. Since $*^2=+1$ in $d=4m+2$ and $*^2=-1$ in $d=4m$, where $*$ 
is the Hodge-dual operator, one can impose the self-duality condition $F=*F$ only in $d=4m+2$ (for $m$ integer); this can be done both for real ($O(N)$) and complex ($U(N)$) fields. In $d=4m$, and only in the non-minimal case with $N$ complex fields, one can impose the 
self-duality condition $F=i*F$. Decomposing $F=F_1+i F_2$ into its real and imaginary parts, this condition implies $F_1=-*F_2$, and selfdual and anti-selfdual 
parts of $F$ are complex conjugate of each other. 

As an example, let us consider $d=4$ and take $N$ complex Maxwell fields with a $U(N)$ singlet constraint. The spectrum of the the single trace operators arising from 
the tensor product $\bar{F}_{\mu\nu}^i \otimes F^{\rho\sigma}_i$ can be found to be \cite{Dolan:2005wy, Beccaria:2014zma}
\begin{equation}
\begin{aligned}
&(2;1,1)_c \otimes (2;1,1)_c = 2(4;0,0)+(4;1,1)_c+(4;2,2)_c\\
&~~~~~~~~~~~~~~~~~~~~~~~~+2\sum_{s=2}^{\infty} (s+2;s,0)+\sum_{s=3}^{\infty} (s+2;s,2)_c
\end{aligned}
\end{equation}
where we use the notation $(2;1,1)_c = (2;1,1)+(2;1,-1)$, corresponding to the sum of the selfdual and anti selfdual 2-form field strength with $\Delta=2$, and similarly 
for the representations appearing on the right-hand side. Note that we use $SO(4)$ notations $[m_1,m_2]$ to specify the representation. The operators in the first 
line are dual to matter fields in AdS$_5$ in the corresponding representations, while the second line corresponds to massless HS gauge fields. Note that a novel feature 
compared to type A and type B is the presence of mixed symmetry representations with two boxes in the second row 
\begin{align}\footnotesize
\overbrace{\ssmall \ydiagram{0+3,0+2}\begin{ytableau}\none[\cdots]\end{ytableau}\ydiagram{0+5}
}^{\mbox{$s$}}
\end{align}
Imposing a reality condition and $O(N)$ singlet constraint, 
one obtains the minimal spectrum \cite{Beccaria:2014zma}
\begin{equation}
\begin{aligned}
&\left[(2;1,1)_c \otimes (2;1,1)_c\right]_{\rm symm} = 2(4;0,0)+(4;2,2)_c\\
&~~~~~~~~~~~~~~~~~~~~~~~~~~+\sum_{s=2}^{\infty} (s+2;s,0)+\sum_{s=4,6,\ldots }^{\infty} (s+2;s,2)_c\,.
\end{aligned}
\end{equation}
Similarly, one may obtain the spectrum in all higher dimensions $d=4m$ and $d=4m+2$, as will be explained in detail in section \ref{subsubsection:typecweights}. As an 
example, in the $d=8$ type C theory we find the representations
\begin{align}\scriptsize
\quad\overbrace{\ssmall \ydiagram{0+3,0+2,0+2,0+2}\begin{ytableau}\none[\cdots]\end{ytableau}\ydiagram{0+3}
}^{\mbox{$s$}}
\quad\&\quad 
\overbrace{\ssmall \ydiagram{0+3,0+2,0+2}\begin{ytableau}\none[\cdots]\end{ytableau}\ydiagram{0+3}
}^{\mbox{$s$}}
\quad\&\quad 
\overbrace{\ssmall \ydiagram{0+3,0+2}\begin{ytableau}\none[\cdots]\end{ytableau}\ydiagram{0+3}
}^{\mbox{$s$}}
\nonumber
\\\small\&\quad
\overbrace{\ssmall \ydiagram{0+3,0+2,0+1,0+1}\begin{ytableau}\none[\cdots]\end{ytableau}\ydiagram{0+3}
}^{\mbox{$s$}}
\quad\&\quad
\overbrace{\ssmall \ydiagram{0+3,0+1,0+1,0+1}\begin{ytableau}\none[\cdots]\end{ytableau}\ydiagram{0+3}
}^{\mbox{$s$}}
\quad\&\quad
\overbrace{\ssmall \ydiagram{0+3,0+1,0+1}\begin{ytableau}\none[\cdots]\end{ytableau}\ydiagram{0+3}
}^{\mbox{$s$}} \\
\& \quad
\quad\overbrace{\ssmall \ydiagram{0+3,0+1,0+1}\begin{ytableau}\none[\cdots]\end{ytableau}\ydiagram{0+3}
}^{\mbox{$s$}}
\quad\&\quad 
\overbrace{\ssmall \ydiagram{0+3}\begin{ytableau}\none[\cdots]\end{ytableau}\ydiagram{0+3}
}^{\mbox{$s$}} \nonumber
\end{align}

Our results for the one-loop calculations in type C theories are summarized in Table~\ref{Table:summary}. We find that the non-minimal $U(N)$ theories have non-zero one-loop contributions, unlike the type A and type B theories (in even $d$). The results can be grouped into two subclasses depending on the spacetime dimension, namely those in $d=4m$ or in $d=4m+2$, where $m$ is an integer. In the minimal type C theories with $O(N)$ singlet constraint, we find that for all $d=4m$ the identification of the bulk coupling constant is $1/G_N\sim N-2$, while in $d=4m+2$, the bulk one-loop free energy vanishes, and therefore no shift is required. In the self-dual $U(N)/O(N)$ theories, the one-loop free energy does not vanish, but can be accounted for by half-integer shifts $1/G_N \sim N\pm 1/2$, as mentioned earlier. We find that all of these results are consistent with calculations of Casimir energies in thermal AdS space, which are collected in the Appendix. 

\begin{table}[t]
\begin{center}{\footnotesize
\begin{tabular}{c|c|c}

\multicolumn{2}{c|}{\textbf{Type of Theory}}             & \textbf{Shift to ${1\over G_N}\sim N$}                \\ \hline
\multicolumn{3}{c}{}\\\hline
\multicolumn{3}{|c|}{\cellcolor[HTML]{000000}{\color[HTML]{FFFFFF}\textbf{Type A Theories}}}                                                      \\ \hline
\multicolumn{2}{c|}{Non-Minimal $U(N)$:}                           & No shift                      \\ \hline
\multicolumn{2}{c|}{Minimal $O(N)$:}                                & $N\rightarrow N-1$                      \\\hline\hline
\multicolumn{3}{c}{}\\\hline

\multicolumn{3}{|c|}{\cellcolor[HTML]{000000}{\color[HTML]{FFFFFF}\textbf{Type B Theories}}}                                                        \\ \hline
\multicolumn{2}{c|}{Non-Minimal $U(N)$:}                           & No shift                      \\ \hline
\multirow{2}{*}{Minimal}             & $O(N)$ in $d=2,4,8$ (mod 8): & $N\rightarrow N-1$            \\ \cline{2-3} 
                                     & $USp(N)$ in $d=6$ (mod 8):     & $N\rightarrow N+1$            \\ \hline
\multicolumn{2}{c|}{Weyl Projection:}                                & No shift                      \\ \hline
Majorana-Weyl:                       & $d=2$ (mod 8):                    & $N\rightarrow N-1$            \\ \hline\hline
\multicolumn{3}{c}{}\\\hline
\multicolumn{3}{|c|}{\cellcolor[HTML]{000000}{\color[HTML]{FFFFFF}\textbf{Type C Theories ($p$-Forms)}}}                                           \\ \hline
\multirow{2}{*}{Non-minimal $U(N)$} & $d=4,\ 8,\ 12,\ldots$:                    & $N\rightarrow N-1$            \\ \cline{2-3} 
                                     & $d=6, 10, 14,\ldots$:                    & $N\rightarrow N+1$            \\ \hline
\multirow{2}{*}{Minimal $O(N)$} & $d=4,\ 8,\ 12,\ldots$:                    & $N\rightarrow N-2$            \\ \cline{2-3} 
                                     & $d=6, 10, 14,\ldots$:                    & No shift                      \\ \hline
\multirow{2}{*}{Self-dual $U(N)$}    & $d=4,\ 8,\ 12,\ldots$:                    & $N\rightarrow N-\dfrac{1}{2}$ \\[1.8ex] \cline{2-3} 
                                     & $d=6, 10, 14,\ldots$:                    & $N\rightarrow N+\dfrac{1}{2}$ \\[1.8ex] \hline
\multirow{2}{*}{Self-dual $O(N)$}    & $d=4,\ 8,\ 12,\ldots$:                    & Not defined                   \\ \cline{2-3} 
                                     & $d=6, 10, 14,\ldots$:                    & $N\rightarrow N-\dfrac{1}{2}$ \\[1.8ex] \hline\hline
\end{tabular}}
\caption{Summary of results of one-loop calculations for even $d>0$. By no shift, we mean that there are no shifts to the relation $G_N \sim 1/N$ due to 
one-loop free energy of the particular theory. Results for Type A theories taken from \cite{Giombi:2014iua}.\label{Table:summary}}
\end{center}
\end{table}

\begin{table}[t]
\begin{center}{\footnotesize
\begin{tabular}{c|c|c}

\multicolumn{2}{c|}{\textbf{Type of Theory}}             & \textbf{Shift to ${1\over G_N}\sim N$}                \\ \hline
\multicolumn{3}{c}{}\\\hline
\multicolumn{3}{|c|}{\cellcolor[HTML]{000000}{\color[HTML]{FFFFFF}\textbf{Type A Theories}}}                                                      \\ \hline
\multicolumn{2}{c|}{Non-Minimal $U(N)$:}                           & No shift                      \\ \hline
\multicolumn{2}{c|}{Minimal $O(N)$:}                                & $N\rightarrow N-1$                      \\\hline\hline
\multicolumn{3}{c}{}\\\hline

\multicolumn{3}{|c|}{\cellcolor[HTML]{000000}{\color[HTML]{FFFFFF}\textbf{Type B Theories}}}                                                        \\ \hline
\multicolumn{2}{c|}{Non-Minimal $U(N)$:}                           & Shifted by \eqref{eqn:typebnonminsphere2}                      \\ \hline
\multirow{2}{*}{Minimal}             & $O(N)$ in $d=3,9$ (mod 8): & See Section~\ref{subsect:oddtypeb}       \\ \cline{2-3} 
                                     & $USp(N)$ in $d=5,7$ (mod 8):     & See Section~\ref{subsect:oddtypeb}        \\ \hline\hline
                                     \multicolumn{3}{c}{}\\\hline
\end{tabular}}
\caption{Summary of results of one-loop calculations for odd $d>0$. Again, by no shift, we mean that there are no shifts to the coupling constant coming from the spectrum of the particular theory. Results for Type A theories taken from \cite{Giombi:2014iua}.\label{Table:summary2}}
\end{center}
\end{table}

\section{Matching the Sphere Free Energy}
\label{sect:ordinaryads}
\subsection{The AdS spectral zeta function}
Let us first review the calculation of the one-loop partition function on the hyperbolic space in the case of the totally symmetric HS fields \cite{Giombi:2013fka,Giombi:2014iua}. After gauge fixing of the linearized gauge invariance, the contribution of a spin $s$ ($s\ge 1$) totally symmetric gauge field 
to the bulk partition function is obtained as \cite{Gaberdiel:2010ar, Gaberdiel:2010xv,Gupta:2012he}
\begin{align}
Z_s = \frac{\left[{\rm det}^{STT}_{s-1}\left(-\nabla^2+(s+d-2)(s-1)\right)\right]^{\frac{1}{2}}}{\left[{\rm det}^{STT}_{s}\left(-\nabla^2+(s+d-2)(s-2)-s\right)\right]^{\frac{1}{2}}}\label{eqn:Zspartition}
\end{align}
where the label $STT$ stands for symmetric traceless transverse tensors, and the numerator corresponds to the contributions of the spin $s-1$ ghosts. The mass-like terms in 
the above kinetic operators are related to the conformal dimension of the dual fields. For a totally symmetric field with kinetic operator $-\nabla^2+\kappa^2$, the dual 
conformal dimension is given by 
\begin{equation}
\Delta(\Delta-d)-s=\kappa^2\,.
\end{equation} 
For the values of $\kappa$ in (\ref{eqn:Zspartition}), one finds for the physical spin $s$ field in the denominator\footnote{We choose the root $\Delta_+$ above the unitarity bound. The alternate root corresponds to gauging the HS symmetry at the boundary \cite{Giombi:2013yva}.}
\begin{equation}
\Delta^{\rm ph}=s+d-2
\end{equation}
which corresponds to the scaling dimension of the dual conserved current in the CFT. Similarly, the conformal dimension obtained from the ghost kinetic operator in 
(\ref{eqn:Zspartition}) is
\begin{equation}
\Delta^{\rm ph}=s+d-1\,.
\end{equation}
From CFT point of view, this is the dimension of the divergence $\partial\cdot J_s$, which is a null state that one has to subtract to obtain the short representation of 
the conformal algebra corresponding to a conserved current.  

The determinants in (\ref{eqn:Zspartition}) can be computed using the heat kernel, or equivalently spectral zeta functions techniques.\footnote{The heat-kernel is related to the spectral zeta-function by a Mellin transformation.} The spectral zeta function for a differential operator on a compact space with discrete eigenvalues $\lambda_n$ and degeneracy $d_n$ is defined as 
\begin{align}
\zeta(z) = \sum_n d_n \lambda_n^{-z}.
\end{align}
In our case, the differential operators in hyperbolic space have continuous spectrum, and the sum over eigenvalues is replaced by an integral. Let us consider a field 
labelled by the representation $\alpha_s=[s,m_2,m_3,\ldots]$ of $SO(d)$\footnote{This can be thought as the representation that specifies the dual CFT operator. From AdS point of view, 
one may view $SO(d)$ as the little group for a massive particle in $d+1$ dimensions.}, where we have denoted by $m_1=s$ the length of the first row in the corresponding Young 
diagram, which we may call the spin of the particle (for example, for a totally symmetric field, we have $\alpha_s = [s,0,0,\ldots,0]$). For a given representation $\alpha_s$, the spectral zeta function takes the form 
\begin{align}
\zeta_{(\Delta;\alpha_s)}(z) & =  \frac{\text{vol}\left({\text{AdS}_{d+1}}\right)}{\text{vol}(S^d)}\frac{2^{d-1}}{\pi}g_{\alpha_s} \int_0^\infty du \frac{\mu_{\alpha_s}(u)}{\left[u^2+\left(\Delta-\frac{d}{2}\right)^2\right]^z},\label{zeta1}
\end{align} 
where $\mu_{\alpha_s}(u)$ is the spectral density of the eigenvalues, which will be given shortly, and $g_{\alpha_s}$ is the dimension of the representation $\alpha_s$ (see eq. (\ref{galpha2}) and (\ref{galpha1}) below). The denominator corresponds to the eigenvalues of the kinetic operator, and $\Delta$ is the dimension of the dual CFT operator.\footnote{For the case of totally symmetric fields, this form of the eigenvalues can be deduced from the results of \cite{Camporesi:1994ga}. See for example the Appendix of \cite{Beccaria:2014xda} and \cite{Beccaria:2014qea} for an explicit derivation in AdS$_5$ and AdS$_7$ for arbitrary representations.} The regularized volume of AdS is given explicitly by \cite{Casini:2010kt,Diaz:2007an,Casini:2011kv}
\begin{align}
\text{vol(AdS}_{d+1}) = \begin{cases}
	\pi^{d/2}\Gamma(-\frac{d}{2}),\qquad &d\text{ odd},\\
	\frac{2(-\pi)^{d/2}}{\Gamma(1+\frac{d}{2})}\log R,\qquad& d\text{ even},
	\end{cases}
	\end{align}
where $R$ is the radius of the boundary sphere. The logarithmic dependence on $R$ in even $d$ is related to the presence of the Weyl anomaly in even dimensional CFTs. Finally, the volume of the round sphere of unit radius is
\begin{align}
\text{vol}(S^{d}) = \frac{2\pi^{(d+1)/2}}{\Gamma[(d+1)/2]}.
\end{align} 
Once the spectral zeta function is known, the contribution of the field labelled by $(\Delta;\alpha_s)$ to the bulk free energy is obtained as
\begin{equation}
F_{(\Delta;\alpha_s)}^{(1)} = \sigma\left[- \frac{1}{2} \zeta'_{(\Delta;\alpha_s)}(0) - \zeta_{(\Delta;\alpha_s)}(0) \log( \ell\, \Lambda)\right],
\label{eqn:freeen}
\end{equation}
where $\sigma=+1$ or $-1$ depending on whether the field is bosonic or fermionic. Here $\ell$ is the AdS curvature, which we will set to 1 henceforth, and $\Lambda$ is a UV cut-off. In general, the coefficient of the logarithmic divergence $ \zeta_{(\Delta;\alpha_s)}(0)$ vanishes for each $\alpha_s$ in even dimension $d$, but it is non-zero for odd $d$.

When the dimension $\Delta=s+d-2$, the field labelled by $\alpha_s$ is a gauge field and one has to subtract the contribution of the corresponding ghosts in the $\alpha_{s-1}$ representation.\footnote{As in the case of totally symmetric fields, the representation labeling the ghosts can be understood from CFT point of view from 
the structure of the character of the short representations of the conformal algebra and the corresponding null states, see \cite{Dolan:2005wy}.} We find it convenient to 
introduce the notation 
\begin{equation}
\mathfrak{Z}_{(\Delta^{\rm ph}=s+d-2;\alpha_s)}(z) \equiv \zeta_{(\Delta^{\rm ph};\alpha_s)}(z) - \zeta_{(\Delta^{\rm ph}+1;\alpha_{s-1})}(z)
\end{equation}
to indicate the spectral zeta function of the HS gauge fields in the $\alpha_s$ representation, with ghost contribution subtracted. The full one-loop free energy may be then obtained by summing over all representations $\alpha_s$ appearing in the spectrum. For instance, in the case of the non-minimal type A theory, 
we may define the ``total'' spectral zeta function
\begin{equation}
\zeta_{\rm type~A}^{\rm HS}(z) = \zeta_{(d-2;[0,\ldots,0])}(z)+\sum_{s=1}^{\infty} \mathfrak{Z}_{(s+d-2;[s,0,\ldots,0])}(z) 
\end{equation}
from which we can obtain the full one-loop free energy
\begin{equation}
F^{(1)}_{\rm type~A} = \left[-\frac{1}{2}(\zeta_{\rm type~A}^{\rm HS})'(0) -\frac{1}{2} \zeta_{\rm type~A}^{\rm HS}(0)\log(\ell^2 \Lambda^2)\right].
\label{eqn:zetaHS2}
\end{equation}
Similarly, one can obtain $\zeta_{\rm total}^{\rm HS}(z)$ and the one-loop free energy in the other higher spin theories we discuss. 
As these calculations requires summing over infinite towers of fields, one has of course to suitably regularized the sums, as discussed in \cite{Giombi:2013fka,Giombi:2014iua} and reviewed in the explicit calculations below. 

\subsubsection{The spectral density for arbitrary representation}
\label{subsect:genspec}

A general formula for the spectral density for a field labelled by the representation $\alpha=[m_1,m_2,\ldots]$ was given in \cite{Camporesi:1994pf}, and 
we summarize their result below. 

In AdS$_{d+1}$, arranging the weights for the irreps of $SO(d)$ as $m_1 \geq m_{2}\geq\cdots\geq |m_p|$, where $p = \frac{d-1}{2}$ for odd $d$ and $p=\frac{d}{2}$ for even $d$, we may define
\begin{align}
\ell_j &= m_{p-j+1} + j - 1,\qquad\qquad \text{\ \!for $d=$ even},\\ 
\ell_j &= m_{p-j+1} + j - \frac{1}{2},\qquad\qquad \text{for $d=$ odd\,.}
\end{align} 
In terms of these, the spectral density takes the form of 
\begin{align}
\label{mu-even}
\mu_\alpha(u) &=\frac{\pi}{\left(2^{d-1}\Gamma\left(\frac{d+1}{2}\right)\right)^2} \prod_{j=1}^p (u^2 + \ell_j^2),\qquad\qquad \ \ \text{for $d=$ even},\\
\label{mu-odd}
\mu_\alpha(u) &=\frac{\pi}{\left(2^{d-1}\Gamma\left(\frac{d+1}{2}\right)\right)^2} f(u)\ \! u\prod_{j=1}^p (u^2 + \ell_j^2),\qquad \text{for $d=$ odd},
\end{align}
where
\begin{align}
f(u) = \begin{cases} \tanh(\pi u),\qquad\ell_j = \text{half-integer},\\
\coth(\pi u),\qquad\ell_j = \text{integer}.
\end{cases}
\end{align}
The pre-factor of $\frac{\pi}{\left(2^{d-1}\Gamma\left(\frac{d+1}{2}\right)\right)^2}$ arises as a normalization constant found by imposing the condition that as we approach flat space from hyperbolic space, the spectral density should approach that of flat space.

The number of degrees of freedom $g_\alpha$ is equal to the dimension of the corresponding representation of $SO(d)$, and is given by \cite{Frappat:1996pb}
\begin{alignat}{3}
\label{galpha2}
g_{\alpha_s} &= \prod_{1\leq i < j \leq p}\frac{m_i-m_j+j-i}{j-i}\prod_{1\leq i < j \leq p}\frac{m_i+m_j+2p-i-j}{2p-i-j}, \qquad &\text{for }d=2p,
\intertext{and}
g_{\alpha_s} &=    \prod_{1\leq i \leq p}\frac{2m_i+2p-2i+1}{2p-2i+1}\prod_{1\leq i < j \leq p}\frac{m_i-m_j+j-i}{j-i}\nonn\\&\quad\times\prod_{1\leq i < j \leq p}\frac{m_i+m_j+2p-i-j+1}{2p-i-j+1}, \qquad&\text{for }d=2p+1, \label{galpha1}
\end{alignat} 
where $\alpha=[m_1,\ldots,m_{p}]$.
As an example, in the Type A case in AdS$_{d+1}$, the only representation we need to consider is $m_1=s$, and for all $j\neq1$, $m_j=0$. This gives us
\begin{align}
\mu_{[s,0,\ldots,0]}(u) & = 
\frac{\pi}{\left(2^{d-1}\Gamma\left(\frac{d+1}{2}\right)\right)^2}\left[u^2+\left(s+\frac{d-2}{2}\right)^2\right]\left|\frac{\Gamma\left(iu + \frac{d-2}{2}\right)}{\Gamma(iu)}\right|^2\nonn\\
&= \begin{cases}\displaystyle\frac{\pi}{\left(2^{d-1}\Gamma\left(\frac{d+1}{2}\right)\right)^2}\left[u^2+\left(s+\frac{d-2}{2}\right)^2\right]\prod_{j=0}^{(d-4)/2}(u^2+j^2),\qquad&d=\text{even},\\
\displaystyle\frac{\pi}{\left(2^{d-1}\Gamma\left(\frac{d+1}{2}\right)\right)^2}u\tanh(\pi u) \left[u^2+\left(s+\frac{d-2}{2}\right)^2\right]\prod_{j=0}^{(d-5)/2}\Bigg[u^2+(j+\frac{1}{2})^2\Bigg],\qquad&d=\text{odd}.
\end{cases}
\end{align}
and 
\begin{align}
g_{[s,0,\ldots,0]}= \frac{(2s+d-2)(s+d-3)!}{(d-2)!s!},\qquad\qquad d\geq 3.
\end{align}
The results agree with the formulas derived in \cite{Camporesi:1991nw} and used in \cite{Giombi:2014iua}.

In type AB theories, we need the spectral density for fermion fields in the $\alpha=[s,1/2,1/2,\ldots,1/2]$ representation. We find that the above general formulas for even and odd $d$ can be expressed in the compact form valid for all $d$
\begin{align}
 \mu_{[s,\frac{1}{2},\ldots,\frac{1}{2}]}(u) =  \frac{\pi}{\left(2^{d-1}\Gamma\left(\frac{d+1}{2}\right)\right)^2} \left[u^2+\left(s+\frac{d-2}{2}\right)^2\right]\left|\frac{\Gamma\left(iu + \frac{d-1}{2}\right)}{\Gamma(iu+\frac{1}{2})}\right|^2,
\end{align}
and
\begin{align}
g_{[s,\frac{1}{2},\ldots,\frac{1}{2}]} = \frac{(s-\frac{5}{2}+d)!}{(s-\frac{1}{2})!(d-2)!}n_F(d),\qquad\qquad n_F(d) = \begin{cases}2^{\frac{d-2}{2}},\qquad&\text{if }d=\text{even},\\
2^{\frac{d-1}{2}},\qquad&\text{if }d=\text{odd}.
\end{cases}
\end{align}
The spectral densities for the mixed symmetry fields appearing in type B and C theories can be obtained in a straightforward way from the above general formulas, and we present the explicit results in the next sections.

\subsection{Calculations in even $d$}\label{sect:calc}
\subsubsection{Type B Theories}
\label{typeB}
\paragraph{Spectrum}\label{subsect:spectrum}
The non-minimal Type B higher spin theory, which is conjectured to be dual to the $U(N)$ singlet sector of the free Dirac fermion theory, contains towers of mixed symmetry gauge fields of all integer spins. From the spectrum given in (\ref{psipsi-even}), we obtain the total spectral zeta function
\begin{align}
\zeta_{\rm type\ B}^{\rm HS}(z) &= 2\zeta_{(\Delta = d-1;[0,0,\ldots,0])}(z) \nonn\\&\quad+ 2\sum_{s=1}^\infty \Big[\mathfrak{Z}_{(\Delta^{\rm ph};[s,1,1,\ldots,1,0])}(z) + \mathfrak{Z}_{(\Delta^{\rm ph};[s,1,1,\ldots,1,0,0])}(z) +\ldots + \mathfrak{Z}_{(\Delta^{\rm ph};[s,1,0,\ldots,0])}(z) \nonn \\
&\qquad\qquad\quad+ \mathfrak{Z}_{(\Delta^{\rm ph};[s,0,0,\ldots,0])}(z)\Big] \nonn\\
& \quad+ \sum_{s=1}^\infty \left[\mathfrak{Z}_{(\Delta^{\rm ph};[s,1,1,\ldots,1,1])}(z) + \mathfrak{Z}_{(\Delta^{\rm ph};[s,1,1,\ldots,1,1,-1])}(z)\right].\label{dolan1} 
\end{align} 
In the third line of \eqref{dolan1}, the representations $[s,1,1,\ldots,1,1]$ and $[s,1,1,\ldots,1,-1]$ 
give the selfdual and anti-selfdual parts of the corresponding fields. At the level of the spectral $\zeta$ functions, 
they yield equal contributions.\footnote{Note that, technically, for all Type B theories 
the field of spin $s=1$ in the tower of spins of representation $[s,1,\ldots]$ is not a gauge field. However, for conciseness we still use 
the symbol $\mathfrak{Z}_{(\Delta^{\rm ph};[s,1,\ldots])}$ for these fields; the corresponding ghost contribution is zero, so it does not make 
a practical difference.\label{footnote:typebgauge}}
Using the 
spectral zeta function formulas listed in Section \ref{subsect:genspec} and summing over all representations given above, we find that for all even $d$ 
\begin{equation}
\zeta_{\rm type~B}^{\rm HS}(z) = {\cal O}(z^2)\,,
\end{equation}
and consequently the one-loop free energy in the non-minimal type B theory in even $d$ exactly vanishes
\begin{equation}
F^{(1)}_{\rm type~B} = 0\,.
\end{equation}

There are various truncations to the non-minimal Type B theory that results in the Weyl, Majorana and Majorana-Weyl projections on the free fermionic CFT. While the Weyl projection can be applied in all even dimensions $d$, the Majorana projection can be applied in dimensions $d=2,3,4,8,9$ (mod 8), and the Majorana-Weyl projection 
only in dimensions $d=2$ (mod 8). An interesting example is $d=10$ (AdS$_{11}$), where we can consider all four types of Type B theories. 

\subparagraph{Weyl projection}
The projection from the non-minimal Type B theory described above is slightly different when the theory is in $d=4m$ or $d=4m+2$. 
Using the results of \cite{Dolan:2005wy} for the product of chiral fermion representations, we find\footnote{To obtain this result, 
we note that in $d=4m$, complex conjugation flips the chirality of a Weyl spinor, while in $d=4m+2$ the Weyl representation is self-conjugate. 
Therefore, in order to obtain $U(N)$ invariant operators, we should use eq. (4.20) of \cite{Dolan:2005wy} for $d=4m$, 
and eq. (4.23) of the same reference for $d=4m+2$.}
	\begin{align}
		\zeta_{\rm type~B~Weyl}^{\rm HS}(z) &=  
		\begin{cases}
			\displaystyle\sum_{s=1}^\infty \left[\mathfrak{Z}_{(\dph;[s,0,0,\ldots,0])}(z) + \mathfrak{Z}_{(\dph;[s,1,1,0,\ldots,0])}(z) +\ldots + \mathfrak{Z}_{(\dph;[s,1,1,\ldots,1])}(z)\right],\qquad&\text{for }d=4m+2,\\
			\displaystyle\sum_{s=1}^\infty \left[\mathfrak{Z}_{(\dph;[s,0,0,\ldots,0])}(z)+ \mathfrak{Z}_{(\dph;[s,1,1,0,\ldots,0])}(z)+\ldots + \mathfrak{Z}_{(\dph;[s,1,1,\ldots,1,0])}(z)\right],\qquad&			\text{for }d=4m,
		\end{cases} 
	\nonumber \\
	& = \begin{cases} 
			\displaystyle\sum_{s=1}^\infty\sum_{\substack{t_i\geq0 \ \ \ \\t_i\geq t_{i+1}}}^1 \!\!\!\mathfrak{Z}_{(\dph;[s,t_1,t_1,\ldots,t_{m},t_m])}(z),\qquad &\text{for }d=4m+2,\\
			\displaystyle\sum_{s=1}^\infty \sum_{\substack{t_i\geq0 \ \ \  \\t_i\geq t_{i+1}}}^1\!\!\!\mathfrak{Z}_{(\dph;[s,t_1,t_1,\ldots,t_{m-1},t_{m-1},0])}(z), \qquad &			\text{for }d=4m,
		\end{cases}\label{eqn:typebweyl}
\end{align}

Note that under this projection, there are no scalars in the spectrum. The case $d=4$ (AdS$_5$) was already discussed in \cite{Giombi:2014yra}. Summing 
over all representations, we find that for all even $d$
\begin{equation}
\zeta_{\rm type~B~Weyl}^{\rm HS}(z) = {\cal O}(z^2)\,,
\end{equation}
and so
\begin{equation}
F^{(1)}_{\rm type~B~Weyl} = 0\,.
\end{equation}


\subparagraph{Minimal Theory (Majorana projection)}
The Majorana condition $\bar\psi = \psi^T {\cal C}$, where ${\cal C}$ is the charge conjugation matrix, 
can be imposed in $d=2,3,4,8,9$ (mod 8), see for instance \cite{VanProeyen:1999ni}. 
In these dimensions, we can consider the theory of $N$ free Majorana fermions and impose an $O(N)$ 
singlet constraint. In $d=6$ (mod 8), provided one has an even number $N$ of fermions, one can impose instead a symplectic Majorana condition 
$\bar\psi^i = \psi_j^T {\cal C} \Omega^{ij}$, where ${\cal C}$ is the charge conjugation matrix and $\Omega^{ij}$ the antisymmetric symplectic metric. 
In this case, we consider the theory of $N$ free symplectic Majoranas with a $USp(N)$ singlet constraint. 

The operator spectrum in the minimal theory can be deduced by working out which operators of the non-minimal theory are projected out by 
the Majorana constraint. The bilinear operators in the non-minimal theory are of the schematic form 
\begin{align}
J_{\mu_1 \cdots\mu_s,\nu_1\cdots\nu_{n-1}} \sim \bar\psi_i (\Gamma^{(n)})_{\nu_1\cdots \nu_{n-1}(\mu_1} \partial_{\mu_2}\partial_{\mu_3}\cdots\partial_{\mu_s)} \psi^i+\ldots
\label{typeB-schematic}
\end{align} 
where $n=0,\ldots,\frac{d}{2}-1$, and $\Gamma^{(n)}$ is the antisymmetrized product of $n$ gamma matrices. For Majorana fermions, we have 
$\bar\psi = \psi^T {\cal C}$, and so the operators are projected out or kept depending on whether $\mathcal{C}\Gamma^{(n)}$ is symmetric or 
antisymmetric. If $\mathcal{C}\Gamma^{(n)}$ is symmetric, then the operators with an even number of derivatives (i.e. odd spin) are projected out; if 
it is antisymmetric, then the operators with an odd number of derivatives (i.e. even spin) are projected out. In addition to (\ref{typeB-schematic}), the non-minimal 
type B theories in even $d$ include two scalars $J_0=\bar\psi_i \psi^i$ and $\tilde J_0=\bar\psi_i \gamma_* \psi^i$, where $\gamma_* \sim \Gamma^{(d)}$ is the chirality matrix. When $\cal C$ is symmetric, $J_0$ is projected out, and when ${\cal C}\gamma_*$ is symmetric, $\tilde J_0$ is projected out.\footnote{As an example, consider the bilinear $\psi^T M \psi$. If $M$ is symmetric, this operator clearly vanishes. On the other hand, 
consider $\psi^T M \partial_\mu \psi$. In this case, if $M$ is an antisymmetric matrix, then this is equal to $+\partial_\mu \psi^T M \psi$. 
In turn, this means that $\psi^T M \partial_\mu \psi = \frac{1}{2} \partial_\mu (\psi^T M\psi)$, and so this operator is a total derivative 
and is not included in the spectrum of primaries.}

For instance, in $d=4$, the non-minimal theory contains the operators given in (\ref{4d-B-sc}), (\ref{4d-B-sym}) and (\ref{4d-B-mixed}). 
In $d=4$, one has that both $\mathcal{C}$ and $\mathcal{C}\gamma_5$ are antisymmetric, so both scalars in (\ref{4d-B-sc}) are retained. 
Then, one has that ${\cal C}\gamma^{\mu}$ is symmetric while $\mathcal{C}\gamma^{\mu}\gamma^5$ antisymmetric, and so we keep the first 
tower in (\ref{4d-B-sym}) for even $s$ and the other tower for odd $s$: together, they make up a single tower in the $[s,0]$ representation with 
all integer spins. Finally, $\mathcal{C}\Gamma_{\mu\nu}$ is symmetric, so we keep the mixed symmetry fields (\ref{4d-B-mixed}) with an 
odd number of derivatives, i.e. the spectrum contains the representations $[s,1]_c = [s,1]+[s,-1]$ for all even $s$. 

Higher dimensions can be analyzed similarly, using the symmetry/antisymmetry properties of ${\cal C}\Gamma^{(n)}$ in various $d$ \cite{VanProeyen:1999ni}. 
The results are summarized in Table \ref{Table:O(n)projection}. 
One finds that under the Majorana projections the operators with the ``heaviest'' weight $[s,1,1,\ldots,1]_c$ 
always form a tower containing all even $s$. The next representation $[s,1,\ldots,1,0]$ form a tower of all integer $s$. Then, $[s,1,\ldots,1,0,0]$ appears 
in two towers of all odd $s$. And finally, $[s,1,\ldots,1,0,0,0]$ form a tower of all integer spins, after which this cycle repeats. 
The number of scalars with $\Delta=d-1$ to be included also changes in a cycle of 4. 
In AdS$_5$, we have 2 scalars; in AdS$_7$, we have 1 (this case, though, should be discussed separately, see below); 
in AdS$_9$, we have 0; in AdS$_{11}$, we have 1, and the cycle repeats. 
In a more compact notation, the total spectral zeta function in the minimal type B theories dual to the $O(N)$ Majorana theories is
	\begin{align}
	&	\zeta_{\rm type\ B\ Maj.}^{\text{HS}}(z)  = 
			\chi(d)\zeta_{(d-1;[0,0,\ldots,0])}(z) \nonn\\
			&\quad + \sum_{s=2,4,6,\ldots}^\infty \!\!\!\sum_{\substack{t_i\geq t_{i+1} \\ t_i \geq 0 \\ \sum_w t_w = w \ \! (\!\!\!\!\!\! \mod 4)}}^1 \!\!\!\!\!\!\!\!\Big(\mathfrak{Z}_{(\Delta;[s,t_1,t_2,\ldots,t_{w-1},t_{w}])}(z)+\mathfrak{Z}_{(\Delta;[s,t_1,t_2,\ldots,t_{w-1},-t_{w}])}(z)\Big) \nonn \\
			&\quad +  \sum_{s=1,2,3,\ldots} \!\!\!\!\!\!\!\sum_{\substack{t_i\geq t_{i+1} \\ t_i \geq 0 \\ \sum_w t_w = (w-1) \ \! (\!\!\!\!\!\! \mod 4)}}^1 \!\!\!\!\!\!\!\!\!\!\!\!\Big(\mathfrak{Z}_{(\Delta;[s,t_1,t_2,\ldots,t_{w-1},t_{w}])}(z)+\mathfrak{Z}_{(\dph;[s,t_1,t_2,\ldots,t_{w-1},-t_{w}])}(z)\Big) \nonn \\
			& \quad +  \sum_{s=1,2,3,\ldots} \!\!\!\!\!\!\!\sum_{\substack{t_i\geq t_{i+1} \\ t_i \geq 0 \\ \sum_w t_w = (w-3) \ \! (\!\!\!\!\!\! \mod 4)}}^1 \!\!\!\!\!\!\!\!\!\!\!\!\Big(\mathfrak{Z}_{(\dph;[s,t_1,t_2,\ldots,t_{w-1},t_{w}])}(z)+\mathfrak{Z}_{(\dph;[s,t_1,t_2,\ldots,t_{w-1},-t_{w}])}(z)\Big) \nonn \\
			& \quad +  \sum_{s=1,3,5,\ldots} \!\!\!\!\!\!\!\sum_{\substack{t_i\geq t_{i+1} \\ t_i \geq 0 \\ \sum_w t_w = (w-2) \ \! (\!\!\!\!\!\! \mod 4)}}^1 \!\!\!\!\!\!\!\!\!\!\!\!\Big(\mathfrak{Z}_{(\dph;[s,t_1,t_2,\ldots,t_{w-1},t_{w}])}(z)+\mathfrak{Z}_{(\dph;[s,t_1,t_2,\ldots,t_{w-1},-t_{w}])}(z)\Big)\label{eqn:typeBOn}
\end{align} where $\chi(d) = 1,2,0$ when $d = 0,2,4\ \!(\!\!\!\mod 8)$ respectively. Explicit illustrations of this formula are given in Table \ref{Table:O(n)projection}. Using these spectra we find, in all even $d$ where the Majorana condition is possible
\begin{equation}
F^{(1)}_{\rm type~B~Maj.} = a_f \log R
\label{F1-Majo}
\end{equation}
where $R$ is the radius of the boundary sphere, and $a_f$ is the $a$-anomaly coefficient of a single Majorana fermion in dimension $d$, given in (\ref{eqn:confreeenergy}). As explained earlier, this is consistent with the duality, provided $G_N^{\rm type~B~Maj.}\sim 1/(N-1)$.

As mentioned above, in $d=6$ (mod 8), i.e. AdS$_{7 ({\rm mod}~8)}$, we should impose a symplectic Majorana condition and consider the $USp(N)$ invariant operators.
In terms of the operators (\ref{typeB-schematic}),  since $\bar\psi = \psi^T {\cal C} \Omega$ with $\Omega$ antisymmetric,  
all this means is that now odd spins are projected out when ${\mathcal C} \Gamma^{(n)}$ is antisymmetric, and even spins are projected out 
when  ${\mathcal C} \Gamma^{(n)}$ is symmetric. Similarly, the scalar operators $\bar\psi_i \psi^i$ and $\psi_i \gamma_*\psi^i$ are now projected out 
when ${\cal C}$ and ${\cal C}\gamma_*$ are antisymmetric, respectively. In $d=6$ (mod 8), one has that ${\cal C}$ is symmetric and ${\cal C}\gamma_*$ is antisymmetric, 
so we retain a single scalar field. On the other hand, ${\cal C}\gamma_{\mu}$ and ${\cal C}\gamma_{\mu}\gamma_*$ are both antisymmetric, and so we have two towers of totally 
symmetric representations of all even $s$.\footnote{Note that, had we tried to impose the standard Majorana condition, we would have retained the totally symmetric 
fields of all odd spins. Then, the spectrum would not include a graviton, i.e. the dual CFT would not have a stress tensor.} The projection of the mixed symmetry representations can be deduced similarly. The total spectral zeta function is given by the formula 
 	\begin{align}
		&\zeta_{\rm type\ B\ Symp.Maj.}^{\rm HS}(z)  = 
			\zeta_{(d-1;[0,0,\ldots,0])}(z) \nonn\\
			&\quad+ \sum_{s=1,3,5,\ldots}^\infty \!\!\!\!\!\!\!\sum_{\substack{t_i\geq t_{i+1} \\ t_i \geq 0 \\ \sum_w t_w = w \ \! (\!\!\!\!\!\! \mod 4)}}^1 \!\!\!\!\!\!\!\!\!\Big(\mathfrak{Z}_{(\dph;[s,t_1,t_2,\ldots,t_{w-1},t_{w}])}(z)+\mathfrak{Z}_{(\dph;[s,t_1,t_2,\ldots,t_{w-1},-t_{w}])}(z)\Big) \nonn \\
			&\quad +  \sum_{s=1,2,3,\ldots} \!\!\!\!\!\!\!\!\!\!\sum_{\substack{t_i\geq t_{i+1} \\ t_i \geq 0 \\ \sum_w t_w = (w-1) \ \! (\!\!\!\!\!\! \mod 4)}}^1 \!\!\!\!\!\!\!\!\!\!\!\!\!\Big(\mathfrak{Z}_{(\dph;[s,t_1,t_2,\ldots,t_{w-1},t_{w}])}(z)+\mathfrak{Z}_{(\dph;[s,t_1,t_2,\ldots,t_{w-1},-t_{w}])}(z)\Big) \nonn \\
			& \quad +  \sum_{s=1,2,3,\ldots} \!\!\!\!\!\!\!\!\!\!\sum_{\substack{t_i\geq t_{i+1} \\ t_i \geq 0 \\ \sum_w t_w = (w-3) \ \! (\!\!\!\!\!\! \mod 4)}}^1 \!\!\!\!\!\!\!\!\!\!\!\!\!\Big(\mathfrak{Z}_{(\dph;[s,t_1,t_2,\ldots,t_{w-1},t_{w}])}(z)+\mathfrak{Z}_{(\dph;[s,t_1,t_2,\ldots,t_{w-1},-t_{w}])}(z)\Big) \nonn \\
			& \quad +  \sum_{s=2,4,6,\ldots} \!\!\!\!\!\!\!\!\!\!\sum_{\substack{t_i\geq t_{i+1} \\ t_i \geq 0 \\ \sum_w t_w = (w-2) \ \! (\!\!\!\!\!\! \mod 4)}}^1 \!\!\!\!\!\!\!\!\!\!\!\!\!\Big(\mathfrak{Z}_{(\dph;[s,t_1,t_2,\ldots,t_{w-1},t_{w}])}(z)+\mathfrak{Z}_{(\dph;[s,t_1,t_2,\ldots,t_{w-1},-t_{w}])}(z)\Big) \label{eqn:typeBUSp}
\end{align}
An illustration of the formula is given in Table~\ref{Table:O(n)projection} for the AdS$_7$ and AdS$_{15}$ cases. Using these spectra, we find that the one loop free energy 
of the minimal type B theory corresponding to the symplectic Majorana projection is given by
\begin{equation}
F^{(1)}_{\rm type~B~sympl.Maj.} = -a_f \log R\,,
\end{equation}
i.e. the opposite sign compared to (\ref{F1-Majo}). This is consistent with the duality, provided $G_N^{\rm type~B~sympl.Maj.} \sim 1/(N+1)$.

\begin{table}[h!]
\begin{center}
{\ssmall
\begin{tabular}{llll}
\textbf{AdS$_{\bf 3}\ O(N)$} &  &  &  \\ \hline\hline
\multicolumn{1}{c|}{$\alpha$} & \multicolumn{3}{c}{$s=$}\\\cline{2-4}
\multicolumn{1}{c|}{}& \multicolumn{1}{c|}{$1,2,3,\ldots$} & \multicolumn{1}{c|}{$2,4,6,\ldots$} & $1,3,5,\ldots$ \\ \hline
\multicolumn{1}{l|}{$[s]$} & \multicolumn{1}{c|}{} & \multicolumn{1}{c|}{\cellcolor[HTML]{EFEFEF}{\color[HTML]{000000} 2}} &   \\ \hline
\multicolumn{1}{c|}{Scalar ($\Delta=1$)} &  \multicolumn{3}{c}{\cellcolor[HTML]{EFEFEF}1}  \\ \hline\hline \multicolumn{1}{c|}{$F^{(1)}$} \rule[-1.5ex]{0pt}{4.2ex}& \multicolumn{3}{c}{$-\frac{1}{6} \log R$} \\\hline \\ \\
\textbf{AdS$_{\bf 7}\ USp(N)$} &  &  &  \\ \hline\hline
\multicolumn{1}{c|}{$\alpha$} & \multicolumn{3}{c}{$s=$}\\\cline{2-4}
\multicolumn{1}{c|}{}& \multicolumn{1}{c|}{$1,2,3,\ldots$} & \multicolumn{1}{c|}{$2,4,6,\ldots$} & $1,3,5,\ldots$ \\ \hline
\multicolumn{1}{l|}{$[s,1,1]_c$} & \multicolumn{1}{c|}{} & \multicolumn{1}{c|}{} &  \multicolumn{1}{c}{\cellcolor[HTML]{99FFFF}{\color[HTML]{000000} $1$}} \\ \hline
\multicolumn{1}{l|}{$[s,1,0]$} & \multicolumn{1}{c|}{\cellcolor[HTML]{EFEFEF}1} & \multicolumn{1}{c|}{} &  \\ \hline
\multicolumn{1}{l|}{$[s,0,0]$} & \multicolumn{1}{c|}{} & \multicolumn{1}{c|}{\cellcolor[HTML]{99FFFF}2} &  \\ \hline
\multicolumn{1}{c|}{Scalar ($\Delta=5$)} &\multicolumn{3}{c}{\cellcolor[HTML]{EFEFEF}1}  \\ \hline\hline \multicolumn{1}{c|}{$F^{(1)}$} \rule[-1.5ex]{0pt}{4.2ex}& \multicolumn{3}{c}{$\frac{191}{7560}\log R$} \\\hline \\ \\
\textbf{AdS$_{\bf 11}\ O(N)$} &  &  &  \\ \hline\hline
\multicolumn{1}{c|}{$\alpha$} & \multicolumn{3}{c}{$s=$}\\\cline{2-4}
\multicolumn{1}{c|}{}& \multicolumn{1}{c|}{$1,2,3,\ldots$} & \multicolumn{1}{c|}{$2,4,6,\ldots$} & $1,3,5,\ldots$ \\ \hline
\multicolumn{1}{l|}{$[s,1,1,1,1]_c$} & \multicolumn{1}{c|}{} & \multicolumn{1}{c|}{\cellcolor[HTML]{EFEFEF}{\color[HTML]{000000} $1$}} &  \\ \hline
\multicolumn{1}{l|}{$[s,1,1,1,0]$} & \multicolumn{1}{c|}{\cellcolor[HTML]{EFEFEF}1} & \multicolumn{1}{c|}{} &  \\ \hline
\multicolumn{1}{l|}{$[s,1,1,0,0]$} & \multicolumn{1}{c|}{} & \multicolumn{1}{c|}{} & \multicolumn{1}{c}{\cellcolor[HTML]{EFEFEF}2} \\ \hline
\multicolumn{1}{l|}{$[s,1,0,0,0]$} & \multicolumn{1}{c|}{\cellcolor[HTML]{EFEFEF}1} & \multicolumn{1}{c|}{} &  \\ \hline
\multicolumn{1}{l|}{$[s,0,0,0,0]$} & \multicolumn{1}{c|}{} & \multicolumn{1}{c|}{\cellcolor[HTML]{EFEFEF}2} &  \\ \hline
\multicolumn{1}{c|}{Scalar ($\Delta=9$)} &  \multicolumn{3}{c}{\cellcolor[HTML]{EFEFEF}1}    \\ \hline\hline 
\multicolumn{1}{c|}{$F^{(1)}$} \rule[-1.5ex]{0pt}{4.2ex}& \multicolumn{3}{c}{$-\frac{14797}{2993760}\log R$} \\\hline \\ \\
\textbf{AdS$_{\bf 15}\ USp(N)$}                   &                                                                                      &                                                                                      &                                                                 \\ \hline\hline
\multicolumn{1}{c|}{$\alpha$} & \multicolumn{3}{c}{$s=$}\\\cline{2-4}
\multicolumn{1}{c|}{}& \multicolumn{1}{c|}{$1,2,3,\ldots$} & \multicolumn{1}{c|}{$2,4,6,\ldots$} & $1,3,5,\ldots$ \\ \hline
\multicolumn{1}{l|}{$[s,1,1,1,1,1,1]_c$}    & \multicolumn{1}{c|}{}                                                                & \multicolumn{1}{c|}{}                & \multicolumn{1}{c}{\cellcolor[HTML]{99FFFF}{\color[HTML]{000000} $1$}}                                                                \\ \hline
\multicolumn{1}{l|}{$[s,1,1,1,1,1,0]$}    & \multicolumn{1}{c|}{\cellcolor[HTML]{EFEFEF}1}                                       & \multicolumn{1}{c|}{}                                                                &                                                                 \\ \hline
\multicolumn{1}{l|}{$[s,1,1,1,1,0,0]$}    & \multicolumn{1}{c|}{}                                                                & \multicolumn{1}{c|}{\cellcolor[HTML]{99FFFF}{\color[HTML]{000000} 2}}                                                                &                                       \\ \hline
\multicolumn{1}{l|}{$[s,1,1,1,0,0,0]$}    & \multicolumn{1}{c|}{\cellcolor[HTML]{EFEFEF}1}                                       & \multicolumn{1}{c|}{}                                                                &                                                                 \\ \hline
\multicolumn{1}{l|}{$[s,1,1,0,0,0,0]$}    & \multicolumn{1}{c|}{}                                                                & \multicolumn{1}{c|}{}                                       & \multicolumn{1}{c}{\cellcolor[HTML]{99FFFF}{\color[HTML]{000000} 2}}                                                                \\ \hline
\multicolumn{1}{l|}{$[s,1,0,0,0,0,0]$}    & \multicolumn{1}{c|}{\cellcolor[HTML]{EFEFEF}1}                                       & \multicolumn{1}{c|}{}                                                                &                                                                 \\ \hline
\multicolumn{1}{l|}{$[s,0,0,0,0,0,0]$}    & \multicolumn{1}{c|}{}                                                                & \multicolumn{1}{c|}{\cellcolor[HTML]{99FFFF}{\color[HTML]{000000} 2}}                                                                &                                     \\ \hline
\multicolumn{1}{c|}{Scalar ($\Delta=13$)}           & \multicolumn{3}{c}{\cellcolor[HTML]{EFEFEF}1}                                              \\ \hline\hline 
\multicolumn{1}{c|}{$F^{(1)}$} \rule[-1.5ex]{0pt}{4.2ex}& \multicolumn{3}{c}{$-\frac{36740617}{35026992000}\log R$} \\\hline \\ \\
\textbf{AdS$_{\bf 19}\ O(N)$} &  &  &  \\ \hline\hline
\multicolumn{1}{c|}{$\alpha$} & \multicolumn{3}{c}{$s=$}\\\cline{2-4}
\multicolumn{1}{c|}{}& \multicolumn{1}{c|}{$1,2,3,\ldots$} & \multicolumn{1}{c|}{$2,4,6,\ldots$} & $1,3,5,\ldots$ \\ \hline
\multicolumn{1}{l|}{$[s,1,1,1,1,1,1,1,1]_c$} & \multicolumn{1}{c|}{} & \multicolumn{1}{c|}{\cellcolor[HTML]{EFEFEF}{\color[HTML]{000000} $1$}} &  \\ \hline
\multicolumn{1}{l|}{$[s,1,1,1,1,1,1,1,0]$} & \multicolumn{1}{c|}{\cellcolor[HTML]{EFEFEF}1} & \multicolumn{1}{c|}{} &  \\ \hline
\multicolumn{1}{l|}{$[s,1,1,1,1,1,1,0,0]$} & \multicolumn{1}{c|}{} & \multicolumn{1}{c|}{} & \multicolumn{1}{c}{\cellcolor[HTML]{EFEFEF}2} \\ \hline
\multicolumn{1}{l|}{$[s,1,1,1,1,1,0,0,0]$} & \multicolumn{1}{c|}{\cellcolor[HTML]{EFEFEF}1} & \multicolumn{1}{c|}{} &  \\ \hline
\multicolumn{1}{l|}{$[s,1,1,1,1,0,0,0,0]$} & \multicolumn{1}{c|}{} & \multicolumn{1}{c|}{\cellcolor[HTML]{EFEFEF}2} &  \\ \hline
\multicolumn{1}{l|}{$[s,1,1,1,0,0,0,0,0]$} & \multicolumn{1}{c|}{\cellcolor[HTML]{EFEFEF}1} & \multicolumn{1}{c|}{} &  \\ \hline
\multicolumn{1}{l|}{$[s,1,1,0,0,0,0,0,0]$} & \multicolumn{1}{c|}{} & \multicolumn{1}{c|}{} & \multicolumn{1}{c}{\cellcolor[HTML]{EFEFEF}2} \\ \hline
\multicolumn{1}{l|}{$[s,1,0,0,0,0,0,0,0]$} & \multicolumn{1}{c|}{\cellcolor[HTML]{EFEFEF}1} & \multicolumn{1}{c|}{} &  \\ \hline
\multicolumn{1}{l|}{$[s,0,0,0,0,0,0,0,0]$} & \multicolumn{1}{c|}{} & \multicolumn{1}{c|}{\cellcolor[HTML]{EFEFEF}2} &  \\ \hline
\multicolumn{1}{c|}{Scalar ($\Delta=15$)} & \multicolumn{3}{c}{\cellcolor[HTML]{EFEFEF}1}  \\ \hline\hline 
\multicolumn{1}{c|}{$F^{(1)}$} \rule[-1.5ex]{0pt}{4.2ex}& \multicolumn{3}{c}{$-\frac{23133945892303}{99786996429120000}\log R$} \\\hline \\ 
\end{tabular}}
{\ssmall
\begin{tabular}{llll}
\textbf{AdS$_{\bf 5}\ O(N)$} &  &  &  \\ \hline\hline
\multicolumn{1}{c|}{$\alpha$} & \multicolumn{3}{c}{$s=$}\\\cline{2-4}
\multicolumn{1}{c|}{}& \multicolumn{1}{c|}{$1,2,3,\ldots$} & \multicolumn{1}{c|}{$2,4,6,\ldots$} & $1,3,5,\ldots$ \\ \hline
\multicolumn{1}{l|}{$[s,1]_c$} & \multicolumn{1}{c|}{} & \multicolumn{1}{c|}{\cellcolor[HTML]{EFEFEF}{\color[HTML]{000000} $1$}} &  \\ \hline
\multicolumn{1}{l|}{$[s,0]$} & \multicolumn{1}{c|}{\cellcolor[HTML]{EFEFEF}1} & \multicolumn{1}{c|}{} &  \\ \hline
\multicolumn{1}{c|}{Scalar ($\Delta=3$)} & \multicolumn{3}{c}{\cellcolor[HTML]{EFEFEF}2}    \\ \hline\hline 
\multicolumn{1}{c|}{$F^{(1)}$} \rule[-1.5ex]{0pt}{4.2ex}& \multicolumn{3}{c}{$\frac{11}{180}\log R$} \\\hline \\
\textbf{AdS$_{\bf 9}\ O(N)$} &  &  &  \\\hline\hline
\multicolumn{1}{c|}{$\alpha$} & \multicolumn{3}{c}{$s=$}\\\cline{2-4}
\multicolumn{1}{c|}{}& \multicolumn{1}{c|}{$1,2,3,\ldots$} & \multicolumn{1}{c|}{$2,4,6,\ldots$} & $1,3,5,\ldots$ \\ \hline
\multicolumn{1}{l|}{$[s,1,1,1]_c$} & \multicolumn{1}{c|}{} & \multicolumn{1}{c|}{\cellcolor[HTML]{EFEFEF}{\color[HTML]{000000} $1$}} &  \\ \hline
\multicolumn{1}{l|}{$[s,1,1,0]$} & \multicolumn{1}{c|}{\cellcolor[HTML]{EFEFEF}1} & \multicolumn{1}{c|}{} &  \\ \hline
\multicolumn{1}{l|}{$[s,1,0,0]$} & \multicolumn{1}{c|}{} & \multicolumn{1}{c|}{} & \multicolumn{1}{c}{\cellcolor[HTML]{EFEFEF}2} \\ \hline
\multicolumn{1}{l|}{$[s,0,0,0]$} & \multicolumn{1}{c|}{\cellcolor[HTML]{EFEFEF}1} & \multicolumn{1}{c|}{} &  \\ \hline
\multicolumn{1}{c|}{Scalar ($\Delta=7$)} & \multicolumn{3}{c}{\cellcolor[HTML]{EFEFEF}0}    \\ \hline\hline 
\multicolumn{1}{c|}{$F^{(1)}$} \rule[-1.5ex]{0pt}{4.2ex}& \multicolumn{3}{c}{$\frac{2497}{226800}\log R$} \\\hline \\
\textbf{AdS$_{\bf 13}\ O(N)$} &  &  &  \\ \hline\hline
\multicolumn{1}{c|}{$\alpha$} & \multicolumn{3}{c}{$s=$}\\\cline{2-4}
\multicolumn{1}{c|}{}& \multicolumn{1}{c|}{$1,2,3,\ldots$} & \multicolumn{1}{c|}{$2,4,6,\ldots$} & $1,3,5,\ldots$ \\ \hline
\multicolumn{1}{l|}{$[s,1,1,1,1,1]_c$} & \multicolumn{1}{c|}{} & \multicolumn{1}{c|}{\cellcolor[HTML]{EFEFEF}{\color[HTML]{000000} $1$}} &  \\ \hline
\multicolumn{1}{l|}{$[s,1,1,1,1,0]$} & \multicolumn{1}{c|}{\cellcolor[HTML]{EFEFEF}1} & \multicolumn{1}{c|}{} &  \\ \hline
\multicolumn{1}{l|}{$[s,1,1,1,0,0]$} & \multicolumn{1}{c|}{} & \multicolumn{1}{c|}{} & \multicolumn{1}{c}{\cellcolor[HTML]{EFEFEF}2} \\ \hline
\multicolumn{1}{l|}{$[s,1,1,0,0,0]$} & \multicolumn{1}{c|}{\cellcolor[HTML]{EFEFEF}1} & \multicolumn{1}{c|}{} &  \\ \hline
\multicolumn{1}{l|}{$[s,1,0,0,0,0]$} & \multicolumn{1}{c|}{} & \multicolumn{1}{c|}{\cellcolor[HTML]{EFEFEF}2} &  \\ \hline
\multicolumn{1}{l|}{$[s,0,0,0,0,0]$} & \multicolumn{1}{c|}{\cellcolor[HTML]{EFEFEF}1} & \multicolumn{1}{c|}{} &  \\ \hline
\multicolumn{1}{c|}{Scalar ($\Delta=11$)} &  \multicolumn{3}{c}{\cellcolor[HTML]{EFEFEF}2} 
 \\ \hline\hline 
\multicolumn{1}{c|}{$F^{(1)}$} \rule[-1.5ex]{0pt}{4.2ex}& \multicolumn{3}{c}{$\frac{92427157}{40864824000}\log R$} \\\hline \\
\textbf{AdS$_{\bf 17}\ O(N)$} &  &  &  \\\hline\hline
\multicolumn{1}{c|}{$\alpha$} & \multicolumn{3}{c}{$s=$}\\\cline{2-4}
\multicolumn{1}{c|}{}& \multicolumn{1}{c|}{$1,2,3,\ldots$} & \multicolumn{1}{c|}{$2,4,6,\ldots$} & $1,3,5,\ldots$ \\ \hline
\multicolumn{1}{l|}{$[s,1,1,1,1,1,1,1]_c$} & \multicolumn{1}{c|}{} & \multicolumn{1}{c|}{\cellcolor[HTML]{EFEFEF}{\color[HTML]{000000} $1$}} &  \\ \hline
\multicolumn{1}{l|}{$[s,1,1,1,1,1,1,0]$} & \multicolumn{1}{c|}{\cellcolor[HTML]{EFEFEF}1} & \multicolumn{1}{c|}{} &  \\ \hline
\multicolumn{1}{l|}{$[s,1,1,1,1,1,0,0]$} & \multicolumn{1}{c|}{} & \multicolumn{1}{c|}{} & \multicolumn{1}{c}{\cellcolor[HTML]{EFEFEF}2} \\ \hline
\multicolumn{1}{l|}{$[s,1,1,1,1,0,0,0]$} & \multicolumn{1}{c|}{\cellcolor[HTML]{EFEFEF}1} & \multicolumn{1}{c|}{} &  \\ \hline
\multicolumn{1}{l|}{$[s,1,1,1,0,0,0,0]$} & \multicolumn{1}{c|}{} & \multicolumn{1}{c|}{\cellcolor[HTML]{EFEFEF}2} &  \\ \hline
\multicolumn{1}{l|}{$[s,1,1,0,0,0,0,0]$} & \multicolumn{1}{c|}{\cellcolor[HTML]{EFEFEF}1} & \multicolumn{1}{c|}{} &  \\ \hline
\multicolumn{1}{l|}{$[s,1,0,0,0,0,0,0]$} & \multicolumn{1}{c|}{} & \multicolumn{1}{c|}{} & \multicolumn{1}{c}{\cellcolor[HTML]{EFEFEF}2} \\ \hline
\multicolumn{1}{l|}{$[s,0,0,0,0,0,0,0]$} & \multicolumn{1}{c|}{\cellcolor[HTML]{EFEFEF}1} & \multicolumn{1}{c|}{} &  \\ \hline
\multicolumn{1}{c|}{Scalar ($\Delta=15$)} & \multicolumn{3}{c}{\cellcolor[HTML]{EFEFEF}0}  \\ \hline\hline 
\multicolumn{1}{c|}{$F^{(1)}$} \rule[-1.5ex]{0pt}{4.2ex}& \multicolumn{3}{c}{$\frac{61430943169}{125046361440000}\log R$} \\\hline \\
\textbf{AdS$_{\bf 21}\ O(N)$} &  &  &  \\ \hline\hline
\multicolumn{1}{c|}{$\alpha$} & \multicolumn{3}{c}{$s=$}\\\cline{2-4}
\multicolumn{1}{c|}{}& \multicolumn{1}{c|}{$1,2,3,\ldots$} & \multicolumn{1}{c|}{$2,4,6,\ldots$} & $1,3,5,\ldots$ \\ \hline
\multicolumn{1}{l|}{$[s,1,1,1,1,1,1,1,1,1]_c$} & \multicolumn{1}{c|}{} & \multicolumn{1}{c|}{\cellcolor[HTML]{EFEFEF}{\color[HTML]{000000} 2}} &  \\ \hline
\multicolumn{1}{l|}{$[s,1,1,1,1,1,1,1,1,0]$} & \multicolumn{1}{c|}{\cellcolor[HTML]{EFEFEF}1} & \multicolumn{1}{c|}{} &  \\ \hline
\multicolumn{1}{l|}{$[s,1,1,1,1,1,1,1,0,0]$} & \multicolumn{1}{c|}{} & \multicolumn{1}{c|}{} & \multicolumn{1}{c}{\cellcolor[HTML]{EFEFEF}2} \\ \hline
\multicolumn{1}{l|}{$[s,1,1,1,1,1,1,0,0,0]$} & \multicolumn{1}{c|}{\cellcolor[HTML]{EFEFEF}1} & \multicolumn{1}{c|}{} &  \\ \hline
\multicolumn{1}{l|}{$[s,1,1,1,1,1,0,0,0,0]$} & \multicolumn{1}{c|}{} & \multicolumn{1}{c|}{\cellcolor[HTML]{EFEFEF}2} &  \\ \hline
\multicolumn{1}{l|}{$[s,1,1,1,1,0,0,0,0,0]$} & \multicolumn{1}{c|}{\cellcolor[HTML]{EFEFEF}1} & \multicolumn{1}{c|}{} &  \\ \hline
\multicolumn{1}{l|}{$[s,1,1,1,0,0,0,0,0,0]$} & \multicolumn{1}{c|}{} & \multicolumn{1}{c|}{} & \multicolumn{1}{c}{\cellcolor[HTML]{EFEFEF}2} \\ \hline
\multicolumn{1}{l|}{$[s,1,1,0,0,0,0,0,0,0]$} & \multicolumn{1}{c|}{\cellcolor[HTML]{EFEFEF}1} & \multicolumn{1}{c|}{} &  \\ \hline
\multicolumn{1}{l|}{$[s,1,0,0,0,0,0,0,0,0]$} & \multicolumn{1}{c|}{} & \multicolumn{1}{c|}{\cellcolor[HTML]{EFEFEF}2} &  \\ \hline
\multicolumn{1}{l|}{$[s,0,0,0,0,0,0,0,0,0]$} & \multicolumn{1}{c|}{\cellcolor[HTML]{EFEFEF}1} & \multicolumn{1}{c|}{} &  \\ \hline
\multicolumn{1}{c|}{Scalar ($\Delta=17$)} & \multicolumn{3}{c}{\cellcolor[HTML]{EFEFEF}2} \\ \hline\hline 
\multicolumn{1}{c|}{$F^{(1)}$} \rule[-1.5ex]{0pt}{4.2ex}& \multicolumn{3}{c}{$\frac{16399688681447}{149003207337600000}\log R$} \\\hline
\end{tabular}}
\caption{Projection of the non-minimal Type B theory to the Majorana/symplectic Majorana mimimal Type B theory in even $d$. Notice that in AdS$_{7}$ and AdS$_{15}$, where 
we impose a symplectic Majorana projection, the pattern does not exactly follow the one seen in the other dimensions, as explained in the text. Instead, they are `inverted', with the swapping of the towers for each weight from being only even integer spins to only odd integer spins. Their shift is highlighted in cyan. As defined earlier, the subscript `$c$' indicates that both selfdual and anti-selfdual parts are included, corresponding to the weights  $[t_1,\ldots,t_{k-1},t_k]$ and $[t_1,\ldots,t_{k-1},-t_k]$. 
}\label{Table:O(n)projection}
\end{center}
\end{table}

\paragraph{Majorana-Weyl projection}
\label{MWsec}
Finally the spectra arising from the Majorana-Weyl projection, which can be imposed in dimensions $d=2 \ \! (\!\!\!\!\!\mod 8)$, is the overlap of the individual Majorana and Weyl projection. The resulting spectrum yields the total zeta function
\begin{equation}	
\begin{aligned}
		\zeta_{\rm Type\ B\ MW}^{\text{HS}}(z) & = 
			 \sum_{s=2,4,6,\ldots}^\infty \!\!\!\sum_{\substack{t_i\geq t_{i+1} \\ t_i \geq 0 \\ \sum_w t_w = w \ \! (\!\!\!\!\!\! \mod 4)}}^1 \!\!\!\!\!\!\!\!\mathfrak{Z}_{(\Delta;[s,t_1,t_2,\ldots,t_{w-1},t_{w}])}(z) \\ 
			 & +  \sum_{s=1,3,5,\ldots} \!\!\!\!\!\!\!\sum_{\substack{t_i\geq t_{i+1} \\ t_i \geq 0 \\ \sum_w t_w = (w-2) \ \! (\!\!\!\!\!\! \mod 4)}}^1 \!\!\!\!\!\!\!\!\!\!\!\!\mathfrak{Z}_{(\dph;[s,t_1,t_2,\ldots,t_{w-1},t_{w}])}(z)\,.
\label{eqn:typeBMin}
\end{aligned} 
\end{equation}
An illustration of this can be seen in Table~\ref{table:majorana-weyl}, where we list the spectra of AdS$_{11}$ and AdS$_{19}$. Summing up over these spectra, we find 
the result
\begin{equation}
F^{(1)}_{\rm type~B~MW} = \frac{1}{2}a_f \log R
\end{equation}
which is the $a$-anomaly coefficient of a single Majorana-Weyl fermion at the boundary.
\begin{table}[h]
\begin{center}
{\scriptsize
\begin{tabular}{llll}
\multicolumn{3}{l}{\textbf{AdS$_{\bf 11}$ (Majorana-Weyl)}}  &   \\ \hline\hline
\multicolumn{1}{c|}{$\alpha$} & \multicolumn{3}{c}{$s=$}\\\cline{2-4}
\multicolumn{1}{c|}{}& \multicolumn{1}{c|}{$1,2,3,\ldots$} & \multicolumn{1}{c|}{$2,4,6,\ldots$} & $1,3,5,\ldots$ \\ \hline
\multicolumn{1}{l|}{$[s,1,1,1,1]$} & \multicolumn{1}{c|}{} & \multicolumn{1}{c|}{\cellcolor[HTML]{EFEFEF}1} &  \\ \hline
\multicolumn{1}{l|}{$[s,1,1,0,0]$} & \multicolumn{1}{c|}{} & \multicolumn{1}{c|}{} & \multicolumn{1}{c}{\cellcolor[HTML]{EFEFEF}1}\\ \hline
\multicolumn{1}{l|}{$[s,0,0,0,0]$} & \multicolumn{1}{c|}{} & \multicolumn{1}{c|}{\cellcolor[HTML]{EFEFEF}1} &  \\ \hline
\hline
\multicolumn{1}{c|}{$F^{(1)}$} \rule[-1.5ex]{0pt}{4.2ex}& \multicolumn{3}{c}{$\frac{-14797}{5987520}\log R$} 
\\\hline \\ \\ \\ \\
\multicolumn{3}{l}{\textbf{AdS$_{\bf 7}$ (Symplectic Majorana-Weyl)}}   &   \\ \hline\hline
\multicolumn{1}{c|}{$\alpha$} & \multicolumn{3}{c}{$s=$}\\\cline{2-4}
\multicolumn{1}{c|}{}& \multicolumn{1}{c|}{$1,2,3,\ldots$} & \multicolumn{1}{c|}{$2,4,6,\ldots$} & $1,3,5,\ldots$ \\ \hline
\multicolumn{1}{l|}{$[s,1,1]$} & \multicolumn{1}{c|}{} & \multicolumn{1}{c|}{} & \multicolumn{1}{c}{\cellcolor[HTML]{EFEFEF}1}  \\ \hline
\multicolumn{1}{l|}{$[s,0,0]$} & \multicolumn{1}{c|}{} & \multicolumn{1}{c|}{\cellcolor[HTML]{EFEFEF}1} &  \\ \hline \hline
\multicolumn{1}{c|}{$F^{(1)}$} \rule[-1.5ex]{0pt}{4.2ex}& \multicolumn{3}{c}{$\frac{191}{15120}\log R$} 
\\\hline \\ \\ \\[3ex] 
\end{tabular}}
{\scriptsize \begin{tabular}{llll}
\multicolumn{2}{l}{\textbf{AdS$_{\bf 19}$ (Majorana-Weyl)}}  &  &   \\ \hline\hline
\multicolumn{1}{c|}{$\alpha$} & \multicolumn{3}{c}{$s=$}\\\cline{2-4}
\multicolumn{1}{c|}{}& \multicolumn{1}{c|}{$1,2,3,\ldots$} & \multicolumn{1}{c|}{$2,4,6,\ldots$} & $1,3,5,\ldots$ \\ \hline
\multicolumn{1}{l|}{$[s,1,1,1,1,1,1,1,1]$} & \multicolumn{1}{c|}{} & \multicolumn{1}{c|}{\cellcolor[HTML]{EFEFEF}{\color[HTML]{000000} 1}} &  \\ \hline
\multicolumn{1}{l|}{$[s,1,1,1,1,1,1,0,0]$} & \multicolumn{1}{c|}{} & \multicolumn{1}{c|}{} & \multicolumn{1}{c}{\cellcolor[HTML]{EFEFEF}1} \\ \hline
\multicolumn{1}{l|}{$[s,1,1,1,1,0,0,0,0]$} & \multicolumn{1}{c|}{} & \multicolumn{1}{c|}{\cellcolor[HTML]{EFEFEF}1} &  \\ \hline
\multicolumn{1}{l|}{$[s,1,1,0,0,0,0,0,0]$} & \multicolumn{1}{c|}{} & \multicolumn{1}{c|}{} & \multicolumn{1}{c}{\cellcolor[HTML]{EFEFEF}1} \\ \hline
\multicolumn{1}{l|}{$[s,0,0,0,0,0,0,0,0]$} & \multicolumn{1}{c|}{} & \multicolumn{1}{c|}{\cellcolor[HTML]{EFEFEF}1} &  \\ \hline
\hline
\multicolumn{1}{c|}{$F^{(1)}$} \rule[-1.5ex]{0pt}{4.2ex}& \multicolumn{3}{c}{$-\frac{23133945892303}{199573992858240000}\log R$} \\\hline \\ \\
\multicolumn{3}{l}{\textbf{AdS$_{\bf 15}$ (Symplectic Majorana-Weyl)}}   &   \\ \hline\hline
\multicolumn{1}{c|}{$\alpha$} & \multicolumn{3}{c}{$s=$}\\\cline{2-4}
\multicolumn{1}{c|}{}& \multicolumn{1}{c|}{$1,2,3,\ldots$} & \multicolumn{1}{c|}{$2,4,6,\ldots$} & $1,3,5,\ldots$ \\ \hline
\multicolumn{1}{l|}{$[s,1,1,1,1,1,1]$} & \multicolumn{1}{c|}{} & \multicolumn{1}{c|}{}   & \multicolumn{1}{c}{\cellcolor[HTML]{EFEFEF}1} \\ \hline
\multicolumn{1}{l|}{$[s,1,1,1,1,0,0]$} & \multicolumn{1}{c|}{} & \multicolumn{1}{c|}{\cellcolor[HTML]{EFEFEF}1}& \\ \hline
\multicolumn{1}{l|}{$[s,1,1,0,0,0,0]$} & \multicolumn{1}{c|}{} &  \multicolumn{1}{c|}{}   & \multicolumn{1}{c}{\cellcolor[HTML]{EFEFEF}1} \\ \hline
\multicolumn{1}{l|}{$[s,0,0,0,0,0,0]$} & \multicolumn{1}{c|}{} & \multicolumn{1}{c|}{\cellcolor[HTML]{EFEFEF}1} &  \\ \hline
\hline
\multicolumn{1}{c|}{$F^{(1)}$} \rule[-1.5ex]{0pt}{4.2ex}& \multicolumn{3}{c}{$-\frac{36740617}{70053984000}\log R$} 
\\\hline \\ \\ 
\end{tabular}}
\caption{Table of weights and their towers of spins for \emph{(top left)} AdS$_{11}$ and \emph{(top right)} AdS$_{19}$ under Majorana-Weyl projection, and for \emph{(bottom left)} AdS$_{7}$ and \emph{(bottom right)} AdS$_{15}$ under the Symplectic Majorana-Weyl projection. There are no subscripts $_c$ for the $[s,1,\ldots,1]$ representations because the dual representations $[s,1,\ldots,1,-1]$ are not included.} 
\label{table:majorana-weyl}
\end{center}
\end{table}

In $d=6$ (mod 8) one may impose a symplectic Majorana-Weyl projection. The resulting spectra are the overlap between the symplectic Majorana and Weyl projections. For instance, 
in $d=6$ we find a minimal theory with a totally symmetric tower $[s,0,0]$ of all even spins, and a tower of the mixed symmetry fields $[s,1,1]$ of all odd spins (see Table~\ref{table:majorana-weyl}). In this case (and similarly for higher $d=14,22,\ldots$), we find
\begin{equation}
F^{(1)}_{\rm type~B~SMW} = -\frac{1}{2}a_f \log R\,.
\end{equation} 
Since the $a$-anomaly of the boundary free theory of $N$ symplectic Majorana-Weyl fermions is $a_{N~{\rm SMW}}=\frac{N}{2} a_f$, this result is 
consistent with a shift $G_N^{\rm type~B~SMW} \sim 1/(N+1)$. 

\paragraph{Sample Calculations}
\subparagraph{AdS$_5$}
Following \eqref{dolan1} for the non-minimal Type B theory,
\begin{align}
\zeta_{\rm type\ B}^{\rm HS}(z) = 2\zeta_{(3;[0,0])}(z) + \sum_{s=1}^\infty\Big( \mathfrak{Z}_{(\dph;[s,1])}(z)+\mathfrak{Z}_{(\dph;[s,-1])}(z)\Big) + 2 \sum_{s=1}^\infty\mathfrak{Z}_{(\dph;[s,0])}(z).
\end{align}
We see that there are two weights to consider in AdS$_5$, corresponding to $[s,0]$ and $[s,\pm 1]$ representation. Using (\ref{zeta1}) and \ref{mu-even}, 
we have
	\begin{align}
		\frac{\zeta_{(\Delta;[s,1])}(z)}{\log R} &= \pi^2\int_0^\infty du \frac{\left(u^2+1\right) \left[(s+1)^2+u^2\right]}{12 \pi ^3}\frac{s(2+s)}{\left[u^2+(\Delta-2)^2\right]^z} \nonn \\
		& =\frac{s(s+2)}{8\sqrt{\pi}\Gamma(z)}\Bigg[ \frac{ (s+1)^2  (\Delta -2)^{1-2 z} \Gamma \left(z-\frac{1}{2}\right)}{3}+\frac{ \left((s+1)^2+1\right) (\Delta -2)^{3-2 z} \Gamma \left(z-\frac{3}{2}\right)}{6}\nonn\\&\qquad\qquad\quad\quad+\frac{ (\Delta -2)^{5-2 z} \Gamma \left(z-\frac{5}{2}\right)}{4}\Bigg]
\end{align}
In the above, we made use of the formula,
	\begin{align}
		\int_0^\infty du \frac{u^{2p}}{\left[u^2+\nu^2\right]^z} = \nu^{2p+1-2z}\int_0^\infty du \frac{u^{2p}}{\left[u^2+1\right]^z} = \nu^{2p+1-2z}\frac{\Gamma(p+\frac{1}{2})}{2}\frac{\Gamma(z-p-\frac{1}{2})}{\Gamma(z)},
	\end{align}%
to go from the first to the second line.

In our regularization scheme, we sum over the physical modes separately from the ghost modes. We introduce $\zeta(k,\nu)$, the Hurwitz zeta function 
(analytically extended to the entire complex plane), which is given by
\begin{align}
\zeta(k,\nu) = \sum_{s=0}^\infty \frac{1}{(s+\nu)^k}. \label{eqn:HurwitzZetafunction}
\end{align}
Then, using $\Delta^{\rm gh} = s+3$,
	\begin{align}
		&\frac{1}{\log R}\sum_{s=1}^\infty \zeta_{(\Delta^{\rm gh};[s-1,1])}(z) \nonn \\
		& =\frac{1}{96\sqrt{\pi}\Gamma(z)}\Bigg[2 \zeta (2 z-7) \Gamma \left(z-\frac{3}{2}\right)+3 \zeta (2 z-7) \Gamma \left(z-\frac{5}{2}\right)+8 \zeta (2 z-6) \Gamma \left(z-\frac{3}{2}\right)\nonn \\
		&\qquad\qquad\qquad+6 \zeta (2 z-6) \Gamma \left(z-\frac{5}{2}\right)+4 \zeta (2 z-5) \Gamma \left(z-\frac{1}{2}\right)+12 \zeta (2 z-5) \Gamma \left(z-\frac{3}{2}\right)\nonn \\
		&\qquad\qquad\qquad+16 \zeta (2 z-4) \Gamma \left(z-\frac{1}{2}\right)+8 \zeta (2 z-4) \Gamma \left(z-\frac{3}{2}\right)+20 \zeta (2 z-3) \Gamma \left(z-\frac{1}{2}\right)\nonn \\
		&\qquad\qquad\qquad+8 \zeta (2 z-2) \Gamma \left(z-\frac{1}{2}\right)\Bigg].\label{AdS5gh}
	\end{align}
Similarly, using $\Delta^{\rm ph}= s+2$,
	\begin{align}
		&\frac{1}{\log R}\sum_{s=1}^\infty \zeta_{(\Delta^{\rm ph};[s,1])}(z) \nonn \\
		& = \frac{1}{96 \sqrt{\pi } \Gamma (z)}\Bigg\{\Gamma \left(z-\frac{5}{2},1\right)\Big[6 \zeta (2z-6,1) +3 \zeta (2 z-7,1) \Big]\nonn\\
			&\qquad\qquad\qquad
				+\Gamma \left(z-\frac{3}{2}\right)\Big[12 \zeta (2 z-5,1) +2 \zeta (2 z-7,1)+8 \zeta (2z-6,1) \Gamma+8 \zeta (2z-4,1) \Big]\nonn\\
			&\qquad\qquad\qquad+ \left(z-\frac{1}{2}\right)\Big[16 \zeta (2z-4,1) \Gamma+8 \zeta (2z-2,1) +20 \zeta (2 z-3,1) +4 \zeta (2 z-5,1)\Big]\Bigg\}\label{AdS5ph}
	\end{align}
Putting \eqref{AdS5gh} and \eqref{AdS5ph} together,
\begin{align}
&\quad \frac{1}{\log R}\sum_{s=1}^\infty\left[ \zeta_{(\Delta^{\rm ph};[s,1])}(z) - \zeta_{(\Delta^{\rm gh};[s-1,1])}(z) \right]  \nonn\\
&= -\frac{1}{90}z+\frac{1}{180}  \left[56 \zeta '(-6)-160 \zeta '(-4)-120 \zeta '(-2)-2 \gamma +3 \psi \left(-\frac{5}{2}\right)-5 \psi \left(-\frac{3}{2}\right)\right]z^2+\mathcal{O}(z^3) \label{ads5mixed}
\end{align}
where $\psi(x)$ is the digamma function, $\gamma$ the Euler-Mascheroni constant, and $\zeta'$ the derivative of the Riemann Zeta function $\zeta(z)$ (which is related to the Hurwitz Zeta function $\zeta(z) = \zeta(z,1)$).
Similarly, for the totally symmetric representation $[s,0]$, we have
\begin{align}
&\quad\frac{1}{\log R}\sum_{s=1}^\infty\mathfrak{Z}_{(\dph;[s,0])}(z) \nonn \\
&=\frac{1}{\log R}\sum_{s=1}^\infty\left[ \zeta_{(\Delta^{\rm ph};[s,0])}(z)-\zeta_{(\Delta^{\rm gh};[s-1,0])}(z)\right] \nonn \\
& = \frac{1}{24\sqrt{\pi}\Gamma(z)}\left[3 \zeta (2 (z-3)) \Gamma \left(z-\frac{5}{2}\right)+4 (\zeta (2 (z-3))+\zeta (2 (z-2))) \Gamma \left(z-\frac{3}{2}\right)\right] \nonn \\
&= \left(\frac{14}{45} \zeta '(-6)+\frac{4}{9} \zeta '(-4)\right)z^2+\mathcal{O}(z^3) \label{ads5symm}
\end{align}
Finally, for the massive scalar with $\Delta = 3$, we have
\begin{align}
\frac{\zeta_{(3;[0,0])}(z)}{\log R} & = \left.\int_0^\infty du \frac{ (s+1)^2 u^2 \left[(s+1)^2+u^2\right] }{12 \pi \left[(\Delta -2)^2+u^2\right]^{z}} \right|_{\substack{\Delta = 3,\\s = 0}}\nonn \\
& = \left[\frac{(s+1)^4 (\Delta -2)^{3-2 z} \Gamma \left(z-\frac{3}{2}\right)}{48 \sqrt{\pi } \Gamma (z)}+\frac{(s+1)^2 (\Delta -2)^{5-2 z} \Gamma \left(z-\frac{5}{2}\right)}{32 \sqrt{\pi } \Gamma (z)} \right]_{\substack{\Delta=3,\\s=0}}\nonn \\
& = \frac{\Gamma \left(z-\frac{3}{2}\right)}{48 \sqrt{\pi } \Gamma (z)}+\frac{\Gamma \left(z-\frac{5}{2}\right)}{32 \sqrt{\pi } \Gamma (z)} \nonn \\
& = \frac{1}{90}z+\frac{1}{180} \left[2 \gamma -3 \psi \left(-\frac{5}{2}\right)+5 \psi \left(-\frac{3}{2}\right)\right]z^2+\mathcal{O}(z^3).\label{ads5scalar}\end{align}
When summing \eqref{ads5mixed}, \eqref{ads5symm}, and \eqref{ads5scalar} together, there are no terms of order $\mathcal{O}(z^0)$ or $\mathcal{O}(z^1)$ in the sum, and hence, taking $z\rightarrow0$, we obtain $F^{(1)} = 0$ for the non-minimal Type B theory.

For the Type B minimal theory, we should evaluate, according to \eqref{eqn:typeBOn}, the following sum
\begin{align}
\zeta_{\rm Total-Type\ B}^{\rm HS}(z) =2\zeta_{(3;[0,0])}(z) + \sum_{s=2,4,6,\ldots}\big(\mathfrak{Z}_{(\dph;[s,1])}(z)+\mathfrak{Z}_{(\dph;[s,-1])}(z)\big)+\sum_{s=1}^\infty\mathfrak{Z}_{(\dph;[s,0])}(z).
\end{align}
The first and third term of the sum have already been evaluated for in the non-minimal theory in \eqref{ads5scalar} and \eqref{ads5symm} respectively. For the second term,
\begin{align}
&\quad\sum_{s=2,4,6,\ldots}\big(\mathfrak{Z}_{(\dph;[s,1])}+\mathfrak{Z}_{(\dph;[s,-1])}\big) \nonn \\
& =2\sum_{s=2,4,6,\ldots} \Bigg[\frac{ (s+1)^2 (s+2) s^{2-2 z} \Gamma \left(z-\frac{1}{2}\right)}{24 \sqrt{\pi } \Gamma (z)}+\frac{ (s+2) \left((s+1)^2+1\right) s^{4-2 z} \Gamma \left(z-\frac{3}{2}\right)}{48 \sqrt{\pi } \Gamma (z)}\nonn \\
&\qquad\qquad+\frac{ (s+2) s^{6-2 z} \Gamma \left(z-\frac{5}{2}\right)}{32 \sqrt{\pi } \Gamma (z)}
\Bigg]\log R \nonn \\
& \quad-2\sum_{s=2,4,6,\ldots} \Bigg[\frac{(s-1) s^2 (s+1)^{2-2 z} \Gamma \left(z-\frac{1}{2}\right)}{24 \sqrt{\pi } \Gamma (z)}+\frac{(s-1) \left(s^2+1\right) (s+1)^{4-2 z} \Gamma \left(z-\frac{3}{2}\right)}{48 \sqrt{\pi } \Gamma (z)}\nonn \\
&\qquad\qquad +\frac{(s-1) (s+1)^{6-2 z} \Gamma \left(z-\frac{5}{2}\right)}{32 \sqrt{\pi } \Gamma (z)}\Bigg]\log R
\end{align}
To illustrate the zeta-regularization, let us consider the last term,
\begin{align}
&\quad\sum_{s=2,4,6,\ldots} \frac{(s-1) (s+1)^{6-2 z} \Gamma \left(z-\frac{5}{2}\right)}{32 \sqrt{\pi } \Gamma (z)} \nonn \\
&=\sum_{s=2,4,6,\ldots}\left[\frac{ (s+1)^{7-2 z} \Gamma \left(z-\frac{5}{2}\right)}{32 \sqrt{\pi } \Gamma (z)}-2\frac{ (s+1)^{6-2 z} \Gamma \left(z-\frac{5}{2}\right)}{32 \sqrt{\pi } \Gamma (z)}\right] \nonn \\
& = \sum_{s=1,2,3,\ldots}\left[\frac{ 2^{7-2z}(s+\frac{1}{2})^{7-2 z} \Gamma \left(z-\frac{5}{2}\right)}{32 \sqrt{\pi } \Gamma (z)}-2\frac{ 2^{6-2z}(s+\frac{1}{2})^{6-2 z} \Gamma \left(z-\frac{5}{2}\right)}{32 \sqrt{\pi } \Gamma (z)}\right] \nonn \\
& = \frac{2^{2-2 z} \zeta \left(2 z-7,\frac{3}{2}\right) \Gamma \left(z-\frac{5}{2}\right)}{\sqrt{\pi } \Gamma (z)}-\frac{2^{2-2 z} \zeta \left(2 z-6,\frac{3}{2}\right) \Gamma \left(z-\frac{5}{2}\right)}{\sqrt{\pi } \Gamma (z)}
\end{align} where on the second line we used the substitution $s\rightarrow 2s$, followed by rewriting $2s+1 = 2(s+\frac{1}{2})$.\footnote{Similar shifts and scaling will be applied in the higher dimensional Type B cases, as well as the Type AB and C cases, and details of transformations to the Hurwitz-zeta function can be found in Appendix~\ref{appendix:zetas}.}
The partial results coming from summing each tower are given in Table~\ref{Table:ads5b}. Putting everything together, 
we obtain $F^{(1)}_{\rm type~B~Maj.} = \frac{11}{180}\log R=a_f^{d=4}\log R$, which agrees with the results of~\cite{Beccaria:2014xda}. 

Finally, for the Weyl truncated theory,
\begin{align}
\zeta_{\rm type\ B\ Weyl}^{\rm HS}(z) =\sum_{s=1}^\infty\mathfrak{Z}_{(\dph;[s,0])} (z)=\mathcal{O}(z^2),
\end{align}
which gives us $F^{(1)} = 0$.

%

%
%
\subparagraph{AdS$_{11}$}
We skip the $d=7,9$ case, whose spectrum for the various theories follow from the discussion in Section~\ref{subsect:oddtypeb}. For reference, the calculated free energy of each weight $F^{(1)}$ is given in Tables~\ref{table:ads7b} and~\ref{appendix:ads9b}. 

Instead, let us consider the $d=11$ case, where we can compare the four different types of fermions: non-minimal ($U(N)$), Weyl, minimal ($O(N)$), and Majorana-Weyl. The calculations of $F^{(1)}$ for each the various weights and their spectra are given in Table~\ref{appendix:ads11b}. In the non-minimal and Weyl projected theories, the bulk $F^{(1)}$ contributions sum to zero, whereas in the minimal and Majorana-Weyl theories, the bulk $F^{(1)}$ contributions are $-14797/2993760 \log R$ and $-14797/ 5987520 \log R$ respectively. The numerical parts of these free energies correspond exactly to the values of the free energy of one real fermion, $-14797/2993760$ and one real Weyl fermion on $S^{10}$, $-14797/ 5987520$. 


\subsubsection{Fermionic Higher Spins in Type AB Theories}
\paragraph{Spectrum}
We described earlier that there is only one irrep of $SO(d)$ of interest here that describes 
the tower of spins corresponding to the fermionic bilinears in type AB theories, namely 
$\alpha_s = [s,\frac{1}{2},\frac{1}{2},\ldots,\frac{1}{2}]$. Therefore, in the non-minimal theories dual to complex scalars and fermions in the 
$U(N)$ singlet sector, the purely fermionic contribution to the 
total zeta function is
\begin{align}
\zeta_{\rm type\ AB\ ferm}^{\rm HS} (z) = 2\zeta_{(\Delta_{1/2};[\frac{1}{2},\frac{1}{2},\frac{1}{2},\ldots,\frac{1}{2}])}^{\rm HS}(z)+2\sum_{s=\frac{3}{2},\frac{5}{2},\ldots}^\infty \mathfrak{Z}_{(\dph;[s,\frac{1}{2},\frac{1}{2},\ldots,\frac{1}{2}])}(z),\label{eqn:typeabspec}
\end{align} where $\Delta_{1/2} = \frac{1}{2}+d-2 = d-\frac{3}{2}$. Thus, the spectrum of spins gives us 
a massive Dirac fermion\footnote{With mass $|m|=(\Delta_{1/2}-d/2)/2=(d-3)/4$.}, and a tower of complex massless higher-spin fermionic fields.\footnote{
The factor of 2 in (\ref{eqn:typeabspec}) just accounts for the fact that the representations are complex.} 
\paragraph{Sample Calculation: AdS$_5$}
After collecting our equations following \eqref{eqn:typeabspec}, we have,
\begin{align}
\zeta_{\rm type~AB~ferm}^{\rm HS} (z) = 2\zeta_{(\Delta_{1/2};[\frac{1}{2},\frac{1}{2}])}^{\rm HS}(z)+2\sum_{s=\frac{3}{2},\frac{5}{2},\ldots}^\infty \mathfrak{Z}_{(\dph;[s,\frac{1}{2}])}(z),
\end{align} with $\dph = 2+s$. For the massive fermion contribution,
	\begin{align}
		\zeta_{(\Delta_{1/2};[s,\frac{1}{2}])}(z) 
		& = \frac{2^{2 z-10} \left(36 \Gamma \left(z-\frac{1}{2}\right)+20 \Gamma \left(z-\frac{3}{2}\right)+3 \Gamma \left(z-\frac{5}{2}\right)\right)}{3 \sqrt{\pi } \Gamma (z)}.
	\end{align}

\noindent Then, 
\begin{align}
\frac{\zeta_{(\Delta;[s,\frac{1}{2}])}(z)}{\log R} & = 
\frac{(s+\frac{1}{2})(s+\frac{3}{2})}{48\pi(\Delta-2)^{2z-1}\Gamma(z)}\nonn\\&\quad\times\left[
(\Delta-2)^4\Gamma\left(\frac{5}{2}\right)\Gamma\left(z-\frac{5}{2}\right) + \Gamma\left(\frac{3}{2}\right)\Gamma\left(z-\frac{3}{2}\right)(\Delta-2)^2\left(\frac{1}{4}+(s+1)^2\right)\right.\nonn\\&\qquad\quad\left.+ \Gamma\left(\frac{1}{2}\right)\Gamma\left(z-\frac{1}{2}\right)\frac{(s+1)^2}{4}
\right].
\end{align} 

This gives us,
\begin{align}
&\quad \sum_{s=\frac{3}{2},\frac{5}{2},\ldots} \zeta_{(\Delta^{\rm ph};[s,\frac{1}{2}])}(z) \nonn\\
& = \frac{1}{{1536 \sqrt{\pi } \Gamma (z)}}\left\{6 \left[4 \zeta \left(2 z-7,\frac{3}{2}\right)+8 \zeta \left(2 z-6,\frac{3}{2}\right)+3 \zeta \left(2 z-5,\frac{3}{2}\right)\right] \Gamma \left(z-\frac{5}{2}\right)\right.\nonn\\
&\quad\left.+\left[16 \zeta \left(2 z-7,\frac{3}{2}\right)+64 \zeta \left(2 z-6,\frac{3}{2}\right)+96 \zeta \left(2 z-5,\frac{3}{2}\right)+64 \zeta \left(2 z-4,\frac{3}{2}\right)\right.\right.\nonn\\&\qquad\left.\left.+15 \zeta \left(2 z-3,\frac{3}{2}\right)\right] \Gamma \left(z-\frac{3}{2}\right)+2 \left[4 \zeta \left(2 z-5,\frac{3}{2}\right)+16 \zeta \left(2 z-4,\frac{3}{2}\right)\right.\right.\nonn\\&\qquad\qquad\left.\left.+23 \zeta \left(2 z-3,\frac{3}{2}\right)+14 \zeta \left(2 z-2,\frac{3}{2}\right)+3 \zeta \left(2 z-1,\frac{3}{2}\right)\right] \Gamma \left(z-\frac{1}{2}\right) \right\}.
\intertext{The technicalities of the shift to the Hurwitz Zeta function in the sum above is similar to the case for the minimal Type B theory in AdS$_5$ which we worked out earlier. More details can be found in Appendix \ref{appendix:zetas}. Similarly,}
&\quad\sum_{s=\frac{3}{2},\frac{5}{2},\ldots} \zeta_{(\Delta^{\rm gh};[s-1,\frac{1}{2}])}(z) \nonn\\
& =\frac{1}{1536 \sqrt{\pi } \Gamma (z)}\Bigg\{6 \left[4 \zeta \left(2 z-7,\frac{5}{2}\right)-8 \zeta \left(2 z-6,\frac{5}{2}\right)+3 \zeta \left(2 z-5,\frac{5}{2}\right)\right] \Gamma \left(z-\frac{5}{2}\right)\nonn\\&\quad+\left[16 \zeta \left(2 z-7,\frac{5}{2}\right)-64 \zeta \left(2 z-6,\frac{5}{2}\right)+96 \zeta \left(2 z-5,\frac{5}{2}\right)-64 \zeta \left(2 z-4,\frac{5}{2}\right)\right.\nonn\\&\quad\left.+15 \zeta \left(2 z-3,\frac{5}{2}\right)\right] \Gamma \left(z-\frac{3}{2}\right)+2 \left[4 \zeta \left(2 z-5,\frac{5}{2}\right)-16 \zeta \left(2 z-4,\frac{5}{2}\right)\right.\nonn\\&\quad\left.+23 \zeta \left(2 z-3,\frac{5}{2}\right)-14 \zeta \left(2 z-2,\frac{5}{2}\right)+3 \zeta \left(2 z-1,\frac{5}{2}\right)\right] \Gamma \left(z-\frac{1}{2}\right) \Bigg\}.
\end{align}
Quite clearly, the Hurwitz-zeta function shifts differently for the physical and ghost modes.
Adding all three contributions and expanding near $z=0$,
\begin{align}
\zeta_{\rm type~AB~ferm}^{\rm HS} (z) = \mathcal{O}(z^2)
\end{align}
which implies that $F^{(1)}_{\rm type~AB~ferm}= 0$, consistently with the duality. 

For reference, we also report the expected expression of $\zeta_{\rm type~AB~ferm}^{\rm HS} (z)$ for AdS$_7$ and AdS$_9$, expanded in $z$ up to the second order, in Appendix~\ref{Appendix:AB}.

\subsubsection{Type C Theories}
\label{typeC-sec}
Calculations for Type C theories are similar to those described above and we will not go through all details explicitly. 
In the following sections, we list the spectrum of fields in these theories, including their various possible truncations.
The free energy contributions in a few explicit examples are collected for reference in Appendix~\ref{appendix:TypeO}.
\paragraph{Spectrum}
\label{subsubsection:typecweights}
The spectrum of the non-minimal type C theories, dual to the free theory of $N$ complex $d/2$-form gauge fields with $U(N)$ singlet constraint, 
can be obtained from the character formulas derived in \cite{Dolan:2005wy}. While the resulting spectra may look complicated, 
they follow a clear pattern that can be rather easily identified if one refers to the tables given in Appendix \ref{appendix:TypeO}. The results 
are split into the cases $d=4m$ and $d=4m+2$.  For $d=4m$, the total spectral zeta function is given by\footnote{\eqref{eqn:typecnonmin4m} and \eqref{eqn:typecnonmin4m2} correspond to equations (4.20)-(4.21) and (4.22)-(4.23) of \cite{Dolan:2005wy} respectively, and the tensorial decomposition in these quoted equations can be further simplified by the formulas on p. 104 of \cite{frappat2000dictionary}.}
\footnote{For all Type C theories, the field of spin $s=2$ in the towers of spins of representation $[s,2,\ldots]$ are not gauge fields, but we will still use the symbol $\mathfrak{Z}$ for conciseness. See footnote~\ref{footnote:typebgauge} for similar remarks.}
\begin{align}
\zeta_{\rm type~C}^{\rm HS} &(z)=2\sum_{\substack{k_i\geq0 \ \ \\k_i\geq k_{i+1}}}^1\zeta_{(4m;[k_1,k_1,\ldots,k_m,k_m])}(z)\nonn \\
&+\sum_{s=2}^\infty\Bigg[ \sum_{\substack{\footnotesize t_i\geq0 \ \ \   \\ \scriptsize t_i\geq t_{i+1}}}^2 \!\!2\mathfrak{Z}_{(\dph;[s,t_1,t_1,t_2,t_2,\ldots,t_{m-1},t_{m-1},0])}(z) \nonn\\
&+\sum_{\substack{\footnotesize j_i\geq0 \ \ \   \\ \scriptsize j_i\geq j_{i+1}}}^2 \!\!\Big( \mathfrak{Z}_{(\dph;[s,2,j_1,j_1,j_2,j_2,\ldots,j_{m-1},+j_{m-1}])}(z) +\mathfrak{Z}_{(\dph;[s,2,j_1,j_1,j_2,j_2,\ldots,j_{m-1},-j_{m-1}])}(z)\Big)\Bigg] \label{eqn:typecnonmin4m}
\intertext{and for $d=4m+2$}
\zeta_{\rm type~C}^{\rm HS} &(z)=2\sum_{\substack{k_i\geq0 \ \ \\k_i\geq k_{i+1}}}^1 \zeta_{(4m+2;[k_1,k_1,\ldots,k_m,k_m,0])}(z)\nonn \\
&+\sum_{s=2}^\infty\Bigg[ \sum_{\substack{\footnotesize t_i\geq0 \ \ \   \\ \scriptsize t_i\geq t_{i+1}}}^2 \!\!2\mathfrak{Z}_{(\dph;[s,2,t_1,t_1,t_2,t_2,\ldots,t_{m-1},t_{m-1},0])}(z) \nonn\\
&+\sum_{\substack{\footnotesize j_i\geq0 \ \ \   \\ \scriptsize j_i\geq j_{i+1}}}^2 \!\!\Big( \mathfrak{Z}_{(\dph;[s,j_1,j_1,j_2,j_2,\ldots,j_{m-1},+j_{m-1}])}(z) +\mathfrak{Z}_{(\dph;[s,j_1,j_1,j_2,j_2,\ldots,j_{m-1},-j_{m-1}])}(z) \Big)\Bigg] \label{eqn:typecnonmin4m2}
\end{align}
Using these spectra and (\ref{mu-even}) to compute the zeta functions, we find the results 
\begin{equation}
\begin{aligned}
&F^{(1)}_{\rm type~C} = 2a_{d/2-{\rm form}} \log R\,,\quad d=4m\\
&F^{(1)}_{\rm type~C} = -2a_{d/2-{\rm form}} \log R\,,\quad d=4m+2
\label{F-typeC}
\end{aligned}
\end{equation}
where $a_{d/2-{\rm form}}$ is the $a$-anomaly coefficient of a single real $(d/2-1)$-form gauge field in dimension $d$. 
The first few values in $d=4,6,8,\ldots$ read \cite{Cappelli:2000fe}
\begin{equation}
a_{d/2-{\rm form}}=\left\{\frac{62}{90},-\frac{221}{210},\frac{8051}{5670},-\frac{1339661}{748440},
\frac{525793111}{243243000},-\frac{3698905481}{1459458000},\ldots  \right\}\,.
\end{equation}
Thus, we see that (\ref{F-typeC}) is consistent with the duality provided $G_N^{\rm type~C}\sim 1/(N-1)$ in $d=4m$, 
and $G_N^{\rm type~C}\sim 1/(N+1)$ in $d=4m+2$.


\subparagraph{Minimal Type C $O(N)$}
The ``minimal type C'' theory corresponds to the $O(N)$ singlet sector of the free theory of $N$ $(d/2-1)$-form gauge fields. Its spectrum 
can be in principle obtained by appropriately ``symmetrizing'' the character formulas given in \cite{Dolan:2005wy} 
and used above to obtain the non-minimal spectrum. The spectra in $d=4$ and $d=6$ were obtained in \cite{Beccaria:2014xda,Beccaria:2014zma,Beccaria:2014qea}. 
Generalizing those results for all $d$, we arrive at the following total spectral zeta functions. In $d=4m$,
\begin{align}
&\zeta_{\rm min.~type~C}^{\rm HS}(z)=2\sum_{\substack{k_i\geq0 \ \ \\k_i\geq k_{i+1}}}^1\zeta_{(4m;[k_1,k_1,k_1,k_1,\ldots,k_{\lfloor\frac{m}{2}\rfloor},k_{\lfloor\frac{m}{2}\rfloor},k_{\lfloor\frac{m}{2}\rfloor},k_{\lfloor\frac{m}{2}\rfloor},0])}(z)\nonn \\
&\quad+\sum_{s=2}^\infty\sum_{\substack{\footnotesize t_i\geq0 \ \ \   \\ \scriptsize t_i\geq t_{i+1}}}^2 \!\!\mathfrak{Z}_{(\dph;[s,t_1,t_1,t_2,t_2,\ldots,t_{m-1},t_{m-1},0])}(z) \nonn\\
&\quad+\sum_{s=2,4,6,\ldots}\!\!\!\!\!\!\sum_{\substack{\footnotesize j_i\geq0 \ \ \   \\ \scriptsize j_i\geq j_{i+1} \\\sum_i j_i = 0\ (\!\!\!\!\!\!\!\mod 2)}}^2 \!\!\!\!\!\!\!\!\!\!\!\!\Big( \mathfrak{Z}_{(\dph;[s,2,j_1,j_1,j_2,j_2,\ldots,j_{m-1},+j_{m-1}])}(z) +\mathfrak{Z}_{(\dph,s;[s,2,j_1,j_1,j_2,j_2,\ldots,j_{m-1},-j_{m-1}])}(z) \Big)\nonn \\
&\quad+\sum_{s=3,5,7,\ldots}\!\!\!\!\!\!\sum_{\substack{\footnotesize j_i\geq0 \ \ \   \\ \scriptsize j_i\geq j_{i+1} \\\sum_i j_i = 1\ (\!\!\!\!\!\!\!\mod 2)}}^2 \!\!\!\!\!\!\!\!\!\!\Big( \mathfrak{Z}_{(\dph;[s,2,j_1,j_1,j_2,j_2,\ldots,j_{m-1},+j_{m-1}])}(z)+\mathfrak{Z}_{(\dph;[s,2,j_1,j_1,j_2,j_2,\ldots,j_{m-1},-j_{m-1}])}(z) \Big)
\intertext{and in $d=4m+2$,}
&\zeta_{\rm min.~type~C}^{\rm HS}(z)=\sum_{\substack{k_i\geq0 \ \ \\k_i\geq k_{i+1}}}^1 \zeta_{(4m+2;[k_1,k_1,\ldots,k_m,k_m,0])}(z)\nonn \\
&\quad+\sum_{s=2}^\infty\sum_{\substack{\footnotesize t_i\geq0 \ \ \   \\ \scriptsize t_i\geq t_{i+1}}}^2 \!\!\mathfrak{Z}_{(\dph;[s,2,t_1,t_1,t_2,t_2,\ldots,t_{m-1},t_{m-1},0])}(z) \nonn\\
&\quad+\sum_{s=2,4,6,\ldots}\!\!\!\!\!\!\sum_{\substack{\footnotesize j_i\geq0 \ \ \   \\ \scriptsize j_i\geq j_{i+1} \\\sum_i j_i = 0\ (\!\!\!\!\!\!\!\mod 2)}}^2 \!\!\!\!\!\!\!\!\!\!\Big( \mathfrak{Z}_{(\dph;[s,j_1,j_1,j_2,j_2,\ldots,j_{m},+j_{m}])}(z) +\mathfrak{Z}_{(\dph;[s,j_1,j_1,j_2,j_2,\ldots,j_{m},-j_{m}])}(z) \Big)\nonn \\
&\quad+\sum_{s=3,5,7,\ldots}\!\!\!\!\!\!\sum_{\substack{\footnotesize j_i\geq0 \ \ \   \\ \scriptsize j_i\geq j_{i+1} \\\sum_i j_i = 1\ (\!\!\!\!\!\!\!\mod 2)}}^2 \!\!\!\!\!\!\!\!\!\!\Big( \mathfrak{Z}_{(\dph;[s,j_1,j_1,j_2,j_2,\ldots,j_{m},+j_{m}])}(z) +\mathfrak{Z}_{(\dph;[s,j_1,j_1,j_2,j_2,\ldots,j_{m},-j_{m}])}(z)\Big),
\end{align} where $\lfloor\frac{m}{2}\rfloor$ denotes the integer part of $\frac{m}{2}$.  

As a consistency check of these spectra, in Appendix ~\ref{sect:thermalads} we computed the corresponding partition functions in thermal AdS with 
$S^1\times S^{d-1}$ boundary. After summing up over all representations appearing in the zeta functions above, the result matches the (symmetrized) square of the 
one-particle partition function of a $(d/2-1)$-form gauge field, see eq. (\ref{thermaltypec3}). 

Evaluating the spectral zeta functions with the help of the formulas in section \ref{subsect:genspec}, we obtain the results
\begin{equation}
\begin{aligned}
F^{(1)}_{\rm min.~type~C~SD} &= 2a_{d/2-{\rm form}}\log R\,,&\qquad d&=4m\\
F^{(1)}_{\rm min.~type~C~SD} &=0\,,&\qquad d&=4m+2
\end{aligned}
\end{equation}
These correspond to the shifts given in Table \ref{Table:summary}. Interestingly, in the minimal type C theory in $d=6,10,\ldots$ the bulk one-loop 
free energy vanishes and no shift of the coupling constant is required.

\subparagraph{Self-dual $U(N)$}
In $d=4m$, we can impose a self-duality constraint $F^i=i*F^i$ in the theory of $N$ complex $p$-forms. The resulting spectrum 
of $U(N)$ invariant bilinears leads to the following total zeta function in the bulk\footnote{This corresponds to eq. (4.20) in \cite{Dolan:2005wy}. 
This is because in $d=4m$ complex conjugation maps self-dual to anti self-dual forms.}
\begin{align}
\zeta_{\rm type~C~SD}^{\rm HS}(z)&=\sum_{s=2}^\infty\sum_{\substack{\footnotesize t_i\geq0 \ \ \   \\ \scriptsize t_i\geq t_{i+1}}}^2 \!\!\mathfrak{Z}_{(\dph;[s,t_1,t_1,t_2,t_2,\ldots,t_{m-1},t_{m-1},0])}(z)\,.
\end{align}
In $d=4m+2$, we can impose the self-duality condition $F^i=*F^i$, and the resulting truncated spectrum gives the following total zeta function\footnote{
This corresponds to eq. (4.23) in \cite{Dolan:2005wy}.}
\begin{align}
\zeta_{\rm type~C~SD}^{\rm HS}(z)&=\sum_{s=2}^\infty \sum_{\substack{\footnotesize j_i\geq0 \ \ \   \\ \scriptsize j_i\geq j_{i+1}}}^2 \!\!\mathfrak{Z}_{(\dph;[s,j_1,j_1,j_2,j_2,\ldots,j_{m-1},+j_{m-1}])}(z)\,.
\end{align}
Using these spectra, we find the results 
\begin{equation}
\begin{aligned}
F^{(1)}_{\rm type~C~SD} =& \quad \ \! \frac{1}{2} a_{d/2-{\rm form}}\log R\,,&\qquad &d=4m\\
F^{(1)}_{\rm type~C~SD} =& -\frac{1}{2} a_{d/2-{\rm form}}\log R\,,&\qquad &d=4m+2
\end{aligned}
\end{equation}
which correspond to the shifts given in Table \ref{Table:summary}. 

\subparagraph{Self-dual $O(N)$}
In $d=4m+2$, we can impose a self-duality condition on the theory of $N$ real forms with $O(N)$ singlt constraint. The spectrum 
is given by the ``overlap" of the minimal type C and self-dual $U(N)$ spectra given above. The resulting total zeta function is given by
\begin{align}
\zeta_{\rm min.~type~C~SD}^{\rm HS}(z)&=\sum_{s=2,4,6,\ldots}\!\!\!\!\!\!\sum_{\substack{\footnotesize j_i\geq0 \ \ \   \\ \scriptsize j_i\geq j_{i+1} \\\sum_i j_i = 0\ \! (\!\!\!\!\!\!\!\mod 2)}}^2 \!\!\!\!\!\!\!\!\mathfrak{Z}_{(\Delta;[s,j_1,j_1,j_2,j_2,\ldots,j_{m},+j_{m}])}(z)  \nonn \\
& \quad+\sum_{s=3,5,7,\ldots}\!\!\!\!\!\!\sum_{\substack{\footnotesize j_i\geq0 \ \ \   \\ \scriptsize j_i\geq j_{i+1} \\\sum_i j_i = 1\ \!(\!\!\!\!\!\!\!\mod 2)}}^2 \!\!\!\!\!\!\!\!\mathfrak{Z}_{(\Delta;[s,j_1,j_1,j_2,j_2,\ldots,j_{m},+j_{m}])}(z)  
\end{align}
from which we find the result
\begin{equation}
\begin{aligned}
F^{(1)}_{\rm min.~type~C~SD} = \frac{1}{4} a_{d/2-{\rm form}}\log R\,,\quad d=4m+2\,.
\end{aligned}
\end{equation}

\subsection{Calculations in odd $d$}

\subsubsection{Preliminaries} 

\paragraph{Alternate Regulators}
In the calculations for even $d$ discussed above, 
we chose to sum over the spins before sending the spectral parameter $z\rightarrow 0$. This analytic continuation in $z$ 
is a natural way to regulate the sums. In practice, this is possible in the even $d$ case because 
the spectral density is polynomial in the integrating variable $u$. In the case of odd $d$, summing before sending $z\rightarrow 0$ 
is not easy to do, and we will instead first send $z\rightarrow 0$ and then evaluate the regularized sums over spins. 
There are two equivalent ways to do this. 
The first involves using exponential factors to suppress the spins
\begin{align}
\sum_{\substack{\text{all spins}\\ \text{in }\alpha_s}} &\Bigg[\left.\mathfrak{Z}_{(\dph;\alpha_s)}(z)\right|_{z=0}\Bigg] \nonn \\& = \lim_{\epsilon\rightarrow0} \sum_{\substack{\text{all spins}\\ \text{in }\alpha_s}} e^{-\epsilon(\Delta^{\rm ph}-\frac{d}{2})} (\zeta_{(\dph;\alpha_s)})(0) - \lim_{\epsilon\rightarrow0} \sum_{\substack{\text{all spins}\\ \text{in }\alpha_s}} e^{-\epsilon(\Delta^{\rm gh}-\frac{d}{2})}(\zeta_{(\Delta^{\rm gh};\alpha_{s-1})})(0),
\intertext{and similarly}
\sum_{\substack{\text{all spins}\\ \text{in }\alpha_s}} &\Bigg[\left.\frac{\partial}{\partial z} \mathfrak{Z}_{(\dph;\alpha_s)}(z)\right|_{z=0}\Bigg] \nonn \\& = \lim_{\epsilon\rightarrow0} \sum_{\substack{\text{all spins}\\ \text{in }\alpha_s}} e^{-\epsilon(\Delta^{\rm ph}-\frac{d}{2})} (\zeta_{(\dph;\alpha_s)})'(0) - \lim_{\epsilon\rightarrow0} \sum_{\substack{\text{all spins}\\ \text{in }\alpha_s}} e^{-\epsilon(\Delta^{\rm gh}-\frac{d}{2})}(\zeta_{(\Delta^{\rm gh};\alpha_{s-1})})'(0),
 \end{align}
where we recall that $\Delta^{\rm ph}=s+d-2$ and $\Delta^{\rm gh}=s+d-1$. In even $d$, one can show that this procedure, with the shifted exponentials 
as above, gives the same result as first summing over all representations and then sending the spectral parameter $z\rightarrow 0$. Equivalently, instead 
of the exponential regulators, one can use the analytic continuation of the Hurwitz zeta function by evaluating
\begin{align}
\sum_{\substack{\text{all spins}\\ \text{in }\alpha_s}} &\Bigg[\left. \mathfrak{Z}_{(\dph;\alpha_s)}(z)\right|_{z=0}\Bigg] \nonn \\& = \lim_{\epsilon\rightarrow0} \sum_{\substack{\text{all spins}\\ \text{in }\alpha_s}} \left(\Delta^{\rm ph}-\frac{d}{2}\right)^{-\epsilon} (\zeta_{(\dph,s;\alpha_s)}^{\rm HS})(0) - \lim_{\epsilon\rightarrow0} \sum_{\substack{\text{all spins}\\ \text{in }\alpha_s}}  \left(\Delta^{\rm gh}-\frac{d}{2}\right)^{-\epsilon} (\zeta_{(\Delta^{\rm gh};\alpha_{s-1})}^{\rm HS})(0),\label{zeta0-HZ}
\intertext{and}
\sum_{\substack{\text{all spins}\\ \text{in }\alpha_s}} &\Bigg[\left.\frac{\partial}{\partial z} \mathfrak{Z}_{(\dph;\alpha_s)}(z)\right|_{z=0}\Bigg] \nonn \\& = \lim_{\epsilon\rightarrow0} \sum_{\substack{\text{all spins}\\ \text{in }\alpha}} \left(\Delta^{\rm ph}-\frac{d}{2}\right)^{-\epsilon} (\zeta_{(\dph;\alpha_s)})'(0) - \lim_{\epsilon\rightarrow0} \sum_{\substack{\text{all spins}\\ \text{in }\alpha_s}}  \left(\Delta^{\rm gh}-\frac{d}{2}\right)^{-\epsilon} (\zeta_{(\Delta^{\rm gh};\alpha_{s-1})}^{\rm HS})'(0),\label{zeta0-HZp}
\end{align}
This method, which is closely related to the one previously used in \cite{Giombi:2014iua},\footnote{In that paper, 
an ``averaged'' regulator of $(\frac{\Delta^{\rm ph}+\Delta^{\rm gh}}{2} -\frac{d}{2})^{-\epsilon}$ was preferred 
for the Type A theory calculations, and it can be shown to give the same result as the regulators (\ref{zeta0-HZ})-(\ref{zeta0-HZp}) that we will use 
in our calculations. In type AB theories, however, it appears that ``averaged'' regulator does not work, 
and we will use the shifts defined in (\ref{zeta0-HZ})-(\ref{zeta0-HZp}) in all theories consistently.}
will be described in the next sections in greater detail. 

Note that, while in even $d$ $\zeta_{(\Delta_s;\alpha_s)}(0)$ vanishes identically for any representation, this is not true in odd $d$. Vanishing of the 
logarithmic divergence in the one-loop free energy requires in this case summing over all the bulk fields, as reviewed below.  

\paragraph{Integrals}
In all odd $d$ calculations, we encounter the integrals of the type
\begin{align}
\int_0^\infty \frac{u^k}{e^{2\pi u}\pm1}\log[u^2+b^2] = \int_0^\infty \frac{u^k\ du}{e^{2\pi u}\pm1}\left[\log(u^2) + \int_0^{b^2} \frac{1}{u^2+x}dx\right]\,.
\label{eqn:evendlogintegral}
\end{align}
We define
\begin{align}
A_k^{\pm}(x)\equiv \int_0^\infty \frac{u^k}{e^{2\pi u}\pm1}\frac{du}{u^2+x}, \qquad\qquad
B_k^\pm\equiv \int_0^\infty \frac{u^k}{e^{2\pi u}\pm1}.
\end{align}
There exists a recursive relation between the various $A_k$'s and $B_k$'s for any odd integer $2k+1$ 
(see Appendix~\ref{appendix:recursive} for a proof):
\begin{align}
A_{2k+1}^\pm(x) &= (-x)^k A_1^{\pm}(x) + \sum_{j=1}^k (-x)^{k-j} B_{2j-1}^\pm. \label{eqn:Apmrecursive}
\end{align} 
As a consequence of this relation, we only need the explicit analytic expressions of the integrals $A_1^\pm$,\footnote{
While not needed, the integral results for $B_k^\pm$, can be identified with the Hurwitz-Lerch Phi function $\Phi(z,s,v)$,
\begin{align}
\int_0^\infty du \frac{u^k}{e^{2\pi u}\pm1} = \int_0^\infty du\frac{1}{2\pi} \frac{(\frac{u}{2\pi})^k e^{- u}}{1 \pm e^{-u}} = \frac{\Gamma(k+1)}{(2\pi)^{k+1}} \Phi(\pm1,k+1,1)
\end{align}
}
which is given by \cite{Camporesi:1991nw}
\begin{align}
A_1^+(x) &= \frac{1}{2}\left[-\log(\sqrt{x}) + \psi\left(\sqrt{x} + \frac{1}{2}\right)\right] \\
A_1^-(x) &= \frac{1}{2} \left[\log(\sqrt{x}) - \frac{1}{2\sqrt{x}} -\psi(\sqrt{x})\right]\,,
\end{align} 
where $\psi(x)$ is the digamma function $\psi(x) = \Gamma'(x)/\Gamma(x)$.

\subsubsection{Calculational method and Type A example}
To illustrate the method of calculation, we first review the calculation in the non-minimal Type A theory \cite{Giombi:2013fka,Giombi:2014iua}. 
The calculations for the various Type B theories are similar and we will not give all details. 
Calculations for the Type AB theory are similar with slight differences that will be discussed below.

Unlike the even $d$ case, the spectral function $\mu_\alpha(u)$ is no longer polynomial in $u$, but a polynomial in $u$ multiplied by a hyperbolic function,
\begin{align}
\mu_{\alpha}(u) = \mu_{\alpha}^{\rm poly}(u)\times f_{\pm} (u),\qquad\text{where }\quad f_{\pm}(u) = \begin{cases} f_{+}(u) = \tanh(\pi u),\qquad\text{for bosons,} \\  f_{-}(u) = \coth(\pi u),\qquad\text{for fermions.}
\end{cases}
\end{align}
Then, for a particular spectral weight $\alpha$, the partition function can be written as
\begin{align}
\zeta_{(\Delta; \alpha)} (z) = \frac{\text{vol}\left({\text{AdS}_{d+1}}\right)}{\text{vol}(S^d)}\frac{2^{d-1}}{\pi} 
\int_0^\infty du \frac{ g_{\alpha} \mu_{\alpha}^{\rm poly}(u)}{\left[u^2+\left(\Delta -\frac{d}{2}\right)^2\right]^z}f_{\pm}(u).
\end{align} 

We will use the example of the Type A theory in AdS$_4$ to walk us through the calculations. In the non-minimal Type A theory in AdS$_4$, 
the only representations are the totally symmetric ones $\alpha=[s]$, $s\ge 0$, and the spectral zeta function for a given spin $s$ is 
\begin{align}
 \zeta_{(\Delta; \alpha_s)}(z)& =  \frac{\text{vol}\left({\text{AdS}_{4}}\right)}{\text{vol}(S^3)}\frac{4}{\pi} 
\int_0^\infty du \frac{ g_{[s]}(s)\mu_{[s]}^{\rm poly}(u)}{\left[u^2+\left(\Delta^{\rm ph} -\frac{3}{2}\right)^2\right]^z}f_{+}(u) \nonn\\
&=  \int_0^\infty du \frac{ (2 s+1) u \left[\left(s+\frac{1}{2}\right)^2+u^2\right]}{6\left[\left(\Delta -\frac{3}{2}\right)^2+u^2\right]^{z}} \tanh(\pi u), \label{eqn:AdS4TypeA}
\end{align}
where $\mu_{[s]}^{\rm poly}(u) = \frac{u \left(u ^2+\left(s+\frac{1}{2}\right)^2\right)}{8 \pi ^2}$ and $ g_{[s]} = 2s+1$. 

To calculate the one-loop free energy, we will need to evaluate $\sum \zeta_{(\Delta; \alpha)}(0)$ and $\sum \zeta_{(\Delta; \alpha)}'(0)$.

\paragraph{\bf Computing $\sum \zeta_{(\alpha;\Delta)} (0)$:} 

Setting $z=0$ in \eqref{eqn:AdS4TypeA}, we find
\begin{align}
\zeta_{(\Delta; [s])}(0)  &= \int_0^\infty du \frac{ (2 s+1) u \left[\left(s+\frac{1}{2}\right)^2+u^2\right]}{6}
\tanh(\pi u).
\end{align} 
Regulating this sum by inserting the prefactor $(\Delta-\frac{d}{2})^{-\epsilon}$ as in (\ref{zeta0-HZ}), we find\footnote{Alternatively, one 
could first write $\tanh(\pi u) = 1-2/(e^{2\pi i u}+1)$, evaluate the integral coming from the first term by analytic continuation 
in $z$, and the one coming from the second term directly at $z=0$, since it converges.} 
 \begin{align}
 \sum_{s=1}^\infty \zeta_{(\Delta^{\rm ph}; [s])}(0) 
  &=
\lim_{\epsilon\rightarrow0}\sum_{s=1}^\infty \int_0^\infty du\frac{(2s+1)u\left[\left(s+\frac{1}{2}\right)^2+u^2\right]}{6}\left(s-\frac{1}{2}\right)^{-\epsilon}\tanh(\pi u)  \nonn \\
& = \int_0^\infty \frac{du}{6} \lim_{\epsilon\rightarrow0}\Bigg[2 \zeta\left(-1+\epsilon,\frac{1}{2}\right) u^3+2 \zeta\left(-3+\epsilon,\frac{1}{2}\right) u+6 \zeta\left(-2+\epsilon,\frac{1}{2}\right) u\nonn \\
&\qquad\qquad\qquad\qquad+6 \zeta\left(-1+\epsilon,\frac{1}{2}\right) u+(2 u^3+2 u)\zeta\left(\epsilon,\frac{1}{2}\right)\Bigg]\tanh(\pi u) \nonn \\
& = \int_0^\infty du\ \left[\frac{u^3}{72}+\frac{113 u}{2880}\right]\tanh(\pi u).
 \end{align}
A similar calculation for the ghost modes using the prefactor $(\Delta^{\rm gh}-\frac{d}{2})^{-\epsilon}$ yields
 \begin{align}
  \sum_{s=1}^\infty \zeta_{(\Delta^{\rm gh}; [s-1])}(0) 
= \int_0^\infty \frac{du}{e^{2\pi i u}+1} \left[\frac{233 u}{2880} +\frac{13 u^3}{72}\right]\tanh(\pi u).
\end{align}
For the bulk scalar, we simply set $s=0$ in $\zeta_{(\Delta^{\rm ph}; [s])}(0)$, and 
obtain $\zeta_{(1; [0])}(0) = \int_0^\infty du\ \left[\frac{ u}{24} +\frac{ u^3}{6}\right]\tanh(\pi u)$. Putting all contributions 
together, the coefficient of the logarithmic divergence in the one-loop free energy is 
 \begin{align}
\zeta_{\rm type\ A}^{\rm HS}(0) &=  \zeta_{(1; [0])}(0) +  \sum_{s=1}^\infty \zeta_{(\Delta^{\rm ph}; [s])}(0) 
- \sum_{s=1}^\infty \zeta_{(\Delta^{\rm gh}; [s-1])}(0) \nonn \\
&= \int_0^\infty du \tanh(\pi u)\left[\frac{u^3}{72}+\frac{113 u}{2880}-\frac{233 u}{2880} -\frac{13 u^3}{72} + \frac{ u}{24} +\frac{ u^3}{6}\right] \nonn \\
& = 0.
\end{align}

It is remarkable that when we sum over the entire spectrum of bulk fields, we get
\begin{align}
\zeta_{\rm Total}^{\rm HS} (0)=0,
\end{align}
which indicates that the one-loop free energies have no logarithmic divergences. We find that this result holds not only 
in type A theories \cite{Giombi:2013fka,Giombi:2014iua}, but also in all of the type B and type AB theories we discuss below. 

\paragraph{\bf Computing $\zeta_{(\Delta;\alpha_s)}' (0)$:}
The evaluation of $\zeta'(0)$ in odd $d$ is considerably more complicated. One may begin by splitting the $f_{\pm}(u)$ term as
$
f_{\pm}(u) = 1 \mp \frac{2}{e^{2\pi i u}\pm 1}
$ so that
\begin{align}
\zeta_{(\Delta;\alpha)} (z) &= \zeta_{(\Delta; \alpha)}^{\rm poly} (z) + \zeta_{(\Delta; \alpha)}^{\rm exp} (z) \label{eqn:zetaHS}
\intertext{where}
 \zeta_{(\Delta; \alpha)}^{\rm poly} (z) &= 
\frac{\text{vol}\left({\text{AdS}_{d+1}}\right)}{\text{vol}(S^d)}\frac{2^{d-1}}{\pi} \int_0^\infty du \frac{ g_{\alpha}\mu_\alpha^{\rm poly}(u)}{\left[u^2+\left(\Delta -\frac{d}{2}\right)^2\right]^z} \\
 \zeta_{(\Delta; \alpha)}^{\rm exp} (z)
& =  \mp \frac{\text{vol}\left({\text{AdS}_{d+1}}\right)}{\text{vol}(S^d)}\frac{2^{d-1}}{\pi} \int_0^\infty du \frac{ g_{\alpha}\mu_\alpha^{\rm poly}(u)}{\left[u^2+\left(\Delta -\frac{d}{2}\right)^2\right]^z}  \frac{2}{e^{2\pi i u}\pm 1} \label{eqn:HSpartexpz}
 \end{align}
And, by differentiating \eqref{eqn:zetaHS},
 \begin{align}
\pd{}{z} \zeta_{(\Delta;\alpha)} (z)\big|_{z=0} &= \pd{}{z}\zeta_{(\Delta; \alpha)}^{\rm poly} (z)\big|_{z=0} + \pd{}{z}\zeta_{(\Delta; \alpha)}^{\rm exp} (z)\big|_{z=0} \label{eqn:diffHSzero}
\end{align}
The integral in $\zeta_{(\Delta; \alpha)}^{\rm poly} (z)$ may be evaluated at arbitrary $z$, and after taking the derivative and summing 
over spins, one finds a zero contribution to the free energy. The evaluation of $\zeta_{(\Delta; \alpha)}^{\rm exp} (z)$ is more involved, and we refer 
the reader to Appendix \ref{zetap0-details} and \cite{Giombi:2013fka, Giombi:2014iua} for more details. The final result is that, 
in the non-minimal theory \cite{Giombi:2013fka}
\begin{equation}
\zeta_{(1;[0])}'(0) + \sum_{s=1}^\infty  \zeta_{(s+1;[s])}'(0) - \sum_{s=1}^\infty  \zeta_{(s+2;[s-1])}'(0)=0\,,
\end{equation}
which implies that the one loop free energy vanishes. In the non-minimal theory, one finds instead 
\begin{equation}
-\frac{1}{2}\left[\zeta_{(1;[0])}'(0) + \sum_{s=2,4,6,\ldots }^\infty  \zeta_{(s+1;[s])}'(0) - \sum_{s=2,4,6,\ldots}^\infty  \zeta_{(s+2;[s-1])}'(0)\right]=
\frac{\log 2}{8}-\frac{3\zeta(3)}{16\pi^2}\,,
\end{equation}
which is the free energy of a single real conformal scalar on $S^3$. An analogous result is found for the type A theory in AdS$_{d+1}$ for all 
$d$ \cite{Giombi:2014iua}.  

\subsubsection{Type B Theories}\label{subsect:oddtypeb}
\paragraph{Non-minimal Theory}
The full spectral zeta function for the non-minimal type B theory in odd $d$ follows from eq. (\ref{psipsi-odd}), and reads 
\begin{align}
&\zeta_{\rm type~B}^{\rm HS}(z) = \zeta_{(d-1;[0,0,\ldots,0])}(z) \nonn\\
&\quad + \sum_{s=1}^\infty \Big(\mathfrak{Z}_{(\Delta^{\rm ph};[s,1,1,\ldots,1,1])}(z) + \mathfrak{Z}_{(\Delta^{\rm ph};[s,1,1,\ldots,1,1,0])}(z) 
+\ldots + \mathfrak{Z}_{(\Delta^{\rm ph};[s,1,0,\ldots,0])}(z) + \mathfrak{Z}_{(\Delta^{\rm ph};[s,0,0,\ldots,0])}(z) \Big)\nonn \\
&\quad = \zeta_{(d-1;[0,0,\ldots,0])}(z)+ \sum_{s=1}^\infty \sum_{\substack{t_i\geq t_{i+1} \\ t_i \geq 0}}^1 \mathfrak{Z}_{(\Delta^{\rm ph};[ s,t_1,t_2,\ldots,t_{w-1},t_{w}])}(z),
\end{align}
Note that instead of two towers, there is only one tower for each representation, due to the lack of the chirality matrix.  
Using this spectrum and the procedure outlined above to regulate the sums, we find that the logarithmic divergence correctly cancels
\begin{equation}
\zeta^{\rm HS}_{\rm type~B}(0)=0\,.
\end{equation}
However, as summarized in section \ref{key-summary}, the evaluation of  $(\zeta^{\rm HS}_{\rm type~B})'(0)$ leads to a surprising result. The one-loop free energy of the 
non-minimal type B theories in all odd $d$ does not vanish, but is given by (\ref{eqn:typebnonminsphere2}), 
or equivalently by (\ref{eqn:typebnonminsphere}). This apparent mismatch with the expected result $F^{(1)}=0$ remains to be understood.  

\paragraph{Minimal theories} 
Majorana fermions in odd $d$ can be defined for $d=3,9$ (mod 8). When the Majorana condition is not possible, one 
can impose the symplectic Majorana (SM) condition and consider the $USp(N)$ singlet sector of $N$ free SM fermions, as explained in the even $d$ case above.  
\begin{table}[h!]
\begin{center}
{\ssmall
\begin{tabular}{llll}
\textbf{AdS$_{\bf 4}\ O(N)$} &  &  &  \\ \hline\hline
\multicolumn{1}{c|}{$\alpha$} & \multicolumn{3}{c}{$s=$}\\\cline{2-4}
\multicolumn{1}{c|}{}& \multicolumn{1}{c|}{$1,2,3,\ldots$} & \multicolumn{1}{c|}{$2,4,6,\ldots$} & $1,3,5,\ldots$ \\ \hline
\multicolumn{1}{l|}{$[s]$} & \multicolumn{1}{c|}{} & \multicolumn{1}{c|}{\cellcolor[HTML]{EFEFEF}{\color[HTML]{000000} 1}} &   \\ \hline
\multicolumn{1}{c|}{Scalar ($\Delta=2$)} &  \multicolumn{3}{c}{\cellcolor[HTML]{EFEFEF}1} \\ \hline\hline \\ \\
\textbf{AdS$_{\bf 8}\ USp(N)$} &  &  &  \\ \hline\hline
\multicolumn{1}{c|}{$\alpha$} & \multicolumn{3}{c}{$s=$}\\\cline{2-4}
\multicolumn{1}{c|}{}& \multicolumn{1}{c|}{$1,2,3,\ldots$} & \multicolumn{1}{c|}{$2,4,6,\ldots$} & $1,3,5,\ldots$ \\ \hline
\multicolumn{1}{l|}{$[s,1,1]$} & \multicolumn{1}{c|}{} & \multicolumn{1}{c|}{} &  \multicolumn{1}{c}{\cellcolor[HTML]{99FFFF}{\color[HTML]{000000} $1$}} \\ \hline
\multicolumn{1}{l|}{$[s,1,0]$} & \multicolumn{1}{c|}{} & \multicolumn{1}{c|}{\cellcolor[HTML]{99FFFF}1} &  \\ \hline
\multicolumn{1}{l|}{$[s,0,0]$} & \multicolumn{1}{c|}{} & \multicolumn{1}{c|}{\cellcolor[HTML]{99FFFF}1} &  \\ \hline
\multicolumn{1}{c|}{Scalar ($\Delta=6$)} &\multicolumn{3}{c}{\cellcolor[HTML]{99FFFF}1}  \\ \hline\hline  \\ \\
\textbf{AdS$_{\bf 12}\ O(N)$} &  &  &  \\ \hline\hline
\multicolumn{1}{c|}{$\alpha$} & \multicolumn{3}{c}{$s=$}\\\cline{2-4}
\multicolumn{1}{c|}{}& \multicolumn{1}{c|}{$1,2,3,\ldots$} & \multicolumn{1}{c|}{$2,4,6,\ldots$} & $1,3,5,\ldots$ \\ \hline
\multicolumn{1}{l|}{$[s,1,1,1,1]$} & \multicolumn{1}{c|}{} & \multicolumn{1}{c|}{\cellcolor[HTML]{EFEFEF}{\color[HTML]{000000} $1$}} &  \\ \hline
\multicolumn{1}{l|}{$[s,1,1,1,0]$} & \multicolumn{1}{c|}{} & \multicolumn{1}{c|}{} &  \multicolumn{1}{c}{\cellcolor[HTML]{EFEFEF}1}  \\ \hline
\multicolumn{1}{l|}{$[s,1,1,0,0]$} & \multicolumn{1}{c|}{} & \multicolumn{1}{c|}{} & \multicolumn{1}{c}{\cellcolor[HTML]{EFEFEF}1} \\ \hline
\multicolumn{1}{l|}{$[s,1,0,0,0]$} & \multicolumn{1}{c|}{} & \multicolumn{1}{c|}{\cellcolor[HTML]{EFEFEF}1} &  \\ \hline
\multicolumn{1}{l|}{$[s,0,0,0,0]$} & \multicolumn{1}{c|}{} & \multicolumn{1}{c|}{\cellcolor[HTML]{EFEFEF}1} &  \\ \hline
\multicolumn{1}{c|}{Scalar ($\Delta=10$)} &  \multicolumn{3}{c}{\cellcolor[HTML]{EFEFEF}1}    \\ \hline\hline 
 \\ \\
\textbf{AdS$_{\bf 16}\ USp(N)$}                   &                                                                                      &                                                                                      &                                                                 \\ \hline\hline
\multicolumn{1}{c|}{$\alpha$} & \multicolumn{3}{c}{$s=$}\\\cline{2-4}
\multicolumn{1}{c|}{}& \multicolumn{1}{c|}{$1,2,3,\ldots$} & \multicolumn{1}{c|}{$2,4,6,\ldots$} & $1,3,5,\ldots$ \\ \hline
\multicolumn{1}{l|}{$[s,1,1,1,1,1,1]$}    & \multicolumn{1}{c|}{}                                                                & \multicolumn{1}{c|}{}                & \multicolumn{1}{c}{\cellcolor[HTML]{99FFFF}{\color[HTML]{000000} $1$}}                                                                \\ \hline
\multicolumn{1}{l|}{$[s,1,1,1,1,1,0]$}    & \multicolumn{1}{c|}{}                                       & \multicolumn{1}{c|}{\cellcolor[HTML]{99FFFF}1}                                                                &                                                                 \\ \hline
\multicolumn{1}{l|}{$[s,1,1,1,1,0,0]$}    & \multicolumn{1}{c|}{}                                                                & \multicolumn{1}{c|}{\cellcolor[HTML]{99FFFF}{\color[HTML]{000000} 1}}                                                                &                                       \\ \hline
\multicolumn{1}{l|}{$[s,1,1,1,0,0,0]$}    & \multicolumn{1}{c|}{}                                       & \multicolumn{1}{c|}{}                                                                &                                          \multicolumn{1}{c}{\cellcolor[HTML]{99FFFF}1}                        \\ \hline
\multicolumn{1}{l|}{$[s,1,1,0,0,0,0]$}    & \multicolumn{1}{c|}{}                                                                & \multicolumn{1}{c|}{}                                       &  \multicolumn{1}{c}{\cellcolor[HTML]{99FFFF}1}                                                               \\ \hline
\multicolumn{1}{l|}{$[s,1,0,0,0,0,0]$}    & \multicolumn{1}{c|}{}                                       & \multicolumn{1}{c|}{ \cellcolor[HTML]{99FFFF}1}                                                           &                                                                 \\ \hline
\multicolumn{1}{l|}{$[s,0,0,0,0,0,0]$}    & \multicolumn{1}{c|}{}                                                                & \multicolumn{1}{c|}{\cellcolor[HTML]{99FFFF}{\color[HTML]{000000} 1}}                                                                &                                     \\ \hline
\multicolumn{1}{c|}{Scalar ($\Delta=14$)}           & \multicolumn{3}{c}{\cellcolor[HTML]{99FFFF}1}                                              \\ \hline\hline 
\\ \\
\end{tabular}}
{\ssmall
\begin{tabular}{llll}
\textbf{AdS$_{\bf 6}\ USp(N)$} &  &  &  \\ \hline\hline
\multicolumn{1}{c|}{$\alpha$} & \multicolumn{3}{c}{$s=$}\\\cline{2-4}
\multicolumn{1}{c|}{}& \multicolumn{1}{c|}{$1,2,3,\ldots$} & \multicolumn{1}{c|}{$2,4,6,\ldots$} & $1,3,5,\ldots$ \\ \hline
\multicolumn{1}{l|}{$[s,1]$} & \multicolumn{1}{c|}{} & \multicolumn{1}{c|}{} &\multicolumn{1}{c}{\cellcolor[HTML]{99FFFF}{\color[HTML]{000000} $1$}}  \\ \hline
\multicolumn{1}{l|}{$[s,0]$} & \multicolumn{1}{c|}{} & \multicolumn{1}{c|}{\cellcolor[HTML]{99FFFF}{\color[HTML]{000000} $1$}}&  \\ \hline
\multicolumn{1}{c|}{Scalar ($\Delta=4$)} & \multicolumn{3}{c}{}    \\ \hline\hline 
\\ \\
\textbf{AdS$_{\bf 10}\ O(N)$} &  &  &  \\\hline\hline
\multicolumn{1}{c|}{$\alpha$} & \multicolumn{3}{c}{$s=$}\\\cline{2-4}
\multicolumn{1}{c|}{}& \multicolumn{1}{c|}{$1,2,3,\ldots$} & \multicolumn{1}{c|}{$2,4,6,\ldots$} & $1,3,5,\ldots$ \\ \hline
\multicolumn{1}{l|}{$[s,1,1,1]$} & \multicolumn{1}{c|}{} & \multicolumn{1}{c|}{\cellcolor[HTML]{EFEFEF}{\color[HTML]{000000} $1$}} &  \\ \hline
\multicolumn{1}{l|}{$[s,1,1,0]$} & \multicolumn{1}{c|}{} & \multicolumn{1}{c|}{} &  \multicolumn{1}{c}{\cellcolor[HTML]{EFEFEF}1}\\ \hline
\multicolumn{1}{l|}{$[s,1,0,0]$} & \multicolumn{1}{c|}{} & \multicolumn{1}{c|}{} & \multicolumn{1}{c}{\cellcolor[HTML]{EFEFEF}1} \\ \hline
\multicolumn{1}{l|}{$[s,0,0,0]$} & \multicolumn{1}{c|}{} & \multicolumn{1}{c|}{\cellcolor[HTML]{EFEFEF}1}&  \\ \hline
\multicolumn{1}{c|}{Scalar ($\Delta=8$)} & \multicolumn{3}{c}{}    \\ \hline\hline 
\\ \\
\textbf{AdS$_{\bf 14}\ USp(N)$} &  &  &  \\ \hline\hline
\multicolumn{1}{c|}{$\alpha$} & \multicolumn{3}{c}{$s=$}\\\cline{2-4}
\multicolumn{1}{c|}{}& \multicolumn{1}{c|}{$1,2,3,\ldots$} & \multicolumn{1}{c|}{$2,4,6,\ldots$} & $1,3,5,\ldots$ \\ \hline
\multicolumn{1}{l|}{$[s,1,1,1,1,1]$} & \multicolumn{1}{c|}{} & \multicolumn{1}{c|}{\cellcolor[HTML]{FFFFFF}{\color[HTML]{000000} }} & \multicolumn{1}{c}{\cellcolor[HTML]{99FFFF}{\color[HTML]{000000} $1$}}  \\ \hline
\multicolumn{1}{l|}{$[s,1,1,1,1,0]$} & \multicolumn{1}{c|}{} & \multicolumn{1}{c|}{\cellcolor[HTML]{99FFFF}1} &  \\ \hline
\multicolumn{1}{l|}{$[s,1,1,1,0,0]$} & \multicolumn{1}{c|}{} & \multicolumn{1}{c|}{\cellcolor[HTML]{99FFFF}1} & \multicolumn{1}{c}{} \\ \hline
\multicolumn{1}{l|}{$[s,1,1,0,0,0]$} & \multicolumn{1}{c|}{} & \multicolumn{1}{c|}{} &  \multicolumn{1}{c}{\cellcolor[HTML]{99FFFF}1}  \\ \hline
\multicolumn{1}{l|}{$[s,1,0,0,0,0]$} & \multicolumn{1}{c|}{} & \multicolumn{1}{c|}{\cellcolor[HTML]{FFFFFF}} &  \multicolumn{1}{c}{\cellcolor[HTML]{99FFFF}1}  \\ \hline
\multicolumn{1}{l|}{$[s,0,0,0,0,0]$} & \multicolumn{1}{c|}{\cellcolor[HTML]{FFFFFF}} & \multicolumn{1}{c|}{\cellcolor[HTML]{99FFFF}1}  &  \\ \hline
\multicolumn{1}{c|}{Scalar ($\Delta=12$)} &  \multicolumn{3}{c}{\cellcolor[HTML]{FFFFFF} }
 \\ \hline\hline 
 \\ \\\\\\\\\\\\\\\\\\\\\\\
\end{tabular}}
\caption{Spectra of the minimal Type B theory dual to the fermionic vector model with Majorana (or symplectic Majorana) projection. 
The corresponding values of $F^{(1)}$ can be found in Table~\ref{Table:TypeBMin}.}\label{Table:O(n)projection2} 
\end{center}
\end{table}
The spectra of the minimal theories can be again deduced from the symmetry/antisymmetry properties of the ${\cal C}\Gamma^{(n)}$ matrices. In the 
Majorana case, if ${\cal C}\Gamma^{(n)}$ is symmetric the operators of the form (\ref{typeB-schematic}) 
are retained for even spins and projected out for odd spins, and vice-versa if ${\cal C}\Gamma^{(n)}$ is antisymmetric. The scalar 
operator $\bar\psi_i \psi^i$ is projected out if ${\cal C}$ is symmetric. For instance, in $d=3$ the 
${\cal C}$ matrix is antisymmetric and ${\cal C}\gamma_{\mu}$ is symmetric, and so the spectrum of the minimal theory includes the $\Delta=2$ (pseudo)-scalar 
and the tower of totally symmetric fields of even spin. Higher dimensional cases can be worked out similarly, and the first few examples are listed 
in Table~\ref{Table:O(n)projection2}. In a compact notation, the total spectral zeta function of the minimal theories dual to the Majorana 
projected fermion model reads
	\begin{align}
		&\zeta_{\rm type~B~Maj.}^{\text{HS}}(z)  = 
			\chi(d)\zeta_{(d-1;[0,0,\ldots,0])}(z) \nonn\\
			& + \sum_{s=2,4,6,\ldots}^\infty \!\!\!\!\!\!\!\!\!\sum_{\substack{t_i\geq t_{i+1} \\ t_i \geq 0 \\ \sum_i t_i = w \ \! (\!\!\!\!\!\! \mod 4)}}^1 \!\!\!\!\!\!\!\!\Big(\mathfrak{Z}_{(\Delta;[s,t_1,t_2,\ldots,t_{w-1},t_{w}])}(z)\Big) +  \sum_{s=1,3,5,\ldots} \!\!\!\!\!\!\!\sum_{\substack{t_i\geq t_{i+1} \\ t_i \geq 0 \\ \sum_i t_i = (w-1) \ \! (\!\!\!\!\!\! \mod 4)}}^1 \!\!\!\!\!\!\!\!\!\!\!\!\!\!\!\Big(\mathfrak{Z}_{(\Delta;[s,t_1,t_2,\ldots,t_{w-1},t_{w}])}(z)\Big) \nonn \\
			& +  \sum_{s=1,3,5,\ldots} \!\!\!\!\!\!\!\!\!\!\!\!\!\sum_{\substack{t_i\geq t_{i+1} \\ t_i \geq 0 \\ \sum_i t_i = (w-2) \ \! (\!\!\!\!\!\! \mod 4)}}^1 \!\!\!\!\!\!\!\!\!\!\!\!\Big(\mathfrak{Z}_{(\dph;[s,t_1,t_2,\ldots,t_{w-1},t_{w}])}(z)\Big) +  \sum_{s=2,4,6,\ldots} \!\!\!\!\!\!\!\!\!\!\!\!\!\sum_{\substack{t_i\geq t_{i+1} \\ t_i \geq 0 \\ \sum_i t_i = (w-3) \ \! (\!\!\!\!\!\! \mod 4)}}^1 \!\!\!\!\!\!\!\!\!\!\!\!\Big(\mathfrak{Z}_{(\dph;[s,t_1,t_2,\ldots,t_{w-1},t_{w}])}(z)\Big)\label{eqn:typeBOnodd}
\end{align} where $\chi(d) = 1,0$ when $d = 3,9$ (mod 8) respectively.

In $d=5,7$ (mod 8) we can impose instead the symplectic Majorana projection. In this case, the condition for which spins are projected out is reversed 
compared to the Majorana case, in a way analogous to what discussed earlier in the even $d$ case. For instance, in $d=5$ (AdS$_6$) one has that 
${\cal C}$ is antisymmetric, and so the scalar operator $\bar\psi_i \psi^i$ is now projected out. Then, ${\cal C}\gamma_{\mu}$ is 
antisymmetric, and so the spectrum includes the totally symmetric $[s,0]$ representations for even $s$ only. Finally, ${\cal C}\gamma_{\mu\nu}$ 
is symmetric, and so we keep the representations $[s,1]$ with odd $s$ only. Higher dimensional cases are worked out similarly, and the first 
few examples are listed in Table \ref{Table:O(n)projection2}. 
The total spectral zeta function can be expressed as 
\begin{align}
		& \zeta_{\rm type~B~Symp.Maj.}^{\text{HS}}(z) = 
			\chi(d)\zeta_{(d-1;[0,0,\ldots,0])}(z) \nonn\\
			&\quad + \sum_{s=1,3,5,\ldots}^\infty \!\!\!\sum_{\substack{t_i\geq t_{i+1} \\ t_i \geq 0 \\ \sum_i t_i = w \ \! (\!\!\!\!\!\! \mod 4)}}^1 \!\!\!\!\!\!\!\!\mathfrak{Z}_{(\Delta;[s,t_1,t_2,\ldots,t_{w-1},t_{w}])}(z) +  \sum_{s=2,4,6,\ldots} \!\!\!\!\!\!\!\sum_{\substack{t_i\geq t_{i+1} \\ t_i \geq 0 \\ \sum_i t_i = (w-1) \ \! (\!\!\!\!\!\! \mod 4)}}^1 \!\!\!\!\!\!\!\!\!\!\!\!\mathfrak{Z}_{(\Delta;[s,t_1,t_2,\ldots,t_{w-1},t_{w}])}(z) \nonn \\
			& \quad +  \sum_{s=2,4,6,\ldots} \!\!\!\!\!\!\!\sum_{\substack{t_i\geq t_{i+1} \\ t_i \geq 0 \\ \sum_i t_i = (w-2) \ \! (\!\!\!\!\!\! \mod 4)}}^1 \!\!\!\!\!\!\!\!\!\!\!\!\mathfrak{Z}_{(\dph;[s,t_1,t_2,\ldots,t_{w-1},t_{w}])}(z) +  \sum_{s=1,3,5,\ldots} \!\!\!\!\!\!\!\sum_{\substack{t_i\geq t_{i+1} \\ t_i \geq 0 \\ \sum_i t_i = (w-3) \ \! (\!\!\!\!\!\! \mod 4)}}^1 \!\!\!\!\!\!\!\!\!\!\!\!\mathfrak{Z}_{(\dph;[s,t_1,t_2,\ldots,t_{w-1},t_{w}])}(z)\label{eqn:typeBsympodd}
\end{align} where $\chi(d) = 0,1$ when $d = 5,7$ (mod 8) respectively.
In both versions of the minimal truncation, we find that the coefficient of the logarithmic divergence still vanishes after summing up over the full 
spectrum. However, similarly to the non-minimal case, the minimal Type B theories in odd $d$ 
appear to have a non-zero one-loop free energy, which we report in Table~\ref{Table:TypeBMin}.  
\begin{table}[t]
\begin{center}
\begin{tabular}{c|l}
$d$& $F_{\rm computed}$ (Minimal Type B) \\ \hline \\
3 & $\displaystyle\frac{\log (2)}{8}-\frac{5 \zeta (3)}{16 \pi ^2}$\\ \\
5 & $\displaystyle\frac{3 \log (2)}{64}+\frac{7 \zeta (3)}{192 \pi ^2}-\frac{49 \zeta (5)}{128 \pi ^4}$ \\ \\
7 & $\displaystyle\frac{5 \log (2)}{128}+\frac{227 \zeta (3)}{3840 \pi ^2}-\frac{5 \zeta (5)}{256 \pi ^4}-\frac{441 \zeta (7)}{512 \pi ^6}$ \\ \\
9 & $\displaystyle -\frac{35 \log (2)}{2048}+\frac{315 \zeta (7)}{2048 \pi ^6}+\frac{3825 \zeta (9)}{4096 \pi ^8}-\frac{617 \zeta (3)}{21504 \pi ^2}-\frac{85 \zeta (5)}{2048 \pi ^4}$\\ \\
11 & $\displaystyle\frac{63 \log (2)}{16384} + \frac{68843 \zeta (3)}{10321920 \pi ^2}+\frac{31033 \zeta (5)}{2211840 \pi ^4}-\frac{29 \zeta (7)}{98304 \pi ^6}-\frac{13579 \zeta (9)}{98304 \pi ^8}-\frac{31745 \zeta (11)}{65536 \pi ^{10}}$ \\ \\
13 & $\displaystyle \frac{231 \log (2)}{131072}+\frac{1933151 \zeta (3)}{619315200 \pi ^2}+\frac{27993331 \zeta (5)}{3715891200 \pi ^4}+\frac{1056541 \zeta (7)}{123863040 \pi ^6}-\frac{285799 \zeta (9)}{11796480 \pi ^8}$ \\[2.5ex] &$\displaystyle -\frac{150541 \zeta (11)}{786432 \pi ^{10}}-\frac{258049 \zeta (13)}{524288 \pi ^{12}}$\\[4ex]
15 & $\displaystyle\frac{429 \log (2)}{524288}+ \frac{2423526031 \zeta (3)}{1653158707200 \pi ^2}+\frac{41124367 \zeta (5)}{10899947520 \pi ^4}+\frac{12837 \zeta (7)}{2097152 \pi ^6}+\frac{47549 \zeta (9)}{66060288 \pi ^8}$ \\[2.5ex] &
$\displaystyle-\frac{104687 \zeta (11)}{2097152 \pi ^{10}}-\frac{503685 \zeta (13)}{2097152 \pi ^{12}}-\frac{2080641 \zeta (15)}{4194304 \pi ^{14}}$ \\[4ex]
\end{tabular}
\end{center}
\caption{One-loop free energy of the minimal Type B HS theory in AdS$_{d+1}$ for odd $d$.}
\label{Table:TypeBMin}
\end{table}
We did not find an analytic formula for these results similar to (\ref{eqn:typebnonminsphere2}). However, we note that 
all these ``anomalous'' values only involve the 
Riemann zeta functions $\zeta(2k+1)$ divided by $\pi^2$, 
and interestingly all other transcendental constants that appear in intermediate steps of the calculation cancel out.

\subsubsection{Type AB Theories}
\paragraph{Spectrum and Results}
As in the even $d$ case, the only irrep of $SO(d)$ describing the tower of 
half-integer spins is $\alpha_s = [s,\frac{1}{2},\frac{1}{2},\ldots,\frac{1}{2}]$. 
Thus, the total spectral zeta function is given by the same equation as in \eqref{eqn:typeabspec}.

The calculation is rather similar to the one we outlined for the Type A theory. The only difference is that 
the spectral density $\mu_\alpha(u)$ includes $\coth(\pi u)$ instead of $\tanh(\pi u)$. 
For example, in the Type AB theory in AdS$_4$, the higher-spin zeta-function is given by
\begin{align}
\zeta_{(\Delta; [s])}(z)  &= \int_0^\infty du \frac{ (2 s+1) u \left[\left(s+\frac{1}{2}\right)^2+u^2\right]}{6\left[\left(\Delta -\frac{3}{2}\right)^2+u^2\right]^{z}} \coth(\pi u).
\end{align}
The calculations for $\sum \zeta_{(\Delta; \alpha_s)}(0)$ are essentially identical to that of Type A theories, and in particular we find that 
the contribution to the logarithmic divergence due to the fermionic fields vanishes after summing over the whole tower. 
Heading straight to the calculation of $\sum \zeta_{(\Delta; \alpha_s)}'(0)$, if we follow the procedure outlined for the Type A case, we have
\begin{align}
\zeta_{(\Delta; [s])}'(0) = -\int_0^\infty du \frac{(2 s+1) u }{3(e^{2\pi u}-1)} \left[u^2+\left(s+\frac{1}{2}\right)^2\right] \log \left[\left(\Delta -\frac{3}{2}\right)^2+u^2\right].
\end{align}
Rewriting the exponential terms using (\ref{eqn:evendlogintegral}), we should use $A_1^-(x)$ instead of $A_1^+(x)$. 
This introduces an extra $\frac{1}{2\sqrt{x}}$ term in $\zeta_{\Delta,s; [s]}^{\rm HS-exp'}(0)$, i.e. 
\begin{align}
\zeta_{(\Delta; [s])}^{\rm exp'}(0) & = 
-\int_0^\infty du \frac{(2 s+1) u }{3(e^{2\pi u}-1)} \left[u^2+\left(s+\frac{1}{2}\right)^2\right] \log(u^2) \nonn \\
&- \int_0^{(s-\frac{1}{2})^2}dx \frac{1}{12} (2 s+1) \left[(2 s+1)^2+4 u^2\right] \left(-\frac{1}{2 \sqrt{x }}+\frac{\log \left(\sqrt{x }\right)}{2}-\frac{\psi\left(\sqrt{x }\right)}{2}\right) \label{eqn:evendtypeAB}\end{align}
In any case, the terms involving $\frac{1}{2\sqrt{x}}$, which we can call $\zeta_{(\Delta; [s])}^{\rm exp-sqrt'}(0)$, will not contribute to the value of $\zeta_{(\Delta; [s])}^{\rm exp'}(0)$. Only the contributions from the terms involving $\psi(\sqrt{x})$, namely the third term inside the bracket of \eqref{eqn:evendtypeAB} will contribute. 
After putting all together, the end result is
\begin{equation}
\zeta_{(\frac{3}{2}; [\frac{1}{2}])}'(0)+\sum_{s=\frac{3}{2},\frac{5}{2},\ldots} \left(\zeta_{(s+1; [s])}'(0)-
\zeta_{(s+2; [s-1])}'(0)\right) =0\,,
\end{equation}
i.e., the tower of fermionic fields in type AB theories yields a vanishing contribution to the bulk one-loop free energy. This result extends to 
all higher $d$. 

\section*{Acknowledgments}

The work of SG was supported in part by the US NSF under Grant No.~PHY-1318681.
The work of IRK was supported in part by the US NSF under Grant No.~PHY-1314198.
The work of ZMT was supported in part by the \'Ecole Normale Sup\'erieure, France under the grant LabEx ENS-ICFP: ANR-10-LABX-0010/ANR-10-IDEX-0001-02 PSL, and the Public Service Commission, Singapore. Parts of this paper are based on ZMT's Princeton University 2014 undergraduate thesis. 
IRK thanks the Niels Bohr International Academy for hospitality during the final stages of his work on this paper.

 \appendix
\section{$S^1\times S^{d-1}$ partition functions and Casimir energies} 
\label{sect:thermalads}
Besides testing the gauge/gravity duality by comparing the partition functions on the Euclidean AdS$_{d+1}$ (hyperbolic space) and CFT$_d$ on $S^d$, we can also compare thermal partition functions of higher spin  theories on thermal AdS$_{d+1}$ and boundary CFTs defined on $ S^1 \times  S^{d-1}$, where the inverse temperature $\beta$ of the thermal AdS space is interpreted as the length of $S^1$. Calculations of the thermal free energy and Casimir energy serve as useful checks on 
our results in hyperbolic space with $S^d$ boundary. The results below follow and generalize \cite{Giombi:2014yra}, which considered type A theories in all $d$ and type B theories in $d=2,3,4$, and \cite{Beccaria:2014zma,Beccaria:2014xda,Beccaria:2014qea}, where type B theories in $d=6$ and type C theories in $d=4,6$ were discussed. 

The free energy on $S^1 \times S^{d-1}$ takes the form
\begin{equation}
F = F_{\beta} +\beta E_c
\end{equation}
where $F_{\beta}$ depends non-trivially on the temperature and goes to zero at large $\beta$, and $E_c$ is the Casimir energy. The latter is related to the ``one-particle'' 
partition function on $S^1\times S^{d-1}$ by (see e.g. \cite{Giombi:2014yra} for a review)
\begin{align}
E_c = \sigma\frac{1}{2}\zeta_E(-1) = \sigma\frac{1}{2\Gamma(z)}\int_0^\infty 	d\beta \beta^{z-1}\mathcal{Z}(\beta)\Bigg|_{z=-1}.
\label{Casi}
\end{align} 
where $\sigma=+1$ for bosonic fields, and $\sigma=-1$ for fermionic ones, and ${\cal Z}(\beta)$ denotes the one-particle partition function. This also determines $F_{\beta}$ 
by
\begin{equation}
\begin{aligned}
&F_{\beta} = -\sum_{m=1}^{\infty} \frac{1}{m}{\cal Z}(m\beta)\,,\quad {\rm boson}\\
&F_{\beta} = \sum_{m=1}^{\infty} \frac{(-1)^m}{m}{\cal Z}(m\beta)\,,\quad {\rm fermion}
\label{Fbeta}
\end{aligned}
\end{equation}
Note that $E_c$ vanishes for a CFT$_d$ in odd $d$, but it is non-zero in even $d$. 

In the vector models restricted to the singlet sector, one finds that $F_{\beta}={\cal O}(N^0)$, due to the integration over the flat connection which enforces the 
gauge singlet constraint \cite{Shenker:2011zf, Giombi:2014yra}. This term should then match the temperature dependent part of the bulk one-loop thermal free energy, obtained by summing over all fields in the AdS spectrum, and the agreement serves as a useful check on the bulk spectra. The Casimir term, on the other hand, is just given by $N$ times the Casimir energy of a single conformal field. If no shift is expected in the map between the bulk coupling constant and $N$, then the CFT Casimir contribution should be reproduced just by a classical calculation in AdS (which we have no access to at present), and bulk loop corrections to the Casimir energy should vanish. However, when a shift $G_N \sim 1/(N-k)$ is expected, the one-loop correction to the Casimir energy should precisely be consistent with such a shift. We will see below that this is the case in all higher spin theories we considered in this paper. 

On the CFT side, the one-particle partition functions of a conformal scalar and Majorana (or Weyl) fermion are given by
\begin{equation}
\begin{aligned}
{\cal Z}_0(q) = \frac{q^{\frac{d}{2}-1}(1+q)}{(1-q)^{d-1}}\,,\qquad 
{\cal Z}_{\frac{1}{2}}(q) = \frac{2^{\lfloor\frac{d}{2}\rfloor}q^{\frac{d-1}{2}}}{(1-q)^{d-1}}\,,
\qquad q = e^{-\beta}
\label{Zscfer}
\end{aligned}
\end{equation}
Using (\ref{Casi}) and the identity $(1-q)^{-b}=\sum_{n=1}^{\infty}\begin{pmatrix}n+b-2\\b-1\end{pmatrix}q^{n-1}$, one then finds the Casimir energies
\begin{equation}
\begin{aligned}
&E_{c,0} = \sum_{n=0}^\infty \frac{(n+d-3)!}{(d-2)!n!}[n+\frac{1}{2}(d-2)]^{1-z}|_{z=-1}\,,\\
&E_{c,1/2} = -2^{\lfloor\frac{d}{2}\rfloor-1}\sum_{n=0}^\infty \frac{(n+d-2)!}{(d-2)!n!}[n+\frac{1}{2}(d-1)]^{1-z}|_{z=-1} \,.
\end{aligned}
\end{equation}
Evaluating this with Hurwitz zeta regularization, one obtains the values in $d=4,6,8,\ldots$:
\begin{equation}
\begin{aligned}
&E_{c,0} = \left\{\frac{1}{240},-\frac{31}{60480},\frac{289}{3628800},-\frac{317}{22809600},\frac{6803477}{2615348736000},\ldots \right\}\\
&E_{c,1/2} = \left\{\frac{17}{960},-\frac{367}{48384},\frac{27859}{8294400},-\frac{1295803}{851558400},\frac{5329242827}{7608287232000},\ldots \right\}
\end{aligned}
\end{equation}
For the real $(d/2-1)$-form gauge field, with no self-duality imposed on the $d/2$-form field strength, 
the one-particle partition function is given by (see for instance Appendix D of \cite{Beccaria:2014qea})  
\begin{equation}
\mathcal{Z}_{\frac{d}{2}\text{-form}}(q) = \frac{2q^{d/2}}{(1-q)^{d-1}}\left(\sum_{j=1}^{\frac{d}{2}} a_{d,j} (-q)^{\frac{d}{2}-j}\right),\qquad\qquad a_{d,j} = \begin{pmatrix}d-1 \\j-1 \end{pmatrix}.
\label{Zpform}
\end{equation}
Note that when we expand $\mathcal{Z}_{\frac{d}{2}\text{-form}}(q)$ around $q=1$, the leading pole term is,
\begin{align}
\mathcal{Z}_{\frac{d}{2}\text{-form}}(q) \sim \frac{2}{(1-q)^{d-1}}n(d),\qquad\qquad \text{where}\quad n(d) = \begin{pmatrix} d-2 \\ \frac{d}{2}-1 \end{pmatrix},
\end{align} 
which gives the correct number of propagating degrees of freedom of a $(d/2-1)$-form gauge field. Inserting (\ref{Zpform}) into (\ref{Casi}), one finds the Casimir energies in $d=4,6,8,\ldots$:
\footnote{The values obtained for $d=4,6$ agree with those in the literature~\cite{Beccaria:2014zma,Beccaria:2014xda,Beccaria:2014qea}, while the values for other dimensions are new as far as we know.}
\begin{equation}
E_{c,d/2-{\rm form}}=\left\{\frac{11}{120},-\frac{191}{2016},\frac{2497}{25920},-\frac{14797}{152064},\frac{92427157}{943488000},\ldots \right\}\,.
\end{equation}

On the AdS side, at the level of the one-particle partition functions, 
the contribution of a bulk field to the thermal free energy is given essentially by the character of the corresponding 
representation of the conformal group. For the representations $\alpha_s$ dual to massless gauge fields, we have 
	\begin{align}
		{\cal Z}_{\alpha_s}(q) &= \frac{q^{\dph}}{(1-q)^d} [g_{\alpha_s}-q g_{\alpha_{s-1}}]\,,
	\end{align} 
where $\dph=s+d-2$ and $g_{\alpha_s}$ is the dimension of the representation $\alpha_s$ (the number of propagating degrees of freedom in the bulk is 
$g_{\alpha_s}-g_{\alpha_{s-1}}$). For the massive fields, the ghost contribution is not present, and one has
\begin{equation}
{\cal Z}_{\alpha}= \frac{q^{\Delta}}{(1-q)^d}g_{\alpha}\,.
\end{equation} 
One may obtain a ``total'' one-particle partition function ${\cal Z}(\beta)$ in the bulk by summing over all representations in the spectrum, and from it one may then find the bulk one-loop Casimir energy by (\ref{Casi}) and $F_{\beta}$ by (\ref{Fbeta}).  In the following we summarize the result of these calculations in the 
various higher spin theories considered in this paper.

\paragraph{Type A Theories}
In \cite{Giombi:2014yra}, it was shown that 
\begin{alignat}{3}
&\text{Non-Minimal Type A:}\qquad&\mathcal{Z}(\beta) = \sum_{\alpha}  {\cal Z}_\alpha(q)&= [\mathcal{Z}_{0}(q)]^2,\label{thermaltypea1}\\
&\text{Minimal Type A:}\qquad&\mathcal{Z}(\beta) =\sum_{\gamma}  {\cal Z}_\gamma(q)&= \frac{1}{2} \Big[ [\mathcal{Z}_{0}(q)]^2+ \mathcal{Z}_{0}(q^2) \Big]\label{thermaltypea2}
\end{alignat} 
where $\alpha$ refers to the spectrum containing the weights $[s,0,\ldots,0]$ with each integer spin $s=0,1,2,\ldots$, and $\gamma$ refers to the spectra containing the weights $[s,0,\ldots,0]$ with each even integer spin $s=0,2,4,\ldots$. The result on the right-hand side, where ${\cal Z}_0(\beta)$ is the scalar one-particle partition function given in (\ref{Zscfer}), precisely agrees with the singlet sector CFT calculation \cite{Shenker:2011zf, Giombi:2014yra}. 

The bulk Casimir energy can be obtained by inserting the right-hand side of (\ref{thermaltypea1}) and (\ref{thermaltypea2}) into (\ref{Casi}) (alternatively, one 
may compute the Casimir contributions spin by spin, and sum up at the end). One finds that $[\mathcal{Z}_{0}(q)]^2$ 
gives zero contribution to the Casimir energy,\footnote{This is because of symmetry under $q\rightarrow 1/q$. Any function symmetric under this exchange gives 
a zero contribution under the integral in (\ref{Casi}).} while $\mathcal{Z}_{0}(q^2)$ gives a contribution equal to $2E_{c,0}$. Then, $E_{c,{\rm type~A}} =0$ and $E_{c,{\rm min.~type~A}} =E_{c,0}$, 
consistently with the expected shift of $G_N$ deduced from the $S^d$ calculations.

\paragraph{Type B Theories}
In the type B theories and their various truncations, we find
\begin{alignat}{3}
&\text{Non-Minimal Type B:}\qquad&\sum_{\alpha} {\cal Z}_\alpha(q)&= [\mathcal{Z}_{\frac{1}{2}}(q)]^2,\label{thermaltypeb1}\\
&\text{Weyl-Projection:}\qquad&\sum_{\gamma}  {\cal Z}_\gamma(q)&= \frac{1}{4} [\mathcal{Z}_{\frac{1}{2}}(q)]^2,\label{thermaltypeb2}\\
&\text{Minimal Type B:}\qquad&\sum_{\delta}   {\cal Z}_\delta(q)&= \begin{cases} \frac{1}{2} \Big[ [\mathcal{Z}_{\frac{1}{2}}(q)]^2- \mathcal{Z}_{\frac{1}{2}}(q^2) \Big], \qquad& \text{for }O(N), \\
\frac{1}{2} \Big[ [\mathcal{Z}_{\frac{1}{2}}(q)]^2+ \mathcal{Z}_{\frac{1}{2}}(q^2) \Big], \qquad & \text{for }U\!Sp(N), \\
\end{cases}\label{thermaltypeb3} \\
&\text{Majorana-Weyl:}\qquad&\sum_{\epsilon}  {\cal Z}_\epsilon(q)&= \frac{1}{2} \left(\frac{1}{4}[\mathcal{Z}_{\frac{1}{2}}(q)]^2-\frac{1}{2}\mathcal{Z}_{\frac{1}{2}}(q^2) \right) \label{thermaltypeb4} \\
&\text{Symplectic Majorana-Weyl:}\qquad&\sum_{\kappa}   {\cal Z}_\kappa(q)&= \frac{1}{2} \left(\frac{1}{4}[\mathcal{Z}_{\frac{1}{2}}(q)]^2+\frac{1}{2}\mathcal{Z}_{\frac{1}{2}}(q^2) \right) \label{thermaltypeb5}
\end{alignat}
\indent where $\alpha, \gamma, \delta$ are the spectra given by \eqref{dolan1},~\eqref{eqn:typebweyl},~\eqref{eqn:typeBOn}-(\ref{eqn:typeBUSp}), 
and $\epsilon$, $\kappa$ the Majorana-Weyl truncations discussed in \ref{MWsec}.
The right-hand side of all the above equations, with ${\cal Z}_{\frac{1}{2}}(q)$ given in (\ref{Zscfer}), is again in precise agreement with the thermal calculations in the singlet sector of the fermionic CFT (with the relevant fermion projection and gauge group).
As an explicit example, in AdS$_{11}$, we have
\begin{alignat}{3}
&\text{Non-Minimal Type B:}\qquad& \sum_\alpha {\cal Z}_\alpha(q)&= \frac{1024 q^9}{(q-1)^{18}} = \left(\frac{32 q^{9/2}}{(1-q)^{9}}\right)^2,\\
&\text{Weyl-Projection:}\qquad& \sum_\gamma {\cal Z}_\delta(q)&= \frac{256 q^9}{(q-1)^{18}},\\
&\text{Minimal Type B:}\qquad& \sum_\delta {\cal Z}_\gamma(q)&= \frac{1}{2} \left(\frac{1024 q^9}{(1-q)^{18}}-\frac{32 q^9}{\left(1-q^2\right)^9}\right), \\
&\text{Majorana-Weyl:}\qquad&\sum_\epsilon {\cal Z}_\epsilon(q) & = \frac{1}{2} \left(\frac{256 q^9}{(1-q)^{18}}-\frac{16 q^9}{\left(1-q^2\right)^9}\right), \label{eqn:weylmajorana11}
\end{alignat}
which all agree with the formulas in \eqref{thermaltypeb1}-\eqref{thermaltypeb5}. For instance, using the spectrum found in Table~\ref{Table:O(n)projection}, the explicit computations for the Majorana-Weyl case is as follows:
\begin{alignat}{3}
[s,1,1,1,1]:&\quad& &\sum_{s=2,4,6,\ldots} \frac{q^{s+8}}{576(1-q)^{10}}\left[\frac{(s+8)!}{(s+4)(s-1)!}-q\frac{(s+7)!}{(s+3)(s-2)!}\right] \nonn \\ && \qquad& = \frac{q^8}{(q-1)^{18} (q+1)^9} \big(-q^{13}-q^{12}+8 q^{11}+134 q^{10}+98 q^9+3914 q^8+2948 q^7\nonn \\
&\quad&&\qquad+12984 q^6+4983 q^5+8799 q^4+924 q^3+1050 q^2\big) \label{eqn:weylmajorana11pt1}\\
[s,1,1,0,0]:&\qquad& &\sum_{s=1,3,5,\ldots} \frac{q^{s+8}}{720(1-q)^{10}}\left[\frac{(s+4)(s+8)!}{(s+2)(s+6)(s-1)!}-q\frac{(s+3)(s+7)!}{(s+1)(s+5)(s-2)!}\right] \nonn
\\ && \qquad& = \frac{q^7}{(q-1)^{18} (q+1)^9} \big(q^{18}+q^{17}-8 q^{16}-8 q^{15}+29 q^{14}+29 q^{13}-64 q^{12}-64 q^{11} +1043 q^{10} \nonn \\
&\qquad&&\qquad+923 q^9+6992 q^8+3760 q^7+10039 q^6+2407 q^5+3352 q^4+120 q^3+120 q^2\big) \label{eqn:weylmajorana11pt2} \\
[s,0,0,0,0]: &\quad& &\sum_{s=2,4,6,\ldots} \frac{q^{s+8}}{20160(1-q)^{10}}\left[\frac{(s+4)(s+7)!}{s!}-q\frac{(s+3)(s+6)!}{(s-1)!}\right] \nonn 
\\ && \quad& = \frac{q^8 }{(q-1)^{18} (q+1)^9}\big(-q^{17}-q^{16}+8 q^{15}+8 q^{14}-28 q^{13}-28 q^{12}+56 q^{11}+66 q^{10}\nonn \\
&\quad&&\qquad-61 q^9+59 q^8+140 q^7+392 q^6+98 q^5+218 q^4+44 q^3+54 q^2\big) \label{eqn:weylmajorana11pt3}
\end{alignat}
Summing \eqref{eqn:weylmajorana11pt1}-\eqref{eqn:weylmajorana11pt3} up, we obtain \eqref{eqn:weylmajorana11}.

The bulk one-loop contribution to the Casimir energy in Type B theories can be obtained by inserting the right-hand sides of (\ref{thermaltypeb1})--(\ref{thermaltypeb5}) into (\ref{Casi}). The only non-zero contribution comes from $\mathcal{Z}_{\frac{1}{2}}(q^2)$, which yields $2E_{c,1/2}$. Then, we 
see that the bulk one-loop Casimir energies in all variants of the type B theories are in precise agreement with the shifts of the coupling constant summarized in Table~\ref{Table:summary}.  Note that in odd $d$ we get zero Casimir energy on both CFT and bulk sides, as it should be, so this calculation does not shed light on the anomalous 
shifts we encountered in type B theories in odd $d$. A few explicit values of the bulk one-loop Casimir energies are collected in Table~\ref{table:typebcasimir}. 
\begin{table}[t!]
  \centering
  \footnotesize
    \begin{tabular}{c|c|c|c|c}
   \footnotesize
 \cellcolor[HTML]{000000}{\color[HTML]{FFFFFF}\textbf{$d$}} & \cellcolor[HTML]{000000}{\color[HTML]{FFFFFF}\textbf{Non-Minimal}}  & \cellcolor[HTML]{000000}{\color[HTML]{FFFFFF}\textbf{Weyl}}& \cellcolor[HTML]{000000}{\color[HTML]{FFFFFF}\textbf{Minimal ($O(N)$/$USp(N)$)}} & \cellcolor[HTML]{000000}{\color[HTML]{FFFFFF}\textbf{Majorana-Weyl}} 
\\\hline\hline & & \cellcolor[HTML]{EFEFEF}& & \cellcolor[HTML]{EFEFEF}\\
     3     & 0     & \cellcolor[HTML]{EFEFEF}      & 0     &   \cellcolor[HTML]{EFEFEF} \\ & & \cellcolor[HTML]{EFEFEF}& & \cellcolor[HTML]{EFEFEF} \\[-0.5ex]\hline & & & & \cellcolor[HTML]{EFEFEF}\\  
    4     & 0     & 0     &  $\displaystyle \frac{17}{960}$    & \cellcolor[HTML]{EFEFEF} \\ & & & &\cellcolor[HTML]{EFEFEF} \\[-0.5ex]\hline & &\cellcolor[HTML]{EFEFEF} & &\cellcolor[HTML]{EFEFEF} \\
    5     & 0     &  \cellcolor[HTML]{EFEFEF}     & 0     &  \cellcolor[HTML]{EFEFEF}  \\& &\cellcolor[HTML]{EFEFEF} & &\cellcolor[HTML]{EFEFEF}  \\[-0.5ex]\hline & & & & \cellcolor[HTML]{EFEFEF}\\
    6     & 0     & 0     &  $\displaystyle\frac{367}{48384}$     &\cellcolor[HTML]{EFEFEF} \\& & & & \cellcolor[HTML]{EFEFEF} \\[0.5ex]\hline & &\cellcolor[HTML]{EFEFEF} & &\cellcolor[HTML]{EFEFEF} \\
    7     & 0     &  \cellcolor[HTML]{EFEFEF}     & 0     &  \cellcolor[HTML]{EFEFEF}  \\& &\cellcolor[HTML]{EFEFEF} & &\cellcolor[HTML]{EFEFEF} \\[-0.5ex]\hline & & & & \\
    8     & 0     & 0     &   $\displaystyle\frac{27859}{8294400}$    & $\displaystyle\frac{27859}{16588800}$  \\& & & &  \\[-0.5ex]\hline & &\cellcolor[HTML]{EFEFEF} & & \cellcolor[HTML]{EFEFEF}\\
    9     & 0     &   \cellcolor[HTML]{EFEFEF}    & 0     & \cellcolor[HTML]{EFEFEF} \\& & \cellcolor[HTML]{EFEFEF}& &\cellcolor[HTML]{EFEFEF} \\[-0.5ex]\hline & & & & \\
    10    & 0     & 0    &   $\displaystyle-\frac{12950803}{851558400}$    & $\displaystyle-\frac{12950803}{1703116800}$  \\& & & & \\[-0.5ex]\hline & & \cellcolor[HTML]{EFEFEF}& &\cellcolor[HTML]{EFEFEF} \\
    11    & 0     &   \cellcolor[HTML]{EFEFEF}    & 0     &  \cellcolor[HTML]{EFEFEF}\\  & & \cellcolor[HTML]{EFEFEF}& &\cellcolor[HTML]{EFEFEF} \\
    \end{tabular}
      \caption{Type B Casimir Energies. The grey boxes indicate that the particular type of fermion is not defined for the given dimension.}
  \label{table:typebcasimir}
\end{table}

\paragraph{Type AB Theories}
In the purely fermionic sector of the type AB theories, the only representations are given by the weights $[s,\frac{1}{2},\ldots,\frac{1}{2}]$, 
which lead to a simple computation that gives for a generic $d$,
\begin{align}
\mathcal{Z}_{\rm type~AB~ferm}(q)&=\frac{q^{d-\frac{3}{2}}}{(1-q)^d} g_{[1/2,1/2,\ldots,1/2]}
+\sum_{s=\frac{3}{2},\frac{5}{2},\cdots}   \frac{q^{s+d-2}}{(1-q)^d} \Big[g_{[s,1/2,\ldots,1/2]}-q g_{[s-1,1/2,\ldots,1/2]}\Big] \nonn\\
&= \frac{2^{\lfloor\frac{d}{2}\rfloor} q^{d-\frac{3}{2}}(1+q)}{(1-q)^{2(d-1)}} = {\cal Z}_0(q) {\cal Z}_{\frac{1}{2}}(q).
\end{align} 
A quick calculation gives us $E_c=0$ for the contribution of the fermionic tower in the Type AB theories, 
which is nicely consistent with what we obtained in the $S^d$ calculations, namely that there are no shifts due to the purely fermionic fields.

\paragraph{Type C Theories}
In type C theories, summing up over the relevant bulk spectra given in Section \ref{typeC-sec}, we find 
\begin{alignat}{3}
&\text{Non-Minimal Type C:}\qquad&\sum_\alpha {\cal Z}_\alpha(q)&= [\mathcal{Z}_{\frac{d}{2}\text{-form}}(q)]^2,\label{thermaltypec1}\\
&\text{$U(N)$ Self-Dual:}\qquad&\sum_\gamma {\cal Z}_\delta(q)&= \frac{1}{4} [\mathcal{Z}_{\frac{d}{2}\text{-form}}(q)]^2,\label{thermaltypec2}\\
&\text{Minimal Type C:}\qquad& \sum_\delta {\cal Z}_\gamma(q)&= \frac{1}{2} \Big[ [\mathcal{Z}_{\frac{d}{2}\text{-form}}(q)]^2+ \mathcal{Z}_{\frac{d}{2}\text{-form}}(q^2) \Big],\label{thermaltypec3} \\
&\text{$O(N)$ Self-Dual:}\qquad& \sum_\epsilon {\cal Z}_\epsilon(q)&= \frac{1}{2} \Big[ \frac{1}{4}\left(\mathcal{Z}_{\frac{d}{2}\text{-form}}(q)\right)^2+ \frac{1}{2}\mathcal{Z}_{\frac{d}{2}\text{-form}}(q^2) \Big],\label{thermaltypec4}
\end{alignat}
where ${\cal Z}_{\frac{d}{2}\text{-form}}(q)$ is the one-particle partition function (\ref{Zpform}) of a single real $(d/2-1)$-form gauge field. The results on the right-hand side have 
the correct structure expected from the CFT thermal free energy in the $U(N)/O(N)$ singlet sector of the theory of $N$ differential form gauge fields. 
This calculation was carried out explicitly in the $S^1\times S^3$ case in \cite{Beccaria:2014zma}, and we expect it to generalize to all $d$.

The one-loop Casimir energies of Type C theories can be obtained by plugging the right-hand side of (\ref{thermaltypec1})--(\ref{thermaltypec4}) into (\ref{Casi}). The calculation can be 
simplified by noting that, due to the symmetry properties under $q\rightarrow 1/q$, the term $[\mathcal{Z}_{\frac{d}{2}\text{-form}}(q)]^2$ contributes 
$2(-1)^{d/2}E_{c,d/2-{\rm form}}$ after the integration in (\ref{Casi}),\footnote{See Appendix D of \cite{Beccaria:2014qea} for a discussion of this.} and $\mathcal{Z}_{\frac{d}{2}\text{-form}}(q^2)$ contributes $2E_{c,d/2-{\rm form}}$. Then we see that in all cases the one-loop Casimir energies in the bulk are consistent with the shifts 
of the coupling constant summarized in Table \ref{Table:summary}. A few explicit values are reported in Table~\ref{table:typeccasimir}.
\begin{table}[t]
\begin{center}
\footnotesize
\begin{tabular}{c|c|c|c|c}
\cellcolor[HTML]{000000}{\color[HTML]{FFFFFF}\textbf{$d$}} & \cellcolor[HTML]{000000}{\color[HTML]{FFFFFF}\textbf{Non-Minimal $U(N)$}}  & \cellcolor[HTML]{000000}{\color[HTML]{FFFFFF}\textbf{Self-dual $U(N)$}}& \cellcolor[HTML]{000000}{\color[HTML]{FFFFFF}\textbf{Minimal $O(N)$}} & \cellcolor[HTML]{000000}{\color[HTML]{FFFFFF}\textbf{Self-dual $O(N)$}} \\ \hline\hline & & & &  \cellcolor[HTML]{EFEFEF} \\
4 & $\displaystyle\frac{11}{60} $ & $\displaystyle\frac{11}{240} $ & $\displaystyle\frac{11}{60} $ &  \cellcolor[HTML]{EFEFEF} \\ & & & & \cellcolor[HTML]{EFEFEF}  \\\hline  & & & &   \\
6 & $\displaystyle\frac{191}{1008} $ & $\displaystyle\frac{191}{4032} $ & 0 & $\displaystyle-\frac{191}{8064} $ \\ & & & & \\\hline  & & & &  \cellcolor[HTML]{EFEFEF} \\
8 & $\displaystyle\frac{2497}{12960} $  & $\displaystyle\frac{2497}{51840} $  & $\displaystyle\frac{2497}{12960} $ &  \cellcolor[HTML]{EFEFEF} \\ & & & &  \cellcolor[HTML]{EFEFEF}\\\hline & & & &  \\
10 & $\displaystyle\frac{14797}{76032} $  & $\displaystyle\frac{14797}{304128} $  & $0 $ & $\displaystyle-\frac{14797}{608256}$ \\ & & & & \\\hline & & & &  \cellcolor[HTML]{EFEFEF} \\
12 & $\displaystyle\frac{92427157}{471744000} $  & $\displaystyle\frac{92427157}{1886976000}$  & $\displaystyle\frac{92427157}{471744000} $ &  \cellcolor[HTML]{EFEFEF} \\ & & & &  \cellcolor[HTML]{EFEFEF}\\\hline & & & &   \\
14 & $\displaystyle\frac{36740617}{186624000} $  & $\displaystyle\frac{36740617}{746496000}$  & $0 $ & $\displaystyle-\frac{36740617}{1492992000}$ \\ & & & & \\\end{tabular}
\end{center}
\caption{Type C Casimir Energies. The grey boxes indicate that the particular type of $p$-form is not defined for the dimension.}
\label{table:typeccasimir}
\end{table}%

\section{Some technical details on the one-loop calculations in Hyperbolic space}
\subsection{Hurwitz Zeta regularization}\label{appendix:zetas}
To implement $\zeta$-function regularization, we identify the conventionally divergent term $\sum_{s=1}^\infty 1/(s+\nu)^k$ as $\sum_{s=0}^\infty 1/(s+\nu+1)^k$, and treating it as the Hurwitz zeta function,
	\begin{align}
		\zeta(k,\beta) \equiv \sum_{n=0}^\infty\frac{1}{(n+\beta)^k},\label{eqn:appenzeta}
	\end{align}
	where we then analytically extend to the full complex plane. This allows us to regulate systematically the sums to obtain their finite contributions.

Suppose we want to start summing all integer spins $s\geq \ell\geq0$, then,
\begin{align}\sum_{s=\ell}^\infty {1\over (s+\nu)^k} = \zeta(k,\ell+\nu).
\end{align} This is the convention we applied in this paper, and avoids potential inconsistencies that can occur with the Hurwitz zeta function.
 We might also consider sums that only incorporate a particular subset of spins, such as either all odd integer spins or all even integer spins. To do so, we can transform the summing variable of the original Hurwitz zeta function appropriately. We give two examples:

To sum over all even spins, consider
\begin{align}
\sum_{s=2,4,6,\ldots}^\infty {1\over (s+\nu)^k} = \sum_{s=1}^\infty {1\over (2s+\nu)^k} = \sum_{s=1}^\infty \frac{2^{-k}}{\left(s+\frac{\nu}{2}\right)^k} = \sum_{s=0}^\infty \frac{2^{-k}}{\left(s+\frac{\nu}{2}+1\right)^k} = 2^{-k}\zeta\left(k,\frac{\nu}{2}+1\right).
\end{align}
A similar scheme for summing over all odd spins is
\begin{align}
\sum_{s=1,3,5,\ldots}^\infty {1\over (s+\nu)^k} = \sum_{s=1}^\infty {1\over (2s-1+\nu)^k} = \sum_{s=1}^\infty \frac{2^{-k}}{\left(s+\frac{\nu-1}{2}\right)^k}  = 2^{-k}\zeta\left(k,\frac{\nu-1}{2}+1\right).
\end{align}

Unlike conventional summation where rearrangement of terms may lead to problems, $\zeta$-function regularization allows for rearrangement. In particular, the $\zeta$-function satisfies,\begin{align}
\sum_{s=2,4,6,\ldots}^\infty {1\over (s+\nu)^k} + \sum_{s=1,3,5,\ldots}^\infty {1\over (s+\nu)^k} = \zeta(k,\nu+1),
\end{align} which allows us to obtain the regularization over both the odd or the even integer spins by just doing one of the two calculation.

\subsection{Identity for odd $d$ free energy calculations} 
\label{appendix:recursive}
The relationship described in \eqref{eqn:Apmrecursive} can be derived by
\begin{align}
A_k^\pm(x)&=\int_0^\infty \frac{u^k}{e^{2\pi u}\pm1}\frac{du}{u^2+x} \nonn\\
& = \pd{}{a}\Bigg[\int_0^\infty du\ \frac{u^{k-2}}{e^{2\pi u}\pm1}\log[au^2+x]\Bigg]_{a=1} \nonn\\
& =  \pd{}{a}\Bigg\{\log(a) \int_0^\infty du\ \frac{u^{k-2}}{e^{2\pi u}\pm1} + \int_0^\infty du\ \frac{u^{k-2}}{e^{2\pi u}\pm1}\log\left[u^2+\frac{x}{a}\right]\Bigg\}_{a=1} \nonn\\
& = \int_0^\infty du\ \frac{u^{k-2}}{e^{2\pi u}\pm1} -x\int_0^\infty \frac{u^{k-2}}{e^{2\pi u}\pm1}\frac{du}{u^2+x} \nonn\\
& = B_{k-2}^{\pm}(x) -xA_{k-2}^\pm(x).
\end{align}

\subsection{Evaluating $\zeta_{\Delta,\alpha}'(0)$}
\label{zetap0-details}
Here we collect some details on the evaluation of the term $\partial_z\zeta_{(\Delta; \alpha_s)}^{\rm exp}(z)|_{z=0}$ in (\ref{eqn:HSpartexpz}), 
in the explicit example of the type A theory in AdS$_4$. The calculations in the other theories studied in this paper go through in a similar way.  
After some integral identities and algebraic manipulations, we may write 
%
\begin{align}
 \pd{}{z}\zeta_{(\Delta; \alpha_s)}^{\rm exp} (z)\big|_{z=0} = \zeta_{(\Delta; \alpha_s)}^{\rm exp-log-1'} (0)+\zeta_{(\Delta; \alpha_s)}^{\rm exp-log-2'} (0) + \zeta_{(\Delta; \alpha_s)}^{\rm exp-const'} (0)+ \zeta_{(\Delta; \alpha_s)}^{\rm exp-\psi'} (0). \label{eqn:diffHSzeroexp}
\end{align}
The only overall non-zero contribution will come from the fourth term, $\zeta_{(\Delta; \alpha_s)}^{\rm exp-\psi'} (0)$, and the contributions of the first three will cancel out, after taking into account the ghost modes and all other particles in the entire spectra of the theory. 

To understand what these three terms are, let's return to the Type A non-minimal theory, the l.h.s. of \eqref{eqn:diffHSzeroexp} is now,
\begin{align}
 \pd{}{z}\zeta_{(\Delta; \alpha_s)}^{\rm exp} (z)\big|_{z=0}=\int_0^\infty du \frac{(2 s+1) u^3 \left((2s+1)^2+u^2\right) \log \left(\left(\Delta -\frac{3}{2}\right)^2+u^2\right)}{12 \left(e^{2 \pi  u}+1\right)}.
\end{align}
Using \eqref{eqn:evendlogintegral}, we can rewrite the above term into:
\begin{align}
\underbrace{\int_0^\infty du \frac{(2 s+1) u^3 \left((2s+1)^2+u^2\right)}{12 \left(e^{2 \pi  u}+1\right)}\log(u^2)}_{=\zeta_{(\Delta; \alpha_s)}^{\rm exp-log-1'} (0) } +  \int_0^\infty du\int_0^{(s -\frac{1}{2})^2} dx \frac{(2 s+1) u^3 \left((2s+1)^2+u^2\right)}{12 \left(e^{2 \pi  u}+1\right)} \frac{1}{u^2+x}
\end{align}
where $\Delta^{\rm ph} -\frac{3}{2} = s-\frac{1}{2}$.
The second term can then be explicitly integrated using the recursive relation for $\int_0^k dx \frac{u^k}{e^{2\pi i u}\pm 1}\frac{1}{u^2+x}$ found in Appendix~\ref{appendix:recursive},
\begin{align}
&\quad\int_0^{(s-\frac{1}{2})^2} dx \int_0^\infty du\ \frac{(2 s+1) u^3 \left((2s+1)^2+u^2\right)}{12 \left(e^{2 \pi  u}+1\right)} \frac{1}{u^2+x} \nonn \\
& = \underbrace{\frac{1}{2}\int_0^{(s-\frac{1}{2})^2} dx \log \left(x\right) \left(\frac{1}{6} x  (2 s+1)-\frac{1}{6} \left(s+\frac{1}{2}\right)^2 (2 s+1)\right)}_{\equiv \zeta_{(\Delta^{\rm ph}; [s])}^{\rm exp-log-2'} (0)}  + \underbrace{\frac{1}{3}B_1^{+}(2 s+1)}_{\equiv \zeta_{(\Delta^{\rm ph}; [s])}^{\rm exp-const'} (0)} \nonn \\
&\quad+ \underbrace{\int_0^{(s-\frac{1}{2})^2} dx\ \psi\left(\sqrt{x }+\frac{1}{2}\right) \left(\frac{1}{6} x  (-2 s-1)+\frac{1}{6} (2 s+1) \left(s+\frac{1}{2}\right)^2\right)}_{\equiv \zeta_{(\Delta^{\rm ph}; [s])}^{\rm exp-\psi'}(0)},
\end{align}
where $B_k^\pm :=\int_0^\infty du \frac{u^k}{e^{2\pi u}\pm1}$, and $\psi(x)$ is the digamma function $\psi(x) = \Gamma'(x)/\Gamma(x)$. We concentrate on the last term including the digamma function, since it is the \emph{only} term that contributes to the final partition function. To integrate the digamma function, we make use of its integral representation
\begin{align}
\psi (x) = \int_0^{\infty} \left(\frac{e^{-t}}{t}-\frac{e^{-xt}}{1-e^{-t}}\right)dt \label{eqn:psiint}
\end{align}
so that we get
\begin{align}
\zeta_{(\Delta^{\rm ph}; [s])}^{\rm exp-\psi'}(0) 
 & = \int_0^\infty dt \int_0^{(s-\frac{1}{2})^2} dx\ \left(\frac{e^{-t}}{t}-\frac{e^{-(\sqrt{x }+\frac{1}{2})}}{1-e^{-t}}\right)  \left(-\frac{1}{6} x  (2 s+1)+\frac{1}{24} (2 s+1)^3 \right) \nonn \\
 & =  \int_0^\infty dt\ \frac{(2 s+1) e^{-t}}{24 \left(e^t-1\right) t}\Bigg\{e^{-st+2t} \left[-\frac{4(4 s^2-8 s+1)}{t}+16 s^2+\frac{24-48 s}{t^2}-8 s-\frac{48}{t^3}\right]\nonn \\
&\qquad\qquad\qquad+ \frac{1}{4} \left(1-4 s^2\right)^2 \left(e^t-1\right)-\frac{2 e^{\frac{3 t}{2}}}{t^3} \left[(2 s t+t)^2-24\right]-\frac{1}{8} (1-2 s)^4 \left(e^t-1\right)\Bigg\}
\end{align}
The terms in the integrand above split into those that include a prefactor of $e^{-st}$, and those that do not. For the terms with the prefactor, we can sum over the spins easily and without a regulator,
\begin{align}
&\quad\sum_{s=1}^\infty\frac{(2 s+1) e^{-t}}{24 \left(e^t-1\right) t}\Bigg\{e^{-st+2t} \left[16 s^2-8 s+\frac{-16 s^2+32 s-4}{t}+\frac{24-48 s}{t^2}-\frac{48}{t^3}\right]\Bigg\}
\nonn \\
&=\frac{e^t}{6 \left(e^t-1\right)^5 t^4}\Bigg[ t^2+3 e^{3 t} \left(2 t^3+3 t^2-6 t-12\right)+e^t \left(6 t^3-17 t^2+42 t-60\right)\nonn \\
&\qquad\qquad+e^{2 t} \left(36 t^3-41 t^2-18 t+84\right)-6 t+12\Bigg]\label{eqn:zetaprimeads4exp}.
\end{align}
For those terms without the prefactor, we sum using the same regulator as in the previous segment,
\begin{align}
&\quad\sum_{s=1}^\infty \left(s-\frac{1}{2}\right)^{-\epsilon}
\frac{(2 s+1) e^{-t}}{24 \left(e^t-1\right) t}\Bigg\{ \frac{1}{4} \left(1-4 s^2\right)^2 \left(e^t-1\right)-\frac{2 e^{\frac{3 t}{2}}}{t^3} \left[(2 s t+t)^2-24\right]-\frac{1}{8} (1-2 s)^4 \left(e^t-1\right)\Bigg\} \nonn \\
& = \frac{e^{t/2}}{6 \left(e^t-1\right) t^4}-\frac{113 e^{t/2}}{1440 \left(e^t-1\right) t^2}+\frac{1609 e^{-t}}{241920 \left(e^t-1\right) t}-\frac{1609}{241920 \left(e^t-1\right) t}. \label{eqn:zetaprimeads4exp2}
\end{align}
Combining \eqref{eqn:zetaprimeads4exp} and \eqref{eqn:zetaprimeads4exp2} under the integrand, we obtain the expression for $\zeta_{\Delta^{\rm ph};[s]}^{\rm exp-\psi'}(0)$.
Then, repeating the calculations for the ghost calculations, we obtain
\begin{align}
\zeta_{(\Delta^{\rm gh}; [s-1])}^{\rm exp-\psi'}(0)  &= \int_0^\infty dt \Bigg[\frac{13 e^{t/2}}{6 \left(e^t-1\right) t^4}+\frac{2}{\left(e^t-1\right)^2 t^4}-\frac{4 e^t}{\left(e^t-1\right)^3 t^4}-\frac{4 e^t \left(e^t+1\right)}{\left(e^t-1\right)^4 t^3}+\frac{1}{\left(e^t-1\right)^2 t^3}\nonn \\
&\qquad\qquad\quad-\frac{2 e^t \left(e^t+1\right)}{\left(e^t-1\right)^4 t^2}-\frac{233 e^{t/2}}{1440 \left(e^t-1\right) t^2}+\frac{1}{6 \left(e^t-1\right)^2 t^2}+\frac{e^t}{\left(e^t-1\right)^3 t^2}\nonn \\
&\qquad\qquad\quad-\frac{4 e^t \left(4 e^t+e^{2 t}+1\right)}{3 \left(e^t-1\right)^5 t^2}+\frac{349 e^{-t}}{241920 \left(e^t-1\right) t}-\frac{349}{241920 \left(e^t-1\right) t}+\frac{e^t}{3 \left(e^t-1\right)^3 t}\nonn \\
&\qquad\qquad\quad-\frac{4 e^t \left(4 e^t+e^{2 t}+1\right)}{3 \left(e^t-1\right)^5 t}\Bigg].
\end{align}
After combining these above with the integral representation for the scalar term, we then make use of the integral representation of the Hurwitz-Lerch transcendental function,
\begin{align}
\Phi(z,s,\nu) = \frac{1}{\Gamma(s)}\int_0^\infty dt \frac{t^{s-1}e^{-\nu t}}{1-ze^{-t}}=\sum_0^\infty (n+\nu )^{-s}z^n,
\end{align}
to transform the expressions into sums of derivatives of Hurwitz-Lerch transcendental functions\footnote{ This makes use of the identity,
\begin{align}
\frac{1}{(1-e^{-t})^{n+1}(1+e^{-t})^{m+1}}=\frac{(-1)^n}{n!m!}\partial_{z_1}^n\partial_{z_2}^m\left[\frac{1}{z_1-z_2}\left(\frac{1}{z_1-e^{-t}}-\frac{1}{z_2-e^{-t}}\right)\right]\Bigg|_{z_1=1,z_2=-1}
\end{align}}.
Finally, the Type A non-minimal theory will give us an expression of
\begin{align}
\zeta_{(1;[0])}^{\rm exp-\psi'}(0) + \sum_{s=1}^\infty  \zeta_{(\Delta^{\rm ph};[s])}^{\rm exp-\psi'}(0) - \sum_{s=1}^\infty  \zeta_{(\Delta^{\rm gh};[s-1])}^{\rm exp-\psi'}(0) = 0.
\end{align}

\section{Spectra of Higher Spin Theories and their Free Energy Contributions}\vspace{-10pt}
\subsection{Type B Theories}
\vspace{-10pt}
\renewcommand*{\arraystretch}{1.6}
\begin{table}[h!]
\begin{center}
{\ssmall
\begin{tabular}{ccccc}
\multicolumn{5}{l}{AdS$_5$}                                                                                         \\ \hline\hline
\multicolumn{2}{c|}{\textbf{Towers of Spins}}                                  & \multicolumn{3}{c}{\textbf{Contribution to $F$ from \underline{one} tower summed over:}}              \\ \hline
\multicolumn{2}{c|}{\textbf{($\dph;\alpha$)}}                 & \multicolumn{1}{c|}{\textbf{$s=1,2,3,\ldots$}} &  \multicolumn{1}{c|}{\textbf{$s=2,4,6,\ldots$}}  & \multicolumn{1}{c}{\textbf{$s=1,3,5,\ldots$}}\\ \hline
\multicolumn{1}{c|}{$(\dph;[s,1])$} & \multicolumn{1}{c|}{$\overbrace{\yng(10,1)}^{\mbox{$s$}}$} & \multicolumn{1}{c|}{$\dfrac{1}{180}\log R$}    & \multicolumn{1}{c|}{$\dfrac{13}{360}\log R$}  &  \multicolumn{1}{c}{$-\dfrac{11}{360}\log R$}  \\[1.5ex] \hline
\multicolumn{1}{c|}{$(\dph;[s,0])$} & \multicolumn{1}{c|}{$\overbrace{\yng(10)}^{\mbox{$s$}}$} & \multicolumn{1}{c|}{0}   & \multicolumn{1}{c|}{$\dfrac{1}{90}\log R$}  &  \multicolumn{1}{c}{$-\dfrac{1}{90}\log R$}  \\[1.5ex] \hline
\multicolumn{2}{c}{}                                   &                                                &        &                    \\ 
\hline \hline
\multicolumn{2}{c|}{\textbf{Scalar}}                                      & \multicolumn{3}{c}{\textbf{Contribution to $F$ by one scalar}}                            \\ \hline
\multicolumn{2}{c|}{$( 3;[0,0])$}          \rule[-3ex]{0pt}{7ex}& \multicolumn{3}{c}{$-\dfrac{1}{180} \log R$}                               \\[1.5ex] \hline
\end{tabular}
\caption{\footnotesize Results for Type B theory in AdS$_5$.}  \label{Table:ads5b}}\vspace{-10pt}
{\ssmall
\begin{tabular}{ccccc}
\multicolumn{5}{l}{AdS$_7$}                                                                                         \\ \hline\hline
\multicolumn{2}{c|}{\textbf{Towers of Spins}}                                  & \multicolumn{3}{c}{\textbf{Contribution to $F$ from \underline{one} tower summed over:}}              \\ \hline
\multicolumn{2}{c|}{\textbf{($\dph;\alpha$)}}                 & \multicolumn{1}{c|}{\textbf{$s=1,2,3,\ldots$}} &  \multicolumn{1}{c|}{\textbf{$s=2,4,6,\ldots$}}  & \multicolumn{1}{c}{\textbf{$s=1,3,5,\ldots$}}\\ \hline
\multicolumn{1}{c|}{$(\dph;[s,1,1])$} & \multicolumn{1}{c|}{$\overbrace{\yng(10,1,1)}^{\mbox{$s$}}$} & \multicolumn{1}{c|}{$\dfrac{1}{1512}\log R$}    &  \multicolumn{1}{c|}{$-\dfrac{211}{15120}\log R$} &  \multicolumn{1}{c}{$\dfrac{221}{15120}\log R$} \\[1.6ex] \hline
\multicolumn{1}{c|}{$(\dph;[s,1,0])$} & \multicolumn{1}{c|}{$\overbrace{\yng(10,1)}^{\mbox{$s$}}$} & \multicolumn{1}{c|}{$\dfrac{4}{945}\log R$}    & \multicolumn{1}{c|}{$-\dfrac{2}{315}\log R$}  &  \multicolumn{1}{c}{$\dfrac{2}{189}\log R$}  \\[1.5ex] \hline
\multicolumn{1}{c|}{$(\dph;[s,0,0])$} & \multicolumn{1}{c|}{$\overbrace{\yng(10)}^{\mbox{$s$}}$} & \multicolumn{1}{c|}{$-\dfrac{1}{1512}\log R$}   & \multicolumn{1}{c|}{$-\dfrac{1}{504}\log R$}  &  \multicolumn{1}{c}{$\dfrac{1}{756}\log R$}  \\[1.5ex] \hline
\multicolumn{2}{c}{}                                   &                                                &        &                    \\ 
\hline \hline
\multicolumn{2}{c|}{\textbf{Scalar}}                                      & \multicolumn{3}{c}{\textbf{Contribution to $F$ by one scalar}}                            \\ \hline
\multicolumn{2}{c|}{$(5;[0,0,0])$}           \rule[-3ex]{0pt}{7ex}& \multicolumn{3}{c}{$-\dfrac{4}{945} \log R$}                               \\[1.5ex] \hline
\end{tabular}
\caption{\footnotesize Results for Type B theory in AdS$_7$.}\label{table:ads7b}} \vspace{-10pt}
{\ssmall
\begin{tabular}{ccccc}
\multicolumn{5}{l}{AdS$_{9}$}                                                                                         \\ \hline\hline
\multicolumn{2}{c|}{\textbf{Towers of Spins}}                                  & \multicolumn{3}{c}{\textbf{Contribution to $F$ from \underline{one} tower summed over:}}              \\ \hline
\multicolumn{2}{c|}{\textbf{($\dph;\alpha$)}}                 & \multicolumn{1}{c|}{\textbf{$s=1,2,3,\ldots$}} &  \multicolumn{1}{c|}{\textbf{$s=2,4,6,\ldots$}}  & \multicolumn{1}{c}{\textbf{$s=1,3,5,\ldots$}}\\ \hline
\multicolumn{1}{c|}{$(\dph;[s,1,1,1])$} & \multicolumn{1}{c|}{$\overbrace{\yng(10,1,1,1)}^{\mbox{$s$}}$} & \multicolumn{1}{c|}{$\dfrac{23}{226800}\log R$}    &  \multicolumn{1}{c|}{$\dfrac{3463}{453600}\log R$} &  \multicolumn{1}{c}{$-\dfrac{1139}{151200}\log R$} \\[1.6ex] \hline
\multicolumn{1}{c|}{$(\dph;[s,1,1,0])$} & \multicolumn{1}{c|}{$\overbrace{\yng(10,1,1)}^{\mbox{$s$}}$} & \multicolumn{1}{c|}{$\dfrac{13}{28350}\log R$}    & \multicolumn{1}{c|}{$\dfrac{133}{16200}\log R$}  &  \multicolumn{1}{c}{$-\dfrac{293}{37800}\log R$}  \\[1.5ex] \hline
\multicolumn{1}{c|}{$(\dph;[s,1,0,0])$} & \multicolumn{1}{c|}{$\overbrace{\yng(10,1)}^{\mbox{$s$}}$} & \multicolumn{1}{c|}{$\dfrac{353}{113400}\log R$}    & \multicolumn{1}{c|}{$-\dfrac{1189}{226800}\log R$}  &  \multicolumn{1}{c}{$-\dfrac{23}{10800}\log R$}  \\[1.5ex] \hline
\multicolumn{1}{c|}{$(\dph;[s,0,0,0])$} & \multicolumn{1}{c|}{$\overbrace{\yng(10)}^{\mbox{$s$}}$} & \multicolumn{1}{c|}{$-\dfrac{13}{28350}\log R$}   & \multicolumn{1}{c|}{$-\dfrac{29}{113400}\log R$}  &  \multicolumn{1}{c}{$-\dfrac{23}{113400}\log R$}  \\[1.5ex] \hline
\multicolumn{2}{c}{}                                   &                                                &        &                    \\ 
\hline \hline
\multicolumn{2}{c|}{\textbf{Scalar}}                                      & \multicolumn{3}{c}{\textbf{Contribution to $F$ by one scalar}}                            \\ \hline
\multicolumn{2}{c|}{$(7;[0,0,0,0])$}  \rule[-3ex]{0pt}{7ex}& \multicolumn{3}{c}{{$-\dfrac{9}{2800} \log R$}}              \\ \hline
\end{tabular}
\caption{\footnotesize Results for Type B theory in AdS$_9$.}\label{appendix:ads9b}}
\vspace{-10pt}
\end{center}
\end{table}

\begin{table}[!h]
\begin{center}
{\ssmall
\begin{tabular}{ccccc}
\multicolumn{5}{l}{AdS$_{11}$}                                                                                         \\ \hline\hline
\multicolumn{2}{c|}{\textbf{Towers of Spins}}                                  & \multicolumn{3}{c}{\textbf{Contribution to $F$ from \underline{one} tower summed over:}}              \\ \hline
\multicolumn{2}{c|}{\textbf{($\dph;\alpha$)}}                 & \multicolumn{1}{c|}{\textbf{$s=1,2,3,\ldots$}} &  \multicolumn{1}{c|}{\textbf{$s=2,4,6,\ldots$}}  & \multicolumn{1}{c}{\textbf{$s=1,3,5,\ldots$}}\\ \hline
\multicolumn{1}{c|}{$(\dph;[s,1,1,1,1])$} & \multicolumn{1}{c|}{$\overbrace{\yng(10,1,1,1,1)}^{\mbox{$s$}}$} & \multicolumn{1}{c|}{$\dfrac{263}{14968800}\log R$}    &  \multicolumn{1}{c|}{$-\dfrac{19771}{4276800}\log R$} &  \multicolumn{1}{c}{$\dfrac{138923}{29937600}\log R$} \\[1.6ex] \hline
\multicolumn{1}{c|}{$(\dph;[s,1,1,1,0])$} & \multicolumn{1}{c|}{$\overbrace{\yng(10,1,1,1)}^{\mbox{$s$}}$} & \multicolumn{1}{c|}{$\dfrac{31}{467775}\log R$}    & \multicolumn{1}{c|}{$-\dfrac{2273}{374220}\log R$}  &  \multicolumn{1}{c}{$\dfrac{11489}{1871100}\log R$}  \\[1.5ex] \hline
\multicolumn{1}{c|}{$(\dph;[s,1,1,0,0])$} & \multicolumn{1}{c|}{$\overbrace{\yng(10,1,1)}^{\mbox{$s$}}$} & \multicolumn{1}{c|}{$\dfrac{311}{1069200}\log R$}    & \multicolumn{1}{c|}{$-\dfrac{6599}{2993760}\log R$}  &  \multicolumn{1}{c}{$\dfrac{37349}{14968800}\log R$}  \\[1.5ex] \hline
\multicolumn{1}{c|}{$((\dph;[s,1,0,0,0])$} & \multicolumn{1}{c|}{$\overbrace{\yng(10,1)}^{\mbox{$s$}}$} & \multicolumn{1}{c|}{$\dfrac{1153}{467775}\log R$}   & \multicolumn{1}{c|}{$-\dfrac{3947}{1871100}\log R$}  &  \multicolumn{1}{c}{$\dfrac{19}{53460}\log R$}  \\[1.5ex] \hline
\multicolumn{1}{c|}{$(\dph;[s,0,0,0,0])$} & \multicolumn{1}{c|}{$\overbrace{\yng(10)}^{\mbox{$s$}}$} & \multicolumn{1}{c|}{$-\dfrac{19}{61600}\log R$}   & \multicolumn{1}{c|}{$-\dfrac{5143}{14968800}\log R$}  &  \multicolumn{1}{c}{$\dfrac{263}{7484400}\log R$}  \\[1.5ex] \hline

\multicolumn{2}{c}{}                                   &                                                &        &                    \\ 
\hline \hline
\multicolumn{2}{c|}{\textbf{Scalar}}                                   & \multicolumn{3}{c}{\textbf{Contribution to $F$ by one scalar}}                            \\ \hline
\multicolumn{2}{c|}{$(9;[0,0,0,0,0])$} \rule[-3ex]{0pt}{7ex}& \multicolumn{3}{c}{{$-\dfrac{1184}{467775} \log R$}}              \\ \hline
\end{tabular}
\caption{\footnotesize Results for Type B theory in AdS$_{11}$.}\label{appendix:ads11b}}
\end{center}
\end{table}
\newpage
\subsection{Calculation of $\mathfrak{Z}_{\rm total}^{\rm HS}(z)$ in Type AB Theories} \label{Appendix:AB}
\subsubsection{AdS$_7$}
In this case,
\begin{align}
\mathfrak{Z}_{\rm total}^{\rm HS}(z) & = \frac{z^2\pi }{{86016}}\Big(-11253  \zeta '(-10)+15300   \zeta '(-8)+119658 \zeta '(-6)-137900 \zeta '(-4)+21735   \zeta '(-2)\Big) \nonn\\&\quad+\mathcal{O}\left(z^3\right)
\end{align} which gives us $F_f^{(1)} = 0$, as we set $z\rightarrow0$.
\subsubsection{AdS$_9$}
In this case,
\begin{align}
\mathfrak{Z}_{\rm total}^{\rm HS}(z) = &\frac{\pi}{{16647192576000}} \left[-136525 \zeta '(-14)+1242150 \zeta '(-12)+2651957 \zeta '(-10)\right.\nonn\\&\quad\left.-42097100 \zeta '(-8)+100665453 \zeta '(-6)-71501850 \zeta '(-4)+9993375 \zeta '(-2)\right]z^2\nonn\\&\quad+\mathcal{O}\left(z^3\right)
\end{align}
which gives us $F_f^{(1)}=0$, as we set $z=0$.
\subsection{Free Energy Values for Type C Theories in AdS$_9$}\label{appendix:TypeO}
\begin{table}[h!]
\begin{center}{\ssmall
\begin{tabular}{ccccc}
\multicolumn{5}{l}{AdS$_{9}$ Type C}                                                                                         \\ \hline\hline
\multicolumn{2}{c|}{\textbf{Towers of Spins}}                                  & \multicolumn{3}{c}{\textbf{Contribution to $F$ from \underline{one} tower summed over:}}              \\ \hline
\multicolumn{2}{c|}{\textbf{$(\dph;\alpha)$}}                 & \multicolumn{1}{c|}{\textbf{$s=2,3,4,\ldots$}} &  \multicolumn{1}{c|}{\textbf{$s=2,4,6,\ldots$}}  & \multicolumn{1}{c}{\textbf{$s=3,5,7,\ldots$}}\\ \hline
\multicolumn{1}{c|}{$(\dph;[s,2,2,2])$} & \multicolumn{1}{c|}{$\overbrace{\yng(10,2,2,2)}^{\mbox{$s$}}$} & \multicolumn{1}{c|}{$\dfrac{23}{1800}\log R$}    &  \multicolumn{1}{c|}{$\dfrac{2213}{3600}\log R$} &  \multicolumn{1}{c}{$-\dfrac{2167}{3600}\log R$} \\[1.6ex] \hline
\multicolumn{1}{c|}{$(\dph;[s,2,0,0])$} & \multicolumn{1}{c|}{$\overbrace{\yng(10,2)}^{\mbox{$s$}}$} & \multicolumn{1}{c|}{$\dfrac{3121}{6300}\log R$}    &  \multicolumn{1}{c|}{$\dfrac{14281}{37800}\log R$} &  \multicolumn{1}{c}{$\dfrac{127}{1080}\log R$} \\[1.6ex] \hline
\multicolumn{1}{c|}{$(\dph;[s,2,1,1])$} & \multicolumn{1}{c|}{$\overbrace{\yng(10,2,1,1)}^{\mbox{$s$}}$} & \multicolumn{1}{c|}{$\dfrac{19409}{37800}\log R$}    &  \multicolumn{1}{c|}{$\dfrac{19679}{75600}\log R$} &  \multicolumn{1}{c}{$-\dfrac{19139}{75600}\log R$} \\[1.6ex] \hline
\multicolumn{1}{c|}{$(\dph;[s,2,2,0])$} & \multicolumn{1}{c|}{$\overbrace{\yng(10,2,2)}^{\mbox{$s$}}$} & \multicolumn{1}{c|}{$\dfrac{329}{2700}\log R$}    & \multicolumn{1}{c|}{$-\dfrac{569}{5400}\log R$}  &  \multicolumn{1}{c}{$\dfrac{409}{1800}\log R$}  \\[1.5ex] \hline
\multicolumn{1}{c|}{$(\dph;[s,1,1,0])$} & \multicolumn{1}{c|}{$\overbrace{\yng(10,1,1)}^{\mbox{$s$}}$} & \multicolumn{1}{c|}{$\dfrac{31399}{113400}\log R$}    & \multicolumn{1}{c|}{$\dfrac{133}{16200}\log R$}  &  \multicolumn{1}{c}{$\dfrac{2539}{9450}\log R$}  \\[1.5ex] \hline
\multicolumn{1}{c|}{$(\dph;[s,0,0,0])$} & \multicolumn{1}{c|}{$\overbrace{\yng(10)}^{\mbox{$s$}}$} & \multicolumn{1}{c|}{$\dfrac{35293}{113400}\log R$}   & \multicolumn{1}{c|}{$-\dfrac{29}{113400}\log R$}  &  \multicolumn{1}{c}{$\dfrac{841}{2700}\log R$}  \\[1.5ex] \hline
\multicolumn{2}{c}{}                                   &                                                &        &                    \\ \hline \hline
\multicolumn{2}{c|}{\textbf{Other Particles}}                                   & \multicolumn{3}{c}{\textbf{Contribution to $F$ by one particle}}                            \\ \hline
\multicolumn{2}{c|}{$(8;[1,1,1,1])$}  \rule[-3ex]{0pt}{7ex}& \multicolumn{3}{c}{$-\dfrac{908}{2835} \log R$}             \\\hline
\multicolumn{2}{c|}{$(8;[1,1,0,0])$} \rule[-3ex]{0pt}{7ex}& \multicolumn{3}{c}{$-\dfrac{1856}{14175} \log R$}             \\\hline
\multicolumn{2}{c|}{$(8;[0,0,0,0])$}   \rule[-3ex]{0pt}{7ex}& \multicolumn{3}{c}{$\dfrac{1978}{14175} \log R$}             \\\hline
\end{tabular}
\caption{\footnotesize Results for Type C theory in AdS$_9$.}}
\end{center}
\end{table}
\vfill
\begin{landscape}
\subsubsection{Spectra of Spins for Type C Theories} 
In these following results, $\alpha=[t_1,t_2,\ldots,t_{k-1},t_k]_c$ denote two towers $\alpha=[t_1,t_2,\ldots,t_{k-1},t_k]$ and $\alpha=[t_1,t_2,\ldots,t_{k-1},-t_k]$, which at the level of computation of the spin factor $g_{\alpha}(s)$ and $\mu_\alpha(s)$ are indistinguishable. Hence, a single tower of $[t_1,t_2,\ldots,t_{k-1},t_k]_c$ encompasses one of each of the towers, and correspondingly, a $1/2$ tower encompasses only the $[t_1,t_2,\ldots,t_{k-1},t_k]$ tower (the one with the positive $t_k$).
\begin{table}[h!]
\begin{center}
{\ssmall
\begin{tabular}{lcclcccclllll}
\multicolumn{3}{l}{\textbf{AdS$_{\bf 5}$}}                                                                                                                                                                              &                                                                                    & \multicolumn{1}{l}{}                                                                        &                                                                                             &                                                                                             &                                                                                             &                                                                                             &                                                                                             &                                                                                             &                                                                                             &                                                                                             \\
\hline
\end{tabular}}
\end{center}
\end{table}
\newpage
\begin{table}[h]{\ssmall
\begin{tabular}{lcclcccclllll}
\multicolumn{3}{l}{\textbf{AdS$_{\bf 9}$}}                                                                                                                                                                              &                                                                                    & \multicolumn{1}{l}{}                                                                        & \multicolumn{1}{l}{}                                                                        & \multicolumn{1}{l}{}                                                                        & \multicolumn{1}{l}{}                                                                        &                                                                                             &                                                                                             &                                                                                             &                                                                                             &                                                                                             \\
\end{tabular}}
\end{table}\newpage
\begin{table}{\ssmall
\begin{tabular}{lccclllllllll}
\multicolumn{3}{l}{\textbf{AdS$_{\bf 13}$}}                                                                                                                                                                             & \multicolumn{1}{l}{}                                                               & \multicolumn{1}{l}{}                                                                        & \multicolumn{1}{l}{}                                                                        & \multicolumn{1}{l}{}                                                                        & \multicolumn{1}{l}{}                                                                        &                                                                                             &                                                                                             &                                                                                             &                                                                                             &                                                                                             \\
\hline
\end{tabular}
}
\end{table}
\newpage
\begin{table}[h]
{\ssmall
\begin{tabular}{lccclllllllll}
\multicolumn{3}{l}{\textbf{AdS$_{\bf 15}$}}                                                                                                                                                                             & \multicolumn{1}{l}{}                                                               & \multicolumn{1}{l}{}                                                                        & \multicolumn{1}{l}{}                                                                        & \multicolumn{1}{l}{}                                                                        & \multicolumn{1}{l}{}                                                                        &                                                                                             &                                                                                             &                                                                                             &                                                                                             &                                                                                             \\
\hline
\end{tabular}
}
\end{table}
\end{landscape}

\bibliographystyle{ssg}
\bibliography{Fermionic1Loop}
\end{document}